\documentclass[11pt]{article}
\pdfoutput=1

\usepackage{braket}
\usepackage{color}
\usepackage{mhchem}
\usepackage{xcolor}
\usepackage{colortbl}
\usepackage{hhline}
\usepackage{cite}
\usepackage[colorlinks, linkcolor=red,anchorcolor=black,citecolor=green]{hyperref}
\usepackage[toc,page]{appendix}
\usepackage{mathtools}
\usepackage{amsfonts}
\usepackage{bbold}
\usepackage{textcomp}
\usepackage[DIV13]{typearea}
\usepackage{amsmath, amsthm, amssymb, mathtools,empheq,latexsym,dsfont}
\usepackage{bbm}
\usepackage{amsmath,amssymb,latexsym,dsfont}
\usepackage{slashed, simplewick}
\usepackage[utf8]{inputenc}
\usepackage{graphicx}
\usepackage{graphicx,placeins}
\usepackage{makeidx}
\usepackage[font=small,labelfont=bf]{caption}
\usepackage{nicefrac}
\usepackage{subfigure}
\usepackage{array, bigdelim,multirow,multicol}
\usepackage{mathrsfs}

\usepackage{changepage}
\usepackage{adjustbox}
\usepackage{booktabs}
\usepackage{indentfirst}
\usepackage{booktabs}
\usepackage{tabulary}
\usepackage{longtable}
\usepackage{fancybox}
\usepackage{graphicx}
\usepackage{bm}
\usepackage{bbm}
\usepackage{tikz}
\usepackage{diagbox}
\usepackage{enumitem}
\usepackage[integrals]{wasysym}
\graphicspath{{immagini/}}

\definecolor{Gray}{gray}{0.92}

\newcommand{\ignore}[1]{}

\newcommand{\be}{\begin{equation}}
        \newcommand{\ee}{\end{equation}}
\newcommand{\bea}{\begin{eqnarray}}
        \newcommand{\eea}{\end{eqnarray}}

\definecolor{lightred}{rgb}{1,0.4,0.4}
\definecolor{lightgreen}{rgb}{0.4,1,0.4}
\definecolor{LightCyan}{rgb}{0.88,1,1}

\newcounter{Thm}[section]
\renewcommand{\theThm}{\arabic{section}.\arabic{Thm}}

\newcounter{nodecount}

\tikzstyle{every picture}+=[remember picture,baseline]
\tikzstyle{every node}+=[inner sep=0pt,anchor=base,
text depth=.25ex,outer sep=1.5pt]
\tikzstyle{every path}+=[thick, rounded corners]
\tikzset{plabel/.style={inner sep=2pt}}

\makeatletter
\@addtoreset{equation}{section}
\makeatother

\begin{document}
\title{\begin{center}
{\Large\bf  Non-holomorphic $S^{\prime}_{4}$ modular symmetry for leptons and leptogenesis}
\end{center}}
\date{}
\author{Cai-Chang Li$^{a,b,c,d}$\footnote{E-mail: {\tt ccli@nwu.edu.cn}}\,, \ Jun-Nan Lu$^{e}$\footnote{E-mail: {\tt
hitman@mail.ustc.edu.cn} (corresponding author)}\,,  \
Xiang-Yan Gao$^{a,b,c,d}$\footnote{E-mail: {\tt 202421359@stumail.nwu.edu.cn}}\,,  \
Ming-Hua Weng$^{f}$\footnote{E-mail: {\tt
mhweng@mju.edu.cn}}\,, \ 
Gui-Jun Ding$^{e,g}$\footnote{E-mail: {\tt
dinggj@ustc.edu.cn}}\ \\*[20pt]
\centerline{\begin{minipage}{\linewidth}
\begin{center}
$^a${\it\small School of Physics, Northwest University, Xi'an 710127, China}\\[2mm]
$^b${\it\small Shaanxi Key Laboratory for Theoretical Physics Frontiers, Xi'an 710127, China}\\[2mm]
$^c${\it\small NSFC-SPTP Peng Huanwu Center for Fundamental Theory, Xi'an 710127, China}\\[2mm]
$^d${\it\small Fundamental Discipline Research Center for  Quantum Science and technology of Shaanxi Province, Xi'an 710127, China}\\[2mm]
$^e${\it \small Department of Modern Physics,  and Anhui Center for fundamental sciences in theoretical physics,\\
University of Science and Technology of China, Hefei, Anhui 230026, China}\\[2mm]
$^f${\it \small College of Physics and Electronic Information Engineering, Minjiang University,\\ Fuzhou 350108, China} \\[2mm]
$^g${\it \small  College of Physics, Guizhou University, Guiyang 550025, China}\\[2mm]
\end{center}
\end{minipage}} \\[10mm]}
\maketitle
        
\thispagestyle{empty}
        
\centerline{\large\bf Abstract}

\begin{quote}
\indent
We perform a comprehensive and systematic investigation of lepton models based on the non-holomorphic $S^{\prime}_{4}$ modular symmetry, by using level 4 polyharmonic  Maa{\ss} forms spanning integer weights from $-4$ to $6$. The light neutrino masses are generated by the type-I seesaw mechanism with two right-handed neutrinos, no flavon fields other than the modulus $\tau$ is introduced, and the generalized CP symmetry is not imposed. An exhaustive numerical analysis  yields 36 viable models with only four real couplings besides the modulus $\tau$ when neutrino masses are normal ordering. They are classified into three categories, each containing  twelve models which yield quite similar predictions for lepton observables and are distinguished by the assignment of $E^c_1$. Furthermore, we perform a detailed numerical analysis for one representative model from each category. These representative models are found to yield very sharp predictions for neutrino masses and mixing parameters, and they are distinguished by the predictions for the atmospheric mixing angle $\theta_{23}$, the Dirac CP phase $\delta_{CP}$ and the Majorana CP phase $\alpha_{21}$. Furthermore, we find that only two of these three representative models accommodate successful thermal leptogenesis in the unflavored regime, reproducing the observed baryon asymmetry with the identical parameter values that satisfy neutrino oscillation data. In these models, the real part of the modulus $\tau$ is the unique source of CP violation in both lepton mixing and leptogenesis.
                
\end{quote}
    
\thispagestyle{empty}
\vfill

\clearpage

{\hypersetup{linkcolor=black}
\tableofcontents
}

\section{Introduction}

The Standard Model (SM) successfully describes electroweak interactions but offers no fundamental explanation for the fermion mass hierarchy, the existence of three generations, or the specific patterns of the CKM and PMNS mixing matrices, including the disparity in CP violation between the quark and lepton sectors~\cite{ParticleDataGroup:2026aaa}. These longstanding puzzles motivate the exploration of beyond the SM (BSM) frameworks that can naturally account for the flavor structure of leptons. It has been recognized that large leptonic mixing angles can be explained by extending the SM with a non‑abelian discrete flavor symmetry~\cite{Altarelli:2010gt,Ishimori:2010au,King:2013eh,King:2014nza,King:2017guk,Petcov:2017ggy,Xing:2020ijf,Feruglio:2019ybq,Almumin:2022rml,Ding:2024ozt}. However, in conventional realizations of such symmetries, scalar flavon fields must acquire vacuum expectation values (VEVs) aligned along specific directions in flavor space. Realizing such vacuum alignment dynamically is technically formidable and typically leads to intricate models that offer little quantitative predictivity for fermion masses. In recent years, modular flavor symmetry has emerged as a powerful and economical approach to address these issues, as it determines Yukawa couplings directly from modular forms without introducing auxiliary flavon fields~\cite{Feruglio:2017spp,Ding:2023htn,Kobayashi:2023zzc}. As a result, the complications associated with vacuum alignment of flavon fields are avoided, and the model structure is significantly streamlined. This mechanism not only enhances theoretical economy but also improves predictivity, offering a promising route toward a more fundamental understanding of fermion masses and mixing.

In the standard modular flavor framework, Yukawa couplings are promoted to holomorphic modular forms of a given level $N$ and weight $k$. These functions depend holomorphically on the complex modulus $\tau$ and transform nontrivially under the finite modular group $\Gamma_{N}$~\cite{Feruglio:2017spp} and its double cover $\Gamma^{\prime}_{N}$~\cite{Liu:2019khw}. This construction is inherently formulated within a holomorphic scheme, in which modular forms exist only for non-negative weights. Consequently, theoretical consistency typically requires the presence of supersymmetry~\cite{Lauer:1989ax,Ferrara:1989bc,Ferrara:1989qb,Feruglio:2017spp}. However, supersymmetry has so far eluded experimental confirmation.

To overcome these limitations, a framework of non-holomorphic modular flavor symmetry has recently been proposed~\cite{Qu:2024rns,Qu:2025ddz}. In this approach, it has been demonstrated that the modular flavor mechanism can be consistently formulated without relying on supersymmetry, by replacing holomorphic modular forms with polyharmonic Maa{\ss} forms. These objects satisfy a Laplacian eigenvalue equation~\cite{Green:1997tv,Green:1997me,Pioline:1998mn,Green:1998by,deHaro:2002vk,Green:2010wi,Basu:2011he,Peeters:2000qj,Sinha:2002zr} and are defined for both positive and negative integer weights $k\in \mathbb{Z}$. The transformation property of polyharmonic Maa{\ss} forms under the modular group is completely analogous to that of holomorphic modular forms, ensuring that the invariant construction of the Lagrangian proceeds in the same systematic fashion. Notably, when the weight satisfies $k\geq3$, they coincide exactly with the conventional holomorphic modular forms at the same level. In contrast, for $k\leq 2$, they constitute genuinely new non-holomorphic structures that can serve as building blocks for Yukawa interactions. This non-holomorphic extension liberates the modular flavor program from its supersymmetric origins and dramatically expands the space of viable models, particularly by allowing negative weight modular multiplets that were previously unavailable. A wide range of phenomenologically successful models has been developed using non-holomorphic finite modular groups, including $\Gamma_2\cong S_3$~\cite{Okada:2025jjo}, $\Gamma_3\cong A_4$~\cite{Qu:2024rns,Kumar:2024uxn,Nomura:2024atp,Nomura:2024nwh,Kobayashi:2025hnc,Loualidi:2025tgw,Kumar:2025bfe,Nomura:2025ovm,Nomura:2025raf,Zhang:2025dsa,Priya:2025wdm,Kumar:2025nut,Nanda:2025lem,Jangid:2025thp,Gao:2025jlw,Nasri:2026nbf,Tapender:2026ets,Majhi:2026jdk,Priya:2026ehe,Abbas:2026siw},  $\Gamma_4\cong S_4$~\cite{Ding:2024inn}  and  $\Gamma_5\cong A_5$~\cite{Li:2024svh}. This approach has been further generalized to incorporate odd weight polyharmonic Maa{\ss} forms, which furnish irreducible representations of the homogeneous finite modular groups $\Gamma^{\prime}_{N}$~\cite{Qu:2025ddz}. In this extended setting, modular invariant constructions based on the associated non-holomorphic groups, such as $\Gamma^{\prime}_{3}\cong T^{\prime}$~\cite{Qu:2025ddz,Loualidi:2026pld} and $\Gamma^{\prime}_{5}\cong A^{\prime}_{5}$~\cite{Li:2025kcr,Zhang:2026kyy}, have also been systematically explored.

Among finite modular groups, the level 4 homogeneous modular group $\Gamma^{\prime}_4\cong S^{\prime}_4$ (the double cover of $S_4$) possesses a rich representation structure and has been extensively studied in the context of lepton masses and mixing~\cite{Liu:2019khw,Novichkov:2020eep,Liu:2020akv,Ding:2022nzn,Abe:2023ilq,Abe:2023qmr,Petcov:2026mdx}. Early studies on $S^{\prime}_{4}$ modular symmetry were restricted to the holomorphic framework. The group $S^{\prime}_4$ contains two inequivalent two-dimensional irreducible representations, which makes it particularly suitable for implementing exactly two right-handed neutrinos (2RHNs) in the minimal type-I seesaw mechanism. Motivated by this structural feature, recent developments in non-holomorphic modular flavor symmetry, and phenomenological constraints from neutrino physics and cosmology, we perform a comprehensive and systematic study of non-holomorphic $S^{\prime}_4$ modular lepton models with 2RHNs. 

Our construction is based on the assignment that the three generations of left-handed (LH) lepton doublets transform as an $S^{\prime}_4$ triplet, while the right-handed (RH) charged leptons are placed in singlet representations, and the 2RHNs form an $S^{\prime}_4$ doublet. Without loss of generality, the Higgs doublet is taken to be a trivial singlet with vanishing modular weight. Modular invariance then fixes the admissible polyharmonic Maa{\ss} multiplets that can couple to the matter fields, leading to a finite set of possible mass matrix textures. By scanning over all combinations of modular weights in the range $-4\leq k\leq 6$ and all consistent $S^{\prime}_{4}$ representation assignments, we identify a total of 7280 distinct minimal models, each depending on just four real couplings besides the modulus $\tau$. We then perform an exhaustive numerical analysis of their parameter spaces and find that only 36 models are compatible with current experimental data in the lepton sector. These viable models further organize into three distinct classes, each containing twelve models in which the predictions for lepton mixing parameters and neutrino masses are nearly degenerate.  Furthermore, we present a detailed study of three representative models, one from each class, to illustrate the strong correlations between input parameters and physical observables. Of particular interest is the interplay between $\Re\tau$ and $\delta_{CP}$, as well as the correlations among the three mixing angles. These sharp predictions render the models testable by forthcoming neutrino oscillation experiments.

Besides accounting for the tiny neutrino masses, the type-I seesaw mechanism naturally realizes thermal leptogenesis, which offers a well motivated resolution to the cosmic baryon asymmetry puzzle~\cite{Fukugita:1986hr}. Therefore we further explore the implications for leptogenesis of the phenomenologically viable models identified previously, and examine their capability to reproduce the observed baryon asymmetry of the Universe. In all these constructions, the real part of the modulus $\tau$ acts as the unique source of CP violation, simultaneously governing low energy flavor mixing and the dynamics of leptogenesis. We focus on thermal leptogenesis in the unflavored regime, applicable for $M_{1}>10^{12}$GeV as implied by the hierarchical RH neutrino spectrum. The baryon asymmetry $Y_{B}$ is computed by numerically solving the relevant Boltzmann equations for three representative models. Remarkably, two of them can successfully reproduce the observed baryon asymmetry $Y_{B}=8.703\times 10^{-11}$~\cite{Planck:2018vyg} for the same set of parameter values compatible with the current neutrino oscillation data. This demonstrates a unified description of lepton flavor structure and the cosmic matter antimatter asymmetry within a single set of modular flavor dynamics.

The structure of this paper is as follows. In section~\ref{sec:NHMF}, we review the essential aspects of non-holomorphic modular flavor symmetry and the polyharmonic Maa{\ss} forms. Section~\ref{sec:model} is devoted to the systematic construction of the minimal $S^{\prime}_{4}$ lepton models, including the classification of charged lepton mass matrices and the neutrino seesaw sector.  Furthermore, we present the numerical results of our comprehensive parameter scan, identify 36 phenomenologically viable models. In section~\ref{sec:example_model}, three benchmark models are presented and a thorough numerical analysis is performed to discuss their phenomenology, and examine the predictions for future experimental sensitivities. Section~\ref{sec:leptogenesis} addresses the leptogenesis mechanism within the benchmark scenarios. Section~\ref{sec:conclusion} summarizes our conclusions. The irreducible representations and the related Clebsch-Gordan (CG) coefficients of the finite modular group $S^{\prime}_{4}$ are given in appendix~\ref{sec:S4p_group_theory}.

\section{\label{sec:NHMF}Non-holomorphic modular flavor symmetry}

The special linear group $\Gamma \equiv SL(2,\mathbb{Z})$, also known as the full modular group, consists of $2\times2$ matrices with integer entries and determinant $1$. It is generated by the matrices
\begin{equation}
S=\begin{pmatrix}0&1\\-1&0
\end{pmatrix},\qquad 
T=\begin{pmatrix}1&1\\0&1
\end{pmatrix}\,,
\end{equation}
which satisfy the relations
\begin{equation}
S^{4}=(ST)^{3}=1,\qquad S^{2}T=TS^{2}\,.
\end{equation}
The group $\Gamma$ acts on the upper half-plane $\mathcal{H} = \left\{ \tau  \in \mathbb{C} \left| \mathop{\rm Im} \left( \tau  \right) > 0 \right. \right\}$ via linear fractional transformations. For an element $\gamma  \in \Gamma$, the modulus $\tau$  transforms as
\begin{equation}
\tau\xmapsto{~\gamma~} \dfrac{a\tau+b}{c\tau+d},\qquad 
\gamma =\begin{pmatrix}
a~&b\\
c~&d
\end{pmatrix}\in \Gamma\,.
\end{equation}
In the framework of modular flavor symmetry~\cite{Feruglio:2017spp,Liu:2019khw}, the homogeneous finite modular group $\Gamma^{\prime}_{N}$, which acts as a flavor symmetry, arises from the quotient
\begin{equation}\label{eq:Gamma_N_Def}
\Gamma^{\prime}_N\equiv \Gamma/\Gamma(N)\,,
\end{equation}
where $\Gamma(N)$ denotes the principal congruence subgroup of level $N$ and $N$ can be any positive integer. When $N=1$, $\Gamma(N)$ coincides with the full modular group $\Gamma$. Note $\Gamma(N)$ is a normal subgroup of $\Gamma$ and it contains the element $T^{N}$. Then  finite modular groups $\Gamma^{\prime}_N$ are generated by the modular transformations $S$ and $T$, and they satisfy the following multiplication rules, 
\begin{equation}
S^4=(ST)^3=T^{N}=1, \qquad S^{2}T=TS^{2}\,,
\end{equation}
for $N = 2, 3, 4$ and $5$. The finite modular groups $\Gamma'_2$, $\Gamma'_3$, $\Gamma'_4$ and $\Gamma'_5$ are isomorphic to $S_3$, $T^{\prime}$, $S^{\prime}_4$ and $A^{\prime}_5$, respectively~\cite{Liu:2019khw}.

The polyharmonic Maa{\ss} form $Y(\tau)$ of weight $k$ and level $N$ is a modular function satisfying both modular condition and the harmonic condition, i.e.
\begin{eqnarray}
\label{eq:polyharmonic-def}\left\{\begin{array}{l}
Y(\gamma\tau)=(c\tau+d)^k Y(\tau)\,,~~~\gamma\in\Gamma(N)\,,\\[0.1in]
\left(-4y^2\frac{\partial}{\partial\tau}\frac{\partial}{\partial\bar{\tau}}+2iky\frac{\partial}{\partial\bar{\tau}}\right)Y(\tau)=0\,,
\end{array}
\right.
\end{eqnarray}
where $\tau=x+iy$ is the complex modulus, with $x$ and $y$ denoting its real and imaginary parts, respectively, and $y>0$.  Moreover, $Y(\tau)$ is at most polynomial divergent $Y(\tau)=\mathcal{O}(y^{\alpha})$ in the large volume limit $y\rightarrow+\infty$, where $\alpha$ is some constant. Taking $\gamma=T^N$ and applying the modularity condition, we obtain $Y(\tau+N)=Y(\tau)$ and the Fourier expansion of $Y(\tau)$ reads as~\cite{Qu:2024rns,Qu:2025ddz}
\begin{eqnarray}
\label{eq:Y-Fourier-expansion}Y(\tau)=\sum_{\substack{n\in\frac{1}{N}\mathbb{Z} \\ n\geq0}} c^+(n)q^n + c^-(0)\beta(y)+ \sum_{\substack{n\in\frac{1}{N}\mathbb{Z} \\ n<0}} c^-(n)\Gamma(1-k,-4\pi n y)q^n \,,
\end{eqnarray}
where $q=e^{2\pi i\tau}$ and the function $\beta(y)$ depends on the modular weight $k$ with
\begin{equation}
\beta(y)=\left\{\begin{array}{cc}
y^{1-k},  ~&~ k\neq 1\\
\ln y, ~&~ k=1
\end{array}
\right.\,.
\end{equation}
$\Gamma(s,z)$ is the incomplete gamma function
\begin{eqnarray}
\label{eq:incomp-gamma-func}\Gamma(s,z)=\int_z^{+\infty} e^{-t}t^{s-1}\,dt\,.
\end{eqnarray}
The analytic expression of $\Gamma(s,z)$ for  $s$ being non-negative integer is provided in~\cite{Qu:2025ddz}. In addition to the holomorphic terms of the coefficients $c^+(n)$, $Y(\tau)$ could contain non-holomorphic terms if the coefficients $c^-(0)$ and $c^-(n)$ are non-zero.  Obviously modular forms are special polyharmonic Maa{\ss} forms without non-holomorphic terms, and they satisfy the conditions in Eq.~\eqref{eq:polyharmonic-def} and growth condition. At weight $k\geq3$, the set of polyharmonic Maa{\ss} forms of level $N$ exactly coincides with that of modular forms. At weight 2, the modified Eisenstein series $\widehat{E}_2(\tau)$ is the unique non-holomorphic polyharmonic Maa{\ss} form,
\begin{equation}
\widehat{E}_2(\tau) = 1-\frac{3}{\pi y}-24\sum_{n=1}^{\infty}\sigma_1(n)q^n=1-\frac{3}{\pi y}-24q-72q^2-96q^3-168q^4-\ldots\,,
\end{equation}
where $\sigma_1(n)=\sum_{d|n}d$  is the divisor function. In comparison with the holomorphic modular forms, the modular weight of polyharmonic Maa{\ss} forms can be negative integer. For weights $k\leq0$, 
the linear space of the polyharmonic Maa{\ss} forms can be spanned by the non-holomorphic Eisenstein series at $s=1-k$~\cite{Qu:2025ddz}. Any polyharmonic Maa{\ss} forms of integer weight $k\leq0$ and level $N$ can be expressed as a linear combination of $E_k(N;\tau;1-k;\overline{A/C})$, where $\overline{A/C}$ stands for the cusp of $\Gamma(N)$. The non-holomorphic Eisenstein series can be defined at each cusp of $\Gamma(N)$ as follows~\cite{Diamond:2005afc}, 
\begin{equation}
E_k(N; \tau; s; \overline{A/C}) = \sum_{\substack{(c,d)\equiv (-C, A) \,({\rm mod}\, N) \\ \gcd(c,d)=1} }  \dfrac{y^s}{(c\tau + d)^k |c\tau + d|^{2s}} \,,
\end{equation}
At the weight $k=1$, the Eisenstein series $E_k(N; \tau; 1-k; \overline{A/C})$ are holomorphic functions of $\tau$, and its derivative $E_1^{(1)}(N;\tau;\overline{A/C})\equiv\left.\frac{\partial}{\partial s}E_1(N;\tau;s;\overline{A/C})\right|_{s=0}$ spans the full linear space of weight $1$ polyharmonic Maa{\ss} forms of level $N$~\cite{Qu:2025ddz}.

The polyharmonic Maa{\ss} forms of integer weight $k$ at level $N$ can be organized into irreducible multiplets of the finite modular group $\Gamma'_N$ up to the automorphic factor. Accordingly, one may choose a basis in which the multiplet $Y^{(k)}_{\bm r}(\tau)$ transforms as
\begin{equation}
Y^{(k)}_{\bm r}(\gamma\tau)=(c\tau+d)^k \rho_{\bm r}(\gamma) Y^{(k)}_{\bm r}(\tau)\,, ~~\quad ~~ \gamma\in\Gamma\,,
\end{equation}
where $\rho_{\bm r}$ denotes the irreducible representation matrix of $\Gamma'_N$. The explicit expressions for the multiplets of polyharmonic Maa{\ss} forms with weights between $-4$ and $6$ at lower levels $N=2, 3, 4, 5$ were derived in Refs.~\cite{Qu:2024rns,Qu:2025ddz}. In the present work, we consider the finite modular group $S'_4$ with $N=4$. The multiplets of level $N=4$ polyharmonic Maa{\ss} forms and their transformation under $S'_4$ are summarized in table~\ref{tab:MF_summary}, their analytical expressions are rather lengthy and consequently are not listed here.

In the framework of the non-holomorphic modular invariant theory~\cite{Qu:2024rns,Qu:2025ddz}, the Yukawa couplings are polyharmonic Maa{\ss} forms. The Lagrangian for the Yukawa interactions can be written as  
\begin{eqnarray}
\label{eq:Lagrange-Yuk}\mathcal{L}_Y = Y^{(k_Y)}(\tau) \psi^c \psi H + \mathrm{h.c.} \,,
\end{eqnarray}
where $\psi$ and $\psi^c$ are the two-component spinors for fermion fields carrying the integer modular weights $k_{\psi}$ and $k_{\psi^c}$ respectively. The modular transformations of $\psi$ and $\psi^{c}$ are characterized by their modular weights and irreducible representations of $\Gamma'_N$, i.e.
\begin{eqnarray}
\nonumber && \psi(x) \xmapsto{~\gamma~} (c\tau + d)^{-k_{\psi}} \rho_{\psi}(\gamma) \psi(x) \,,\\
&& \psi^c(x) \xmapsto{~\gamma~} (c\tau + d)^{-k_{\psi^c}} \rho_{\psi^c}(\gamma) \psi^c(x) \,.
\end{eqnarray}
where $\rho_{\psi}$ and $\rho_{\psi^c}$ are the irreducible representations of the finite modular group $\Gamma'_N$. Moreover, the Higgs field $H$ transforms under modular symmetry as, 
\begin{eqnarray}
H(x) \xmapsto{~\gamma~} (c\tau + d)^{-k_H} \rho_H(\gamma) H(x)\,,
\end{eqnarray}
where $k_H$ is its modular weight and $\rho_H$ is a one-dimensional representation of $\Gamma'_N$. Finally $Y^{(k_Y)}(\tau)$ is a polyharmonic Maa{\ss} form multiplet of weight $k_Y$ and level $N$ satisfying  
\begin{eqnarray}
Y^{(k_Y)}(\tau) \xmapsto{~\gamma~}Y^{(k_Y)}(\gamma \tau) = (c\tau + d)^{k_Y} \rho_Y(\gamma) Y^{(k_Y)}(\tau)\,.
\end{eqnarray}
The Yukawa interaction in Eq.~\eqref{eq:Lagrange-Yuk} would be modular invariant if the modular weights and irreducible representations satisfy the following matching conditions,  
\begin{eqnarray}
k_Y = k_{\psi^c} + k_{\psi} + k_{H}\,,~~~ \rho_{Y}\otimes\rho_{\psi^c}\otimes\rho_{\psi}\otimes\rho_H \ni \bm{1}\,,
\end{eqnarray}
where $\bm{1}$ denotes the identity representation of $\Gamma'_N$.

\begin{table}[t!]
\renewcommand{\arraystretch}{1.3}
\centering
\begin{tabular}{|c|c||c|c|}\hline  \hline

Weight $k$ & Polyharmonic Maa{\ss} forms $Y^{(k)}_{\bm{r}}$ & Weight $k$ & Polyharmonic Maa{\ss} forms $Y^{(k)}_{\bm{r}}$ \\ \hline
                        
$-4$ & $Y^{(-4)}_{\bm{1_{0}}}$,\;$Y^{(-4)}_{\bm{2_{0}}}$,\; $Y^{(-4)}_{\bm{3_{0}}}$ & $2$ & $Y^{(2)}_{\bm{1_{0}}}$,\;$Y^{(2)}_{\bm{2_{0}}}$,\; $Y^{(2)}_{\bm{3_{0}}}$\\ \hline
                        
$-3$ & $Y^{(-3)}_{\bm{3_{1}}}$,\; $Y^{(-3)}_{\bm{3_{3}}}$ & $3$ & $Y^{(3)}_{\bm{1_{3}}}$,\; $Y^{(3)}_{\bm{3_{1}}}$,\; $Y^{(3)}_{\bm{3_{3}}}$\\ \hline
                        
$-2$ & $Y^{(-2)}_{\bm{1_{0}}}$,\;$Y^{(-2)}_{\bm{2_{0}}}$,\; $Y^{(-2)}_{\bm{3_{0}}}$  & $4$ & $Y^{(4)}_{\bm{1_{0}}}$,\;$Y^{(4)}_{\bm{2_{0}}}$,\; $Y^{(4)}_{\bm{3_{0}}}$,\; $Y^{(4)}_{\bm{3_{2}}}$\\ \hline
                        
$-1$ & $Y^{(-1)}_{\bm{3_{1}}}$,\; $Y^{(-1)}_{\bm{3_{3}}}$ & $5$ & $Y^{(5)}_{\bm{2_{1}}}$,\; $Y^{(5)}_{\bm{3_{1}}}$,\; $Y^{(5)}_{\bm{3_{3}}\bm{I}}$,\; $Y^{(5)}_{\bm{3_{3}}\bm{II}}$\\ \hline
                        
$0$ & $Y^{(0)}_{\bm{1_{0}}}$,\;$Y^{(0)}_{\bm{2_{0}}}$,\; $Y^{(0)}_{\bm{3_{0}}}$ & $6$ & $Y^{(6)}_{\bm{1_{0}}}$,\;$Y^{(6)}_{\bm{1_{2}}}$,\;$Y^{(6)}_{\bm{2_{0}}}$,\; $Y^{(6)}_{\bm{3_{0}I}}$,\; $Y^{(6)}_{\bm{3_{0}II}}$,\; $Y^{(6)}_{\bm{3_{2}}}$\\ \hline
                        
$1$ & $Y^{(1)}_{\bm{3_{1}}}$,\; $Y^{(1)}_{\bm{3_{3}}}$ &  & \\ \hline \hline
                        
\end{tabular}
\caption{\label{tab:MF_summary} The polyharmonic Maa{\ss} form multiplets $Y^{(k)}_{\bm{r}}$ of level $4$ and weights from $-4$ to $6$, where the subscript $\bm{r}$ denotes the transformation property under $S^{\prime}_{4}$. Here $Y^{(5)}_{\bm{3_{3}I}}$ and $Y^{(5)}_{\bm{3_{3}II}}$ ($Y^{(6)}_{\bm{3_{0}I}}$ and $Y^{(6)}_{\bm{3_{0}II}}$) stand for two linearly independent triplets of Maa{\ss} forms at weight 5 (6). The  explicit forms of these polyharmonic  Maa{\ss} form multiplets can be found in Ref.~\cite{Qu:2025ddz}. }
\end{table}
        
\section{\label{sec:model} Lepton models based on non-holomorphic $S^{\prime}_{4}$ modular flavor symmetry}

In this section, we shall construct lepton models with the smallest number of free parameters based on the non-holomorphic $S^\prime_4$ modular symmetry, and no flavon fields are introduced. Neutrinos are assumed to be Majorana particles and their masses are considered to arise from the type-I seesaw mechanism with 2RHNs. The existence of two inequivalent doublet representations in $S^\prime_{4}$ makes it particularly suitable for realizing minimal seesaw models with 2RHNs in a non-holomorphic modular framework. Without loss of generality, the Higgs doublet $H$ is assumed to be trivial singlet $\bm{1_0}$ under  $S^\prime_4$ and is taken to have zero modular weight. To construct lepton models with a minimal number of input parameters, the three left‑handed lepton doublets $L$ are taken to form an $S^{\prime}_4$ triplet $\bm{3_{p}}$ with modular weight $k_{L}$, the right‑handed charged leptons $E^c_{1,2,3}$ are  singlets $\bm{1_{q_{1,2,3}}}$ with weights $k_{E^c_{1,2,3}}$, and the two RH neutrinos $N^c$ are assigned to a doublet $\bm{2_a}$ with weight $k_{N^c}$, where $p, q_{1,2,3} \in {0,1,2,3}$ and $a \in {0,1}$. Our convention for the irreducible representations and Kronecker tensor products of $S^{\prime}_{4}$ modular group is collected in Appendix~\ref{sec:S4p_group_theory}.

\subsection{\label{sec:ch_mass}Charged lepton sector }

Under the above assumptions, modular invariance fixes the modular forms associated with the charged lepton Yukawa terms $E^{c}_{i}L$ to transform as $\bm{3_{p_{i}}}$ of $S^{\prime}_{4}$ with modular weights $k_{i}=k_{L}+k_{E^{c}_{i}}$, where $i=1,2,3$ and $p_{i}=[-p-q_{i}]\equiv-p-q_{i}$ (\text{mod} 4). Hence the most general Lagrangian governing charged lepton masses reads
\begin{equation}\label{eq:Le1}
-\mathcal{L}_l\,= y_{e} \left(E^c_1 L Y^{(k_{1})}_{\bm{3_{p_{1}}}}\right)_{\bm{1_0}}H^{*} +y_{\mu}\left( E^c_2 L Y^{(k_{2})}_{\bm{3_{p_{2}}}}\right)_{\bm{1_0}}H^{*} +y_{\tau} \left(E^c_3L Y^{(k_{3})}_{\bm{3_{p_{3}}}}\right)_{\bm{1_0}}H^{*}+\text{h.c.}\,,
\end{equation}
where $Y^{(k_{i})}_{\bm{3_{p_{i}}}}$ denotes the triplet polyharmonic Maa{\ss} form transforming as $\bm{3_{p_{i}}}$ at level $N=4$. The three terms in Eq.~\eqref{eq:Le1} share an identical structure and separately populate the three rows of the charged lepton mass matrix. If a given weight $k_i$ admits several linearly independent triplets of polyharmonic Maa{\ss} forms in the same representation, each would enter with an analogous term; the corresponding expressions can be read off directly from the general formulas provided later. Guided by the principle of minimality and simplicity, we focus on the cases
in which there is only one linear independent polyharmonic Maa{\ss} forms triplet $Y^{(k_{i})}_{\bm{3_{p_{i}}}}$ ($i=1, 2, 3$) at the weight $k_i$. Using the tensor product decomposition $\bm{3_{p}} \otimes \bm{3_{q}}=\bm{1_{[p+q]}} \oplus \bm{2_{\langle p+q\rangle}} \oplus \bm{3_{[p+q]}} \oplus \bm{3_{[p+q+2]}}$ given in Eq.~\eqref{eq:Kronecker_products}, one extracts the charged lepton mass matrix
\begin{equation}\label{eq:MCh1}
M_{l}(k_1,p_{1};k_2,p_{2};k_3,p_{3})=\begin{pmatrix}
y_{e} Y^{(k_{1})}_{\bm{3_{p_{1}}},1} & y_{e} Y^{(k_{1})}_{\bm{3_{p_{1}}},3} & y_{e} Y^{(k_{1})}_{\bm{3_{p_{1}}},2} \\
y_{\mu} Y^{(k_{2})}_{\bm{3_{p_{2}}},1} & y_{\mu} Y^{(k_{2})}_{\bm{3_{p_{2}}},3} & y_{\mu} Y^{(k_{2})}_{\bm{3_{p_{2}}},2} \\
y_{\tau} Y^{(k_{3})}_{\bm{3_{p_{3}}},1} & y_{\tau} Y^{(k_{3})}_{\bm{3_{p_{3}}},3} & y_{\tau} Y^{(k_{3})}_{\bm{3_{p_{3}}},2} 
\end{pmatrix}v\,,
\end{equation}
where $v\equiv \braket{H}\approx174$ GeV is the VEV of the SM Higgs field and $Y^{(k_{i})}_{\bm{3_{p_{i}}}}=(Y^{(k_{i})}_{\bm{3_{p_{i}}},1},Y^{(k_{i})}_{\bm{3_{p_{i}}},2},Y^{(k_{i})}_{\bm{3_{p_{i}}},3})^{T}$. 

From the structure of the polyharmonic Maa{\ss} forms at level $N=4$ summarized in table~\ref{tab:MF_summary}, we see that the modular triplet  $Y^{(k_{i})}_{\bm{3_{p_{i}}}}$ can be chosen from the following 16 possibilities:
\begin{eqnarray}\label{eq:single_triplet}
\nonumber && Y^{(-4)}_{\bm{3_{0}}}, ~~ Y^{(-3)}_{\bm{3_{1}}},~~ Y^{(-3)}_{\bm{3_{3}}}, ~~ Y^{(-2)}_{\bm{3_{0}}}, ~~ Y^{(-1)}_{\bm{3_{1}}},~~ Y^{(-1)}_{\bm{3_{3}}}, ~~ Y^{(0)}_{\bm{3_{0}}}, ~~ Y^{(1)}_{\bm{3_{1}}},~~ Y^{(1)}_{\bm{3_{3}}},  ~~ Y^{(2)}_{\bm{3_{0}}}, ~~ Y^{(3)}_{\bm{3_{1}}},~~ Y^{(3)}_{\bm{3_{3}}}, \\
&& Y^{(4)}_{\bm{3_{0}}},~~ Y^{(4)}_{\bm{3_{2}}}, ~~ Y^{(5)}_{\bm{3_{1}}}, ~~ Y^{(6)}_{\bm{3_{2}}}\,,
\end{eqnarray}
Among these multiplets, we find the relations
\begin{equation}
Y^{(4)}_{\bm{3_{2}}}=\sqrt{6}Y^{(3)}_{\bm{1_{3}}}Y^{(1)}_{\bm{3_{3}}}, \qquad Y^{(6)}_{\bm{3_{2}}}=-2\sqrt{6}Y^{(3)}_{\bm{1_{3}}}Y^{(3)}_{\bm{3_{3}}}\,.
\end{equation}
which imply that $Y^{(4)}_{\bm{3_{2}}}$ and $Y^{(6)}_{\bm{3_{2}}}$ are not independent of the others. The overall factor $Y^{(3)}_{\bm{1_{3}}}$ can be absorbed by the free parameters $y_{e}$, $y_{\mu}$ or $y_{\tau}$ whose phases can be removed by rephasing the lepton fields, allowing these parameters to be chosen real and positive.  To avoid a massless charged lepton, no two rows of the charged lepton mass matrix $M_{l}$ are  proportional to each other in Eq.~\eqref{eq:MCh1}.  As a consequence, the polyharmonic Maa{\ss} triplets $Y^{(k_{i})}_{\bm{3_{p_{i}}}}$ should satisfy
\begin{equation}
Y^{(k_{1})}_{\bm{3_{p_{1}}}}\neq Y^{(k_{2})}_{\bm{3_{p_{2}}}}\neq Y^{(k_{3})}_{\bm{3_{p_{3}}}}\,.
\end{equation}
Redefinitions of the RH charged lepton states lead to permutations of the three rows of the charged lepton mass matrix, while the predictions for lepton masses and flavor mixing remain unchanged. Hence, up to row permutations, there exist exactly $C^{3}_{14}=364$ independent charged lepton mass matrices, each uniquely labeled by the weights $k_i$ and indices $p_i$ as $C^{(k_{1},k_{2},k_{3})}_{(p_{1},p_{2},p_{3})}$.

\subsection{\label{sec:nu_mass}Neutrino sector}

The LH doublet $L$ carries modular weight $k_{L}$ and transforms as the $\bm{3_{p}}$ representation of $S^{\prime}_{4}$ ($p=0,1,2,3$). The RH neutrino doublet $N^c$ has weight $k_{N^{c}}$ and belongs to the representation $\bm{2_a}$ ($a=0,1$). According to the Kronecker product rules in Eq.~\eqref{eq:Kronecker_products}, one finds the decompositions $\bm{2_{a}}\otimes\bm{3_{p}}=\bm{3_{[a+p]}}\oplus\bm{3_{[a+p+2]}}$ and $\bm{2_{a}}\otimes\bm{2_{a}}=\bm{1_{2a}}\oplus\bm{1_{[2+2a]}}\oplus\bm{2_{0}}$. Consequently, the modular invariant Dirac and Majorana mass terms for neutrinos take the form:
\begin{eqnarray}
\nonumber\hskip-0.2in -\mathcal{L}_{\nu}&=&g_{1}\left(N^cLY^{(k_{m})}_{\bm{3_{[-p-a]}}}H\right)_{\bm{1_{0}}}+g_{2}\left(N^cLY^{(k_{m})}_{\bm{3_{[2-p-a]}}}H\right)_{\bm{1_{0}}}\\
\label{eq:Lagrangian_nu_gen}\hskip-0.2in&&+\frac{\Lambda}{2}\left(N^cN^cY^{(k_{n})}_{\bm{2_{0}}}\right)_{\bm{1_{0}}}+\frac{\Lambda^{\prime}}{2}\left(N^cN^cY^{(k_{n})}_{\bm{1_{2a}}}\right)_{\bm{1_{0}}}+\text{h.c.}\,,
\end{eqnarray}
where $k_{m}=k_{L}+k_{N^{c}}$ and $k_{n}=2k_{N^{c}}$. Note that the term proportional to $\left(N^cN^cY^{(2k_{N^{c}})}_{\bm{1_{2-2a}}}\right)_{\bm{1_{0}}}$ gives a vanishing contribution and consequently it is omitted. The resulting Dirac and Majorana neutrino mass matrices reads: 
\begin{eqnarray}
\nonumber  M_{D}(k_{m})&=&g_{1}v\left(
\begin{array}{ccc}
 -2  Y^{(k_{m})}_{\bm{3_{[-p-a]}},1} & Y^{(k_{m})}_{\bm{3_{[-p-a]}},3} & Y^{(k_{m})}_{\bm{3_{[-p-a]}},2} \\
 0 & -\sqrt{3} Y^{(k_{m})}_{\bm{3_{[-p-a]}},2} & -\sqrt{3}  Y^{(k_{m})}_{\bm{3_{[-p-a]}},3} \\
\end{array}
\right) \\
\nonumber &&+g_{2}v\left(
\begin{array}{ccc}
0 & \sqrt{3}  Y^{(k_{m})}_{\bm{3_{[2-p-a]}},2} & \sqrt{3}  Y^{(k_{m})}_{\bm{3_{[2-p-a]}},3} \\
 -2  Y^{(k_{m})}_{\bm{3_{[2-p-a]}},1} &  Y^{(k_{m})}_{\bm{3_{[2-p-a]}},3} &  Y^{(k_{m})}_{\bm{3_{[2-p-a]}},2} \\
\end{array}
\right)\,, \\
\label{eq:nu_masses_gen}M_{N}(k_{n})&=&\left\{\begin{array}{lll}
\left(
\begin{array}{cc}
\Lambda^{\prime}Y^{(k_{n})}_{\bm{1_{2a}}}-\Lambda Y^{(k_{n})}_{\bm{2_{0}},1} &\Lambda Y^{(k_{n})}_{\bm{2_{0}},2} \\
\Lambda Y^{(k_{n})}_{\bm{2_{0}},2} & \Lambda^{\prime}Y^{(k_{n})}_{\bm{1_{2a}}}+\Lambda Y^{(k_{n})}_{\bm{2_{0}},1} \\
\end{array}
\right) & \text{for} & a=0 \\[6mm]
\left(
\begin{array}{cc}
\Lambda^{\prime}Y^{(k_{n})}_{\bm{1_{2a}}}+\Lambda Y^{(k_{n})}_{\bm{2_{0}},2} &\Lambda Y^{(k_{n})}_{\bm{2_{0}},1} \\
\Lambda Y^{(k_{n})}_{\bm{2_{0}},1} & \Lambda^{\prime}Y^{(k_{n})}_{\bm{1_{2a}}}-\Lambda Y^{(k_{n})}_{\bm{2_{0}},2} \\
\end{array}
\right) & \text{for} & a=1 
\end{array}
\right.\,.
\end{eqnarray}

We now discuss the Majorana neutrino mass matrix for different choices of the representation and modular weight of the RH neutrinos $N^{c}$. For $N^{c}\sim\bm{2_{0}}$ ($a=0$), the modular multiplets coupling to $N^{c}N^{c}$ are the singlets $Y^{(k_n)}_{\bm{1_0}}$ and doublets $Y^{(k_n)}_{\bm{2_0}}$, both of which exist for every even weight $k_n$. As a result, the two Majorana terms in Eq.~\eqref{eq:Lagrangian_nu_gen} are generally allowed. The corresponding Majorana neutrino mass matrix therefore contains at least two independent couplings and takes the general form given in Eq.~\eqref{eq:nu_masses_gen}. For $N^{c}\sim\bm{2_{1}}$ ($a=1$), the Majorana term $\left(N^cN^cY^{(k_n)}_{\bm{1_2}}\right)_{\bm{1_0}}$ is forbidden for $k_{n}\leq 4$, since no modular singlet transforming as $\bm{1_2}$ exists for these weights. Consequently, only the term $\left(N^cN^cY^{(2k_{N^{c}})}_{\bm{2_0}}\right)_{\bm{1_0}}$ is allowed, and the Majorana mass matrix depends on a single coupling. The resulting Majorana mass matrix is
\begin{equation}\label{eq:nu_masses_MN}
M_{N}(k_{n})=\Lambda\left(
\begin{array}{cc}
Y^{(k_{n})}_{\bm{2_{0}},2} &Y^{(k_{n})}_{\bm{2_{0}},1} \\
Y^{(k_{n})}_{\bm{2_{0}},1} & -Y^{(k_{n})}_{\bm{2_{0}},2} \\
\end{array}
\right)\,, \qquad \text{for} \qquad k_n\leq 4\,.
\end{equation}

As regards the neutrino Yukawa couplings, when $[a+p]=1$ or $3$, the modular form multiplets required for $S^{\prime}_{4}$ invariance are $Y^{(k_{m})}_{\bm{3_{3}}}$ or $Y^{(k_{m})}_{\bm{3_{1}}}$. These modular triplets exist only for odd modular weights and always appear simultaneously. Therefore, the two Dirac terms in Eq.~\eqref{eq:Lagrangian_nu_gen} are either both allowed for odd $k_m$ or both forbidden for even $k_m$. In this case, the viable Dirac mass matrix contains at least two independent parameters, with the allowed odd weights satisfying $k_m\leq 3$, and takes the general form given in Eq.~\eqref{eq:nu_masses_gen}. For $[a+p]=0$ or $2$, $S^{\prime}_{4}$ invariance instead requires the relevant modular multiplets to transform as $\bm{3_0}$ or $\bm{3_2}$. However, for even modular weights with $k_m\leq 2$, only the triplet $Y^{(k_m)}_{\bm{3_0}}$ exists. Under this condition, the Dirac mass matrix depends on a single free parameter and is given by
\begin{eqnarray}
\nonumber \hskip-0.2in M_{D1}(k_{m})&=&gv\left(
\begin{array}{ccc}
 -2 Y^{(k_{m})}_{\bm{3_{0}},1} & Y^{(k_{m})}_{\bm{3_{0}},3} & Y^{(k_{m})}_{\bm{3_{0}},2} \\
 0 & -\sqrt{3} Y^{(k_{m})}_{\bm{3_{0}},2} & -\sqrt{3} Y^{(k_{m})}_{\bm{3_{0}},3} \\
\end{array}
\right), \quad \text{for} \quad [a+p]=0\,, \\
\label{eq:nu_masses_MD} \hskip-0.2in M_{D2}(k_{m})&=&gv\left(
\begin{array}{ccc}
 0 & \sqrt{3} Y^{(k_{m})}_{\bm{3_{0}},2} & \sqrt{3} Y^{(k_{m})}_{\bm{3_{0}},3} \\
 -2 Y^{(k_{m})}_{\bm{3_{0}},1} & Y^{(k_{m})}_{\bm{3_{0}},3} & Y^{(k_{m})}_{\bm{3_{0}},2} \\
\end{array}
\right)=P_{2}M_{D1}(k_{m}), \quad \text{for} \quad [a+p]=2\,,
\end{eqnarray}
where
$P_{2}=\begin{pmatrix}
0 & 1\\-1 & 0
\end{pmatrix}$.

In this work, we aim to construct lepton models with the minimal number of free parameters. To this end, we restrict ourselves to scenarios in which the neutrino Yukawa sector and the heavy Majorana mass sector each contain only a single independent coupling constant. This can be achieved for the assignments $N^{c}\sim \bm{2_{1}}$ and $L\sim \bm{3_{3}}$ or $\bm{3_{1}}$, with even modular weights satisfying $k_{m}\leq 2$ and $k_{n}\leq 4$. The corresponding Majorana mass matrix is given by Eq.~\eqref{eq:nu_masses_MN}, while the Dirac mass matrices are $M_{D1}$ for $L\sim \bm{3_{3}}$ and $M_{D2}$ for $L\sim \bm{3_{1}}$, as defined in Eq.~\eqref{eq:nu_masses_MD}. The resulting light neutrino mass matrices are generated through the seesaw mechanism:
\begin{equation}\label{eq:seesaw}
M_{\nu1}(k_{m},k_{n})=-M^{T}_{D1}(k_{m})M^{-1}_{N}(k_{n})M_{D1}(k_{m}), \quad M_{\nu2}(k_{m},k_{n})=-M^{T}_{D2}(k_{m})M^{-1}_{N}(k_{n})M_{D2}(k_{m})\,.
\end{equation}
It is straightforward to verify that $M_{\nu1}(k_{m},k_{n})=-M_{\nu2}(k_{m},k_{n})$, implying that the two assignments $L\sim \bm{3_{3}}$ and $L\sim \bm{3_{1}}$  are physically equivalent.  Therefore, independent light neutrino mass matrices with a single free parameter are obtained by assigning the LH lepton doublets $L$ to $\bm{3_{3}}$ and the RH neutrinos $N^{c}$ to $\bm{2_{1}}$. 
Since our analysis considers polyharmonic Maa{\ss} forms with integer weights ranging from $-4$ to $6$, this construction gives rise to 20 distinct modular weight assignments, denoted by $S^{(k_{m},k_{n})}$ with $k_{m}\in\{-4,-2,0,2\}$ and $k_{n}\in\{-4,-2,0,2,4\}$.

\subsection{\label{sec:num_analysis}Numerical analysis}
        
\begin{table}[t!]
\centering
\renewcommand{\arraystretch}{1.2}
\begin{tabular}{|c|c|c||c|c|c|c|c|}
\hline \hline
Observable  &  $\text{bf}\pm1\sigma$   & $3\sigma$ region  &   Observable &  $\text{bf}\pm1\sigma$  & $3\sigma$ region     \\ \hline

$\sin^2\theta_{13}$ & $0.02248^{+0.00055}_{-0.00059}$ & $[0.02064,0.02418]$ & $\frac{\Delta m^2_{21}}{10^{-5}\text{eV}^2}$ & $7.537^{+0.094}_{-0.10}$ & $[7.236,7.823]$  \\ [0.050in]

$\sin^2\theta_{12}$ & $0.3088^{+0.0067}_{-0.0066}$ & $[0.2893,0.3295]$ & $\frac{\Delta m^2_{31}}{10^{-3}\text{eV}^2}$ & $2.511^{+0.021}_{-0.020}$ & $[2.450,2.576]$ \\ [0.050in]

$\sin^2\theta_{23}$  & $0.470^{+0.017}_{-0.014}$  & $[0.435,0.584]$ & $m_e/m_{\mu}$ & $0.004737$ & --- \\ [0.050in]

$\delta_{CP}/\pi$  & $1.178^{+0.144}_{-0.2}$ & $[0.694,2.028]$  &  $m_{\mu}/m_{\tau}$ & $0.05882$ & --- \\ [0.050in]

--- & --- & ---  &  $m_{e}/\text{MeV}$ & $0.469652$ & --- \\ [0.050in]
\hline  \hline

\end{tabular}
\caption{\label{tab:bf_13sigma_data} The central values and $1\sigma$ as well as $3\sigma$ allowed intervals of the lepton masses and mixing parameters adopted from the \texttt{NuFIT}-v6.1 global analysis~\cite{Esteban:2024eli}, which incorporates the Super-Kamiokande atmospheric neutrino data and assumes a NO of the neutrino mass spectrum. The charged lepton mass ratios are taken from~\cite{Xing:2007fb} where the uncertainties are very small. We set the uncertainties of the charged lepton mass ratios to be $0.1\%$ of their central value when scanning the parameter space of our models.}
\end{table}

In sections~\ref{sec:ch_mass} and \ref{sec:nu_mass}, we have discussed the possible assignments and the resulting mass matrices for charged leptons and neutrinos within the framework of non-holomorphic $S_{4}^{\prime}$ modular symmetry without imposing gCP symmetry. We focus on the scenario in which the three LH lepton doublets $L$ transform as a triplet of $S_{4}^{\prime}$, the RH charged leptons $E^{c}_{1,2,3}$ transform as singlets, and two generations of RH neutrinos $N^{c}$ are introduced as a doublet of $S_{4}^{\prime}$. The $S'_4$ modular models for lepton with the minimal number of free parameters arise from the specific assignment $L \sim \bm{3_{3}}$ and $N^{c} \sim \bm{2_{1}}$. Restricting the modular weights of the modular forms $Y_{\bm{r}}^{(k)}$ to the range $-4 \leq k \leq 6$, we obtain 364 distinct charged lepton mass matrices of the general form given in Eq.~\eqref{eq:MCh1} and 20 light neutrino mass matrices, see Eq.~\eqref{eq:nu_masses_MN} and Eq.~\eqref{eq:nu_masses_MD}. Combining these possible forms of the charged lepton and neutrino sectors, we obtain a total of $364 \times 20 = 7280$ lepton models, denoted by $C^{(k_{1},k_{2},k_{3})}_{(p_{1},p_{2},p_{3})}-S^{(k_{m},k_{n})}$. In  each of these models, the charged lepton mass matrix depends on the modulus $\tau$ and three real couplings $y_{e}$, $y_{\mu}$ and $y_{\tau}$ which could be tuned to accommodate the measured charged lepton masses, while the light neutrino mass matrix only depends on the complex modulus $\tau$ up to an overall scale $g^{2}v^{2}/\Lambda$. Since all 11 lepton observables—namely, the three charged lepton masses, three neutrino masses, three lepton mixing angles, the Dirac CP-violating phase, and one Majorana CP-violating phase\footnote{The lightest neutrino is massless in our models so that only one Majorana phase is physical and another one can be removed by field redefinition.} are determined by only six independent real input parameters in the absence of gCP symmetry, these non-holomorphic $S_{4}^{\prime}$ modular models exhibit a high degree of predictivity.

For each model, we test its compatibility with experimental data by performing a $\chi^{2}$ analysis. The $\chi^{2}$ function is defined as
\begin{equation}\label{eq:chisq_def}
\chi^{2}(\vec{o}) =
\sum_{i=1}^{2} \left( \frac{o_{i}(x) - \mu_{i}}{\sigma_{i}} \right)^{2}
\, + \,
\sum_{i=3}^{8} \chi_{i}^{2}\!\left(o_{i}(x)\right)\,,
\end{equation}
where
\begin{equation}
\vec{o} = \left(
m_{e}/m_{\mu},\, m_{\mu}/m_{\tau},\,
\sin^2\theta_{12},\, \sin^2\theta_{13},\, \sin^2\theta_{23},\,
\delta_{CP},\, \Delta m^2_{21},\, \Delta m^2_{31}
\right)\,.
\end{equation}
Here, $x$ denotes the set of model parameters, and 
$o_{i}(x)$ are the corresponding theoretical predictions. 
For $i=1,2$, $\mu_{i}$ and $\sigma_{i}$ represent the central values and standard deviations of the charged lepton mass ratios $m_{e}/m_{\mu}$ and $m_{\mu}/m_{\tau}$, taken from~\cite{Xing:2007fb} and listed in table~\ref{tab:bf_13sigma_data}. The mass of the electron $m_{e}$ can be reproduced by the overall mass scale $y_{\tau} v$ of the charged lepton mass matrix. For $i=3,\dots,8$, the functions $\chi_{i}^{2}(o_{i})$ are obtained from the one-dimensional projections of the neutrino observables provided by \texttt{NuFIT}~\cite{Esteban:2024eli}.

We adopt the standard parametrization of the Pontecorvo, Maki, Nakagawa and Sakata (PMNS) lepton mixing matrix~\cite{ParticleDataGroup:2026aaa}:
\begin{equation}\label{eq:PMNS}
U=\left(\begin{array}{ccc}
c_{12}c_{13}  &   s_{12}c_{13}   &   s_{13}e^{-i\delta_{CP}}  \\
-s_{12}c_{23}-c_{12}s_{13}s_{23}e^{i\delta_{CP}}   &  c_{12}c_{23}-s_{12}s_{13}s_{23}e^{i\delta_{CP}}  &  c_{13}s_{23}  \\
s_{12}s_{23}-c_{12}s_{13}c_{23}e^{i\delta_{CP}}   & -c_{12}s_{23}-s_{12}s_{13}c_{23}e^{i\delta_{CP}}  &  c_{13}c_{23}
\end{array}\right)\text{diag}(1,e^{i\frac{\alpha_{21}}{2}},e^{i\frac{\alpha_{31}}{2}})\,,
\end{equation}
where $c_{ij}\equiv \cos\theta_{ij}$, $s_{ij}\equiv \sin\theta_{ij}$, $\delta_{CP}$ is the Dirac CP violation (CPV) phase, and $\alpha_{21,31}$ are Majorana CPV phases \cite{Bilenky:1980cx}. If the lightest neutrino is massless, there is a single Majorana phase $\alpha_{21}$, and another Majorana phase $\alpha_{31}$ can be absorbed by field redefinition.

For any given set of input parameters, the model yields definite predictions for the lepton masses, mixing parameters, and the corresponding $\chi^{2}$ value. By performing a comprehensive scan of the parameter space, the minimum of $\chi^{2}$ can be identified. In this analysis, it is sufficient to restrict the complex modulus $\tau$ to the standard fundamental domain $\mathcal{D} = \left\{ \tau \in \mathcal{H} \,\big|\, |\Re(\tau)| \leq \tfrac{1}{2},\ |\tau| \geq 1 \right\}$. The coupling constants are varied within the range $[0,10^{5}]$. A lepton model is considered phenomenologically viable if, at the $\chi^{2}$ minimum, its predictions for the neutrino masses and mixing parameters lie within the corresponding $3\sigma$ ranges of \texttt{NuFIT}-v6.1 listed in table~\ref{tab:bf_13sigma_data}. In addition, the predicted charged lepton masses are required to agree with the experimental central values within $0.3\%$. After comprehensive numerical analysis, we find that only 36 among the 7280 minimal littlest seesaw models can successfully reproduce the experimental data for normal ordering (NO) neutrino mass spectrum. However, none of the models can accommodate experimental data in case of inverted ordering (IO).

For these viable cases, the best-fit values of the input 
parameters and the corresponding predictions for the mixing 
angles, CP-violating phases, neutrino masses, the effective 
Majorana mass $m_{\beta\beta}$ relevant for neutrinoless 
double beta decay ($0\nu\beta\beta$-decay), and the kinematic 
mass parameter $m_{\beta}$ probed in beta decay are summarized 
in table~\ref{tab:best_fit_noGCP_input} and table~\ref{tab:best_fit_noGCP_output}, respectively. Here, the effective Majorana neutrino mass $m_{\beta\beta}$ is defined as
\begin{equation}
  m_{\beta\beta}=|m_{2}\sin^{2}\theta_{12}\cos^{2}\theta_{13}e^{i\alpha_{21}}+m_{3}\sin^{2}\theta_{13}e^{-2i\delta_{CP}}|\,,\quad  m_{1}=0\,.
\end{equation}
The kinematical mass $m_{\beta}$ is defined as:
\begin{equation}
m_{\beta}=\left(m_{2}^{2}\sin^{2}\theta_{12}\cos^{2}\theta_{13}+m_{3}^{2}\sin^{2}\theta_{13}\right)^{1/2}\,.
\end{equation}

From table~\ref{tab:best_fit_noGCP_input} and table~\ref{tab:best_fit_noGCP_output}, one finds that the 36 phenomenologically viable models can be grouped into three categories of twelve, each characterized by nearly identical predictions for the lepton observables and by minimum $\chi^{2}$ values around $8.8$ (first set), $15$ (second set), and $4.8$ (third set). Within each category, the assignments of representations and modular weights for the lepton fields are almost identical, differing only in the assignment of the first generation RH charged lepton $E^{c}_{1}$, leading to different structures in the first row of $M_{l}$. The models in the first, second, and third sets can be summarized as $C_{(x_{1},0,3)}^{(k_{1},2,-1)}-S^{(-2,-4)}$, $C_{(x_{2},1,3)}^{(k_{2},-3,-1)}-S^{(-2,4)}$, and $C_{(x_{3},0,1)}^{(k_{3},0,1)}-S^{(-2,4)}$, respectively, where the allowed values of $x_{1,2,3}$ and $k_{1,2,3}$ are given in table~\ref{tab:best_fit_noGCP_input}. The reason why the twelve models within each category yield nearly identical predictions for the lepton observables is that, at the best-fit points, the contribution from the first row of the charged lepton mass matrix $M_{l}$ is numerically suppressed by the coupling $y_{e}/y_{\tau}\sim 10^{-4}$ as can be seen from table~\ref{tab:best_fit_noGCP_input}. Note that there exist clear hierarchies among the coupling parameters $y_{e}$, $y_{\mu}$ and $y_{\tau}$, which are responsible for the observed hierarchies in the charged lepton masses. Such hierarchies can be naturally realized via the so-called weighton mechanism~\cite{King:2020qaj,Ding:2025mar}, in which a SM singlet weighton field is introduced to generate the fermion mass hierarchy while maintaining Yukawa couplings of comparable magnitude.
     
\begin{table}[t!]
\centering
\small
\renewcommand{\arraystretch}{1.2}
\begin{tabular}{|c|c|c|c|c|c|c|c|}  \hline \hline

\texttt{Models} &   $\Re\langle \tau \rangle$ & $\Im\langle \tau \rangle$  &   $10^{4}y_{e}/y_{\tau}$ &   $10^{2}y_{\mu}/y_{\tau}$ &    $(g^2v^2/\Lambda)/$meV  &  $y_{\tau} v$/GeV & $\chi^2_{\mathrm{min}}$  \\ \hline

$C^{(-4,2,-1)}_{(0,0,3)}-S^{(-2,-4)}$  &    $-$0.3006 & 1.020 & 6.309 & 0.8998 & 7.617 & 8.781 & 8.887 \\ \hline
$C^{(-3,2,-1)}_{(1,0,3)}-S^{(-2,-4)}$  &    $-$0.3006 & 1.020 & 2.717 & 0.9001 & 7.617 & 8.839 & 8.819 \\ \hline
$C^{(-3,2,-1)}_{(3,0,3)}-S^{(-2,-4)}$  &   $-$0.3006 & 1.020 & 1.888 & 0.8976 & 7.617 & 8.883 & 8.848 \\ \hline
$C^{(-2,2,-1)}_{(0,0,3)}-S^{(-2,-4)}$ &   $-$0.3006 & 1.020 & 1.703 & 0.8989 & 7.617 & 8.847 & 8.868 \\ \hline
$C^{(-1,2,-1)}_{(1,0,3)}-S^{(-2,-4)}$ &    $-$0.3006 & 1.020 & 2.044 & 0.8974 & 7.617 & 8.849 & 8.769 \\ \hline
$C^{(0,2,-1)}_{(0,0,3)}-S^{(-2,-4)}$ &    $-$0.3006 & 1.020 &  1.341 & 0.8974 & 7.617 & 8.842 & 8.824   \\ \hline

$C^{(1,2,-1)}_{(1,0,3)}-S^{(-2,-4)}$ & $-$0.3006 & 1.020 &  4.687 & 0.8976 & 7.617 & 8.842 & 8.824 \\ \hline
$C^{(1,2,-1)}_{(3,0,3)}-S^{(-2,-4)}$ &  $-$0.3006 & 1.020 & 0.6344 & 0.9010 & 7.617 & 8.842 & 8.863 \\ \hline
$C^{(3,2,-1)}_{(1,0,3)}-S^{(-2,-4)}$ &   $-$0.3006 & 1.020 & 0.7357 & 0.8974 & 7.617 & 8.842 & 8.813 \\ \hline
 $C^{(3,2,-1)}_{(3,0,3)}-S^{(-2,-4)}$ &   $-$0.3006 & 1.020 & 0.5806 & 0.8978 & 7.617 & 8.842 & 8.827 \\ \hline
 $C^{(4,2,-1)}_{(0,0,3)}-S^{(-2,-4)}$&   $-$0.3007 & 1.020 & 0.8764 & 0.8976 & 7.617 & 8.842 & 8.841 \\ \hline
$C^{(5,2,-1)}_{(1,0,3)}-S^{(-2,-4)}$ &   $-$0.3007 & 1.020 & 1.585 & 0.8978 & 7.617 & 8.842 & 8.841 \\ \hline

$C^{(-4,-3,-1)}_{(0,1,3)}-S^{(-2,4)}$ &  $-$0.2930 & 0.9936 & 4.784 & 4.870 & 44.38 & 9.463 & 15.03 \\ \hline
$C^{(-3,-3,-1)}_{(3,1,3)}-S^{(-2,4)}$ & $-$0.2930 & 0.9935 & 2.680 & 4.859 & 44.39 & 9.488 & 15.02 \\ \hline
 $C^{(-2,-3,-1)}_{(0,1,3)}-S^{(-2,4)}$ & $-$0.2930 & 0.9936 & 1.153 & 4.859 & 44.39 & 9.488 & 15.03 \\ \hline
$C^{(-1,-3,-1)}_{(1,1,3)}-S^{(-2,4)}$ &    $-$0.2930 & 0.9936 & 1.167 & 4.858 & 44.38 & 9.489 & 15.03 \\ \hline
$C^{(0,-3,-1)}_{(0,1,3)}-S^{(-2,4)}$ &   $-$0.2930 & 0.9936 & 6.073 & 4.858 & 44.39 & 9.488 & 15.07 \\ \hline
$C^{(1,-3,-1)}_{(1,1,3)}-S^{(-2,4)}$ &   $-$0.2930 & 0.9936 & 9.092 & 4.858 & 44.39 & 9.489 & 15.05 \\ \hline

$C^{(1,-3,-1)}_{(3,1,3)}-S^{(-2,4)}$  &    $-$0.2930 & 0.9936 & 0.5314 & 4.859 & 44.39 & 9.488 & 15.03 \\ \hline
$C^{(2,-3,-1)}_{(0,1,3)}-S^{(-2,4)}$ &   $-$0.2930 & 0.9936 & 0.5350 & 4.858 & 44.39 & 9.490 & 15.03 \\ \hline
$C^{(3,-3,-1)}_{(1,1,3)}-S^{(-2,4)}$ &   $-$0.2930 & 0.9936 & 3.076 & 4.857 & 44.39 & 9.486 & 14.96   \\ \hline
$C^{(3,-3,-1)}_{(3,1,3)}-S^{(-2,4)}$ &  $-$0.2930 & 0.9935 & 2.035 & 4.859 & 44.39 & 9.486 & 15.01 \\ \hline
$C^{(4,-3,-1)}_{(0,1,3)}-S^{(-2,4)}$ &  $-$0.2930 & 0.9936 & 0.6907 & 4.858 & 44.38 & 9.490 & 15.04 \\ \hline
$C^{(5,-3,-1)}_{(1,1,3)}-S^{(-2,4)}$ &   $-$0.2930 & 0.9936 & 1.049 & 4.858 & 44.38 & 9.490 & 15.02 \\ \hline

$C^{(-4,0,1)}_{(0,0,1)}-S^{(-2,4)}$ &  0.3373 & 1.080 & 6.021 & 2.728 & 33.83 & 6.743 & 4.815 \\ \hline
$C^{(-3,0,1)}_{(1,0,1)}-S^{(-2,4)}$ &  0.3373 & 1.080 & 6.882 & 2.731 & 33.84 & 6.701 & 4.799 \\ \hline
$C^{(-3,0,1)}_{(3,0,1)}-S^{(-2,4)}$ & 0.3373 & 1.080 & 2.628 & 2.721 & 33.84 & 6.683 & 4.806 \\ \hline
$C^{(-2,0,1)}_{(0,0,1)}-S^{(-2,4)}$ & 0.3373 & 1.080 & 1.615 & 2.723 & 33.83 & 6.714 & 4.809 \\ \hline
$C^{(-1,0,1)}_{(1,0,1)}-S^{(-2,4)}$ & 0.3373 & 1.080 & 1.397 & 2.730 & 33.84 & 6.705 & 4.804 \\ \hline
$C^{(-1,0,1)}_{(3,0,1)}-S^{(-2,4)}$ &  0.3373 & 1.080 & 8.379 & 2.726 & 33.84 & 6.670 & 4.801 \\ \hline

 $C^{(1,0,1)}_{(3,0,1)}-S^{(-2,4)}$  &  0.3373 & 1.080 & 0.7895 & 2.723 & 33.83 & 6.737 & 4.808 \\ \hline
 $C^{(2,0,1)}_{(0,0,1)}-S^{(-2,4)}$ &   0.3373 & 1.080 & 0.8622 & 2.723 & 33.84 & 6.738 & 4.808 \\ \hline
 $C^{(3,0,1)}_{(1,0,1)}-S^{(-2,4)}$ &  0.3373 & 1.080 & 1.436 & 2.723 & 33.84 & 6.737 & 4.798 \\ \hline
$C^{(3,0,1)}_{(3,0,1)}-S^{(-2,4)}$ &  0.3373 & 1.080 & 1.688 & 2.724 & 33.84 & 6.734 & 4.800 \\ \hline
$C^{(4,0,1)}_{(0,0,1)}-S^{(-2,4)}$  &  0.3373 & 1.080 & 1.216 & 2.724 & 33.84 & 6.736 & 4.807 \\ \hline
 $C^{(5,0,1)}_{(1,0,1)}-S^{(-2,4)}$ &  0.3373 & 1.080 & 1.635 & 2.720 & 33.84 & 6.745 & 4.807 \\ \hline \hline
\end{tabular}
\caption{\label{tab:best_fit_noGCP_input}
The best fit values of the input parameters at the minimum values of $\chi^2$ for the 36 viable models.}
\end{table}

\begin{table}[t!]
\centering
\small
\renewcommand{\arraystretch}{1.2}
\renewcommand{\tabcolsep}{1.mm}
\begin{tabular}{|c|c|c|c|c|c|c|c|c|c|c|c|}  \hline \hline

\texttt{Models} & $\sin^2\theta_{13}$ & $\sin^2\theta_{12}$ & $\sin^2\theta_{23}$ & $\delta_{CP}/\pi$  &   $\alpha_{21}/\pi$ &   $m_2$/meV & $m_3$/meV &    $m_{\beta\beta}$/meV & $m_{\beta}$/meV \\ \hline
 
$C^{(-4,2,-1)}_{(0,0,3)}-S^{(-2,-4)}$ &   0.02099 &  0.3169 &  0.4615 &  1.086 &  1.302 &  8.710 &  50.75 &  2.823 &  8.809  \\ \hline
$C^{(-3,2,-1)}_{(1,0,3)}-S^{(-2,-4)} $ &  0.02100 &  0.3169 &  0.4615 &  1.086 &  1.302 &  8.709 &  50.75 &  2.823 &  8.810  \\ \hline
$C^{(-3,2,-1)}_{(3,0,3)}-S^{(-2,-4)} $ &  0.02099 &  0.3169 &  0.4615 &  1.086 &  1.302 &  8.709 &  50.75 &  2.823 &  8.810  \\ \hline
$C^{(-2,2,-1)}_{(0,0,3)}-S^{(-2,-4)} $ &   0.02099 &  0.3170 &  0.4615 &  1.086 &  1.302 &  8.709 &  50.75 &  2.823 &  8.809  \\ \hline
$C^{(-1,2,-1)}_{(1,0,3)}-S^{(-2,-4)} $ &   0.02100 &  0.3169 &  0.4615 &  1.086 &  1.302 &  8.709 &  50.75 &  2.822 &  8.811  \\ \hline
$C^{(0,2,-1)}_{(0,0,3)}-S^{(-2,-4)} $ &  0.02099 &  0.3170 &  0.4615 &  1.086 &  1.302 &  8.709 &  50.75 &  2.823 &  8.810  \\ \hline

$C^{(1,2,-1)}_{(1,0,3)}-S^{(-2,-4)}$ &  0.02100&  0.3170 &  0.4615 &  1.086 &  1.302 &  8.709 &  50.75 &  2.823 &  8.810  \\ \hline
$C^{(1,2,-1)}_{(3,0,3)}-S^{(-2,-4)} $ &   0.02099 &  0.3170 &  0.4615 &  1.086 &  1.302 &  8.709 &  50.75 &  2.823 &  8.810  \\ \hline
 $C^{(3,2,-1)}_{(1,0,3)}-S^{(-2,-4)} $ &   0.02100 &  0.3169 &  0.4615 &  1.086 &  1.302 &  8.709 &  50.75 &  2.823 &  8.810  \\ \hline
$C^{(3,2,-1)}_{(3,0,3)}-S^{(-2,-4)} $ &    0.02100 &  0.3170 &  0.4615 &  1.086 &  1.302 &  8.709 &  50.75 &  2.823 &  8.810  \\ \hline
 $C^{(4,2,-1)}_{(0,0,3)}-S^{(-2,-4)} $ &   0.02099 &  0.3170 &  0.4615 &  1.086 &  1.302 &  8.709 &  50.75 &  2.824 &  8.810  \\ \hline
$C^{(5,2,-1)}_{(1,0,3)}-S^{(-2,-4)} $ &  0.02099 &  0.3170 &  0.4615 &  1.086 &  1.302 &  8.709 &  50.75 &  2.823 &  8.810  \\ \hline

 $C^{(-4,-3,-1)}_{(0,1,3)}-S^{(-2,4)}$ &   0.02180 &  0.3285 &  0.5309 &  1.717 &  0.9832 &  8.688 &  50.75 &  3.254 &  8.967  \\ \hline
$C^{(-3,-3,-1)}_{(3,1,3)}-S^{(-2,4)}$ &  0.02180 &  0.3284 &  0.5309 &  1.718 &  0.9833 &  8.689 &  50.75 &  3.253 &  8.967  \\ \hline
 $C^{(-2,-3,-1)}_{(0,1,3)}-S^{(-2,4)}$ &   0.02180 &  0.3285 &  0.5309 &  1.717 &  0.9832 &  8.690 &  50.75 &  3.254 &  8.968  \\ \hline
 $C^{(-1,-3,-1)}_{(1,1,3)}-S^{(-2,4)} $ &  0.02180 &  0.3285 &  0.5309 &  1.717 &  0.9832 &  8.689 &  50.75 &  3.255 &  8.968  \\ \hline
 $C^{(0,-3,-1)}_{(0,1,3)}-S^{(-2,4)} $ &   0.02180 &  0.3285 &  0.5309 &  1.718 &  0.9832 &  8.690 &  50.75 &  3.254 &  8.968  \\ \hline
$C^{(1,-3,-1)}_{(1,1,3)}-S^{(-2,4)} $ &  0.02180 &  0.3285 &  0.5309 &  1.718 &  0.9832 &  8.690 &  50.75 &  3.254 &  8.967  \\ \hline

 $C^{(1,-3,-1)}_{(3,1,3)}-S^{(-2,4)} $ &   0.02180 &  0.3285 &  0.5309 &  1.717 &  0.9832 &  8.689 &  50.75 &  3.254 &  8.967  \\ \hline
 $C^{(2,-3,-1)}_{(0,1,3)}-S^{(-2,4)} $ &  0.02179 &  0.3284 &  0.5309 &  1.717 &  0.9832 &  8.689 &  50.75 &  3.254 &  8.967  \\ \hline
 $C^{(3,-3,-1)}_{(1,1,3)}-S^{(-2,4)} $ &   0.02180 &  0.3284 &  0.5309 &  1.717 &  0.9833 &  8.689 &  50.75 &  3.254 &  8.967  \\ \hline
$C^{(3,-3,-1)}_{(3,1,3)}-S^{(-2,4)} $ &   0.02180 &  0.3284 &  0.5308 &  1.718 &  0.9833 &  8.689 &  50.75 &  3.252 &  8.967  \\ \hline
$C^{(4,-3,-1)}_{(0,1,3)}-S^{(-2,4)} $ &  0.02180 &  0.3285 &  0.5309 &  1.717 &  0.9832 &  8.689 &  50.75 &  3.254 &  8.968  \\ \hline
 $C^{(5,-3,-1)}_{(1,1,3)}-S^{(-2,4)} $ &   0.02181 &  0.3285 &  0.5309 &  1.717 &  0.9832 &  8.689 &  50.75 &  3.255 &  8.968  \\ \hline

$C^{(-4,0,1)}_{(0,0,1)}-S^{(-2,4)}$ &   0.02274 &  0.3045 &  0.5487 &  1.564 &  0.2245 &  8.656 &  50.75 &  2.305 &  8.993  \\ \hline
 $C^{(-3,0,1)}_{(1,0,1)}-S^{(-2,4)} $ &   0.02273 &  0.3045 &  0.5487 &  1.564 &  0.2245 &  8.657 &  50.75 &  2.307 &  8.992  \\ \hline
 $C^{(-3,0,1)}_{(3,0,1)}-S^{(-2,4)} $ &   0.02274 &  0.3045 &  0.5487 &  1.564 &  0.2244 &  8.657 &  50.75 &  2.307 &  8.994  \\ \hline
$C^{(-2,0,1)}_{(0,0,1)}-S^{(-2,4)} $ &   0.02274 &  0.3045 &  0.5487 &  1.564 &  0.2245 &  8.656 &  50.75 &  2.306 &  8.992  \\ \hline
 $C^{(-1,0,1)}_{(1,0,1)}-S^{(-2,4)} $ &   0.02274 &  0.3046 &  0.5486 &  1.564 &  0.2243 &  8.656 &  50.75 &  2.306 &  8.993  \\ \hline
$C^{(-1,0,1)}_{(3,0,1)}-S^{(-2,4)} $ &  0.02274 &  0.3046 &  0.5486 &  1.564 &  0.2244 &  8.656 &  50.75 &  2.307 &  8.994  \\ \hline

$C^{(1,0,1)}_{(3,0,1)}-S^{(-2,4)} $ &   0.02274 &  0.3045 &  0.5487 &  1.564 &  0.2245 &  8.656 &  50.75 &  2.306 &  8.993  \\ \hline
$C^{(2,0,1)}_{(0,0,1)}-S^{(-2,4)} $&  0.02274 &  0.3045 &  0.5487 &  1.564 &  0.2244 &  8.657 &  50.75 &  2.306 &  8.993  \\ \hline
 $C^{(3,0,1)}_{(1,0,1)}-S^{(-2,4)} $ &  0.02274 &  0.3045 &  0.5487 &  1.564 &  0.2244 &  8.657 &  50.75 &  2.306 &  8.993  \\ \hline
 $C^{(3,0,1)}_{(3,0,1)}-S^{(-2,4)} $ &   0.02274 &  0.3046 &  0.5487 &  1.564 &  0.2245 &  8.656 &  50.75 &  2.307 &  8.993  \\ \hline
 $C^{(4,0,1)}_{(0,0,1)}-S^{(-2,4)} $  &  0.02274 &  0.3045 &  0.5487 &  1.564 &  0.2244 &  8.656 &  50.75 &  2.306 &  8.993  \\ \hline
$C^{(5,0,1)}_{(1,0,1)}-S^{(-2,4)} $ &    0.02273 &  0.3045 &  0.5487 &  1.564 &  0.2245 &  8.656 &  50.75 &  2.306 &  8.992  \\ \hline  \hline
\end{tabular}
\caption{\label{tab:best_fit_noGCP_output}The best fit values of the  neutrino masses and mixing parameters at the minimum values of $\chi^2$ for the 36 viable models.}
\end{table}

\section{\label{sec:example_model}Lepton masses and mixings in three representative models}

In section~\ref{sec:num_analysis}, we identify 36 minimal lepton flavor models that satisfy all phenomenological constraints as shown in table~\ref{tab:best_fit_noGCP_input} and table~\ref{tab:best_fit_noGCP_output}. These viable models can be partitioned into three distinct categories, classified according to the modular transformations of lepton fields. Each category contains twelve models differing in the assignment of the first generation RH charged lepton field $E^c_1$. Moreover, the models within the same category lead to nearly identical predictions for the charged lepton masses, neutrino masses and lepton mixing parameters. Consequently, it is sufficient to choose one representative model from each category to illustrate the predictive power of the non-holomorphic $S_4^\prime$ framework. For this purpose, we present detailed numerical analyses for the three benchmark models $C^{(-4,2,-1)}_{(0,0,3)}-S^{(-2,-4)}$, $C^{(-4,-3,-1)}_{(0,1,3)}-S^{(-2,4)}$, and $C^{(-4,0,1)}_{(0,0,1)}-S^{(-2,4)}$. The explicit assignments of the representations and modular weights of lepton fields are given as: 
\begin{eqnarray}
\nonumber C^{(-4,2,-1)}_{(0,0,3)}-S^{(-2,-4)} &:&L\sim\bm{3_{3}},\qquad E_1^c\sim\bm{1_{1}},\qquad E_2^c\sim\bm{1_{1}},\qquad E_3^c\sim\bm{1_{2}} ,\qquad N^{c}\sim\bm{2_{1}}\,, \\
\nonumber  && k_{L}=0, \qquad k_{E^{c}_{1}}=-4 \qquad k_{E^{c}_{2}}=2, \qquad k_{E^{c}_{3}}=-1, \qquad k_{N^{c}}=-2\,, \\
\nonumber C^{(-4,-3,-1)}_{(0,1,3)}-S^{(-2,4)}&:&L\sim\bm{3_{3}},\qquad E_1^c\sim\bm{1_{1}},\qquad E_2^c\sim\bm{1_{0}},\qquad E_3^c\sim\bm{1_{2}},\qquad N^{c}\sim\bm{2_{1}}\,, \\
\nonumber  && k_{L}=-4, \qquad k_{E^{c}_{1}}=0 \qquad k_{E^{c}_{2}}=1, \qquad k_{E^{c}_{3}}=3, \qquad k_{N^{c}}=2\,, \\
\nonumber C^{(-4,0,1)}_{(0,0,1)}-S^{(-2,4)}  &:& L\sim\bm{3_{3}},\qquad E_1^c\sim\bm{1_{1}},\qquad E_2^c\sim\bm{1_{1}},\qquad E_3^c\sim\bm{1_{0}},\qquad N^{c}\sim\bm{2_{1}}\,, \\
 && k_{L}=-4, \qquad k_{E^{c}_{1}}=0 \qquad k_{E^{c}_{2}}=4, \qquad k_{E^{c}_{3}}=5, \qquad k_{N^{c}}=2\,.
\end{eqnarray}
From these assignments, the charged lepton mass matrix $M_{l}$, the heavy Majorana neutrino mass matrix $M_{N}$, and the Dirac neutrino mass matrix $M_{D}$ can be read from Eqs.~\eqref{eq:MCh1}, \eqref{eq:nu_masses_MN} and~\eqref{eq:nu_masses_MD}, and they are summarized in table~\ref{tab:3Ms_mass}.

\begin{table}[t!]
\centering
\renewcommand{\tabcolsep}{0.5mm}
\begin{tabular}{|c|c|c|c|c|c|c|c|}
\hline \hline
& & &  \\[-0.16in]
\texttt{Models} &  $C^{(-4,2,-1)}_{(0,0,3)}-S^{(-2,-4)}$   & $C^{(-4,-3,-1)}_{(0,1,3)}-S^{(-2,4)}$ & $C^{(-4,0,1)}_{(0,0,1)}-S^{(-2,4)}$ \\[0.050in] \hline
& & &  \\[-0.14in]

$M_{l}/v$     & $\begin{pmatrix}
y_{e} Y^{(-4)}_{\bm{3_{0}},1} & y_{e} Y^{(-4)}_{\bm{3_{0}},3} & y_{e} Y^{(-4)}_{\bm{3_{0}},2} \\
 y_{\mu} Y^{(2)}_{\bm{3_{0}},1} &  y_{\mu} Y^{(2)}_{\bm{3_{0}},3} &  y_{\mu} Y^{(2)}_{\bm{3_{0}},2}  \\
y_{\tau} Y^{(-1)}_{\bm{3_{3}},1} & y_{\tau} Y^{(-1)}_{\bm{3_{3}},3} & y_{\tau} Y^{(-1)}_{\bm{3_{3}},2} 
\end{pmatrix}$ & $\begin{pmatrix}
y_{e} Y^{(-4)}_{\bm{3_{0}},1} & y_{e} Y^{(-4)}_{\bm{3_{0}},3} & y_{e} Y^{(-4)}_{\bm{3_{0}},2} \\
y_{\mu} Y^{(-3)}_{\bm{3_{1}},1} & y_{\mu} Y^{(-3)}_{\bm{3_{1}},3} & y_{\mu} Y^{(-3)}_{\bm{3_{1}},2} \\
y_{\tau} Y^{(-1)}_{\bm{3_{3}},1} & y_{\tau} Y^{(-1)}_{\bm{3_{3}},3} & y_{\tau} Y^{(-1)}_{\bm{3_{3}},2} 
\end{pmatrix} $ & $\begin{pmatrix}
y_{e} Y^{(-4)}_{\bm{3_{0}},1} & y_{e} Y^{(-4)}_{\bm{3_{0}},3} & y_{e} Y^{(-4)}_{\bm{3_{0}},2} \\
y_{\mu} Y^{(0)}_{\bm{3_{0}},1} & y_{\mu} Y^{(0)}_{\bm{3_{0}},3} & y_{\mu} Y^{(0)}_{\bm{3_{0}},2} \\
y_{\tau} Y^{(1)}_{\bm{3_{1}},1} & y_{\tau} Y^{(1)}_{\bm{3_{1}},3} & y_{\tau} Y^{(1)}_{\bm{3_{1}},2} 
\end{pmatrix}$ \\ [0.30in]\hline
& & \multicolumn{2}{c|}{}  \\[-0.14in]

$M_{N}$ & $\Lambda\left(
\begin{array}{cc}
Y^{(-4)}_{\bm{2_{0}},2} &Y^{(-4)}_{\bm{2_{0}},1} \\
Y^{(-4)}_{\bm{2_{0}},1} & -Y^{(-4)}_{\bm{2_{0}},2} \\
\end{array}
\right)$ & \multicolumn{2}{c|}{$\Lambda\left(
\begin{array}{cc}
Y^{(4)}_{\bm{2_{0}},2} &Y^{(4)}_{\bm{2_{0}},1} \\
Y^{(4)}_{\bm{2_{0}},1} & -Y^{(4)}_{\bm{2_{0}},2} \\
\end{array}
\right)$} \\ [0.18in]\hline

& \multicolumn{3}{c|}{}  \\[-0.14in] 
$M_{D}$ & \multicolumn{3}{c|}{$gv\left(
\begin{array}{ccc}
 -2 Y^{(-2)}_{\bm{3_{0}},1} & Y^{(-2)}_{\bm{3_{0}},3} & Y^{(-2)}_{\bm{3_{0}},2} \\
 0 & -\sqrt{3} Y^{(-2)}_{\bm{3_{0}},2} & -\sqrt{3} Y^{(-2)}_{\bm{3_{0}},3} \\
\end{array}
\right)$} \\ [0.18in]\hline \hline
\end{tabular}
\caption{\label{tab:3Ms_mass} The charged lepton, Majorana neutrino and Dirac neutrino mass matrices of the three benchmark models $C^{(-4,2,-1)}_{(0,0,3)}-S^{(-2,-4)}$, $C^{(-4,-3,-1)}_{(0,1,3)}-S^{(-2,4)}$ and $C^{(-4,0,1)}_{(0,0,1)}-S^{(-2,4)}$. }
\end{table}

\begin{figure}[t!]
\centering
\includegraphics[width=2.8in]{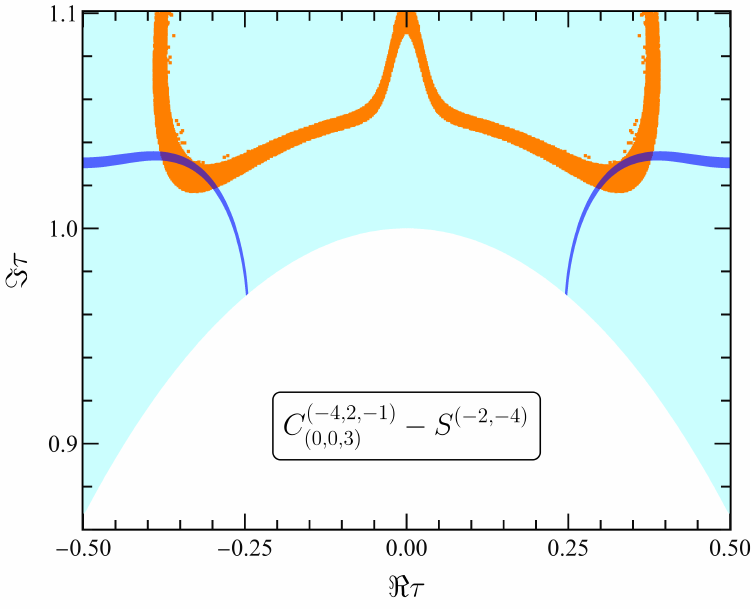}
\includegraphics[width=2.8in]{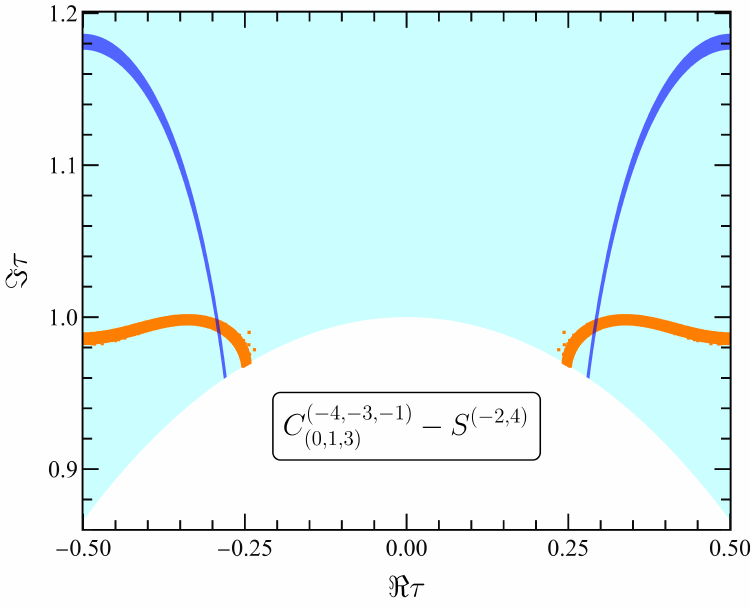}
\includegraphics[width=2.8in]{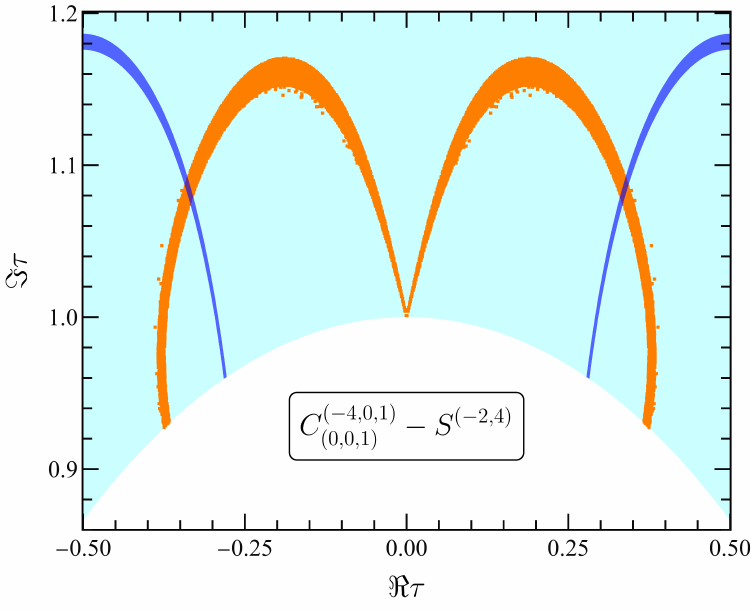}
\caption{Regions of the modulus $\tau$ compatible with experimental data. 
The cyan area denotes the fundamental domain of $\tau$. The blue regions correspond to values of $\tau$ consistent with the 
experimentally allowed range of $\Delta m_{21}^{2}/\Delta m_{31}^{2}$~\cite{Esteban:2024eli} for the benchmark models, respectively. The orange area denotes the viable region of $\tau$ limited only by the measured values of the reactor mixing angle $\theta_{13}$ ~\cite{Esteban:2024eli,Xing:2007fb}. }
\label{fig:mass_ratio}
\end{figure}

For these three benchmark models, the best-fit values of the input parameters and the corresponding predictions for the physical observables are presented in tables~\ref{tab:best_fit_noGCP_input} and~\ref{tab:best_fit_noGCP_output}, respectively. One finds that the predicted lepton observables lie within the $3\sigma$ ranges of the latest global analysis, \texttt{NuFIT} v6.1 with SK atmospheric neutrino data~\cite{Esteban:2024eli}. It is worth noting that the effective neutrino mass matrices in these models depend only on the VEV of the modulus $\tau$ and an overall mass scale $g^{2}v^{2}/\Lambda$. Consequently, the experimentally measured ratio of neutrino mass squared differences, $\Delta m_{21}^{2}/\Delta m_{31}^{2}$, can be used to constrain the allowed range of $\tau$. The corresponding regions are shown in blue in figure~\ref{fig:mass_ratio}. In addition, the orange regions in figure~\ref{fig:mass_ratio} indicate the parameter space of $\tau$ consistent with the experimental constraints on the reactor mixing angle $\theta_{13}$.

To further explore the phenomenological implications of these models, we employ the sampler \texttt{MultiNest}~\cite{Feroz:2007kg,Feroz:2008xx} to scan the parameter space, imposing that all measured lepton observables lie within their experimentally preferred $3\sigma$ ranges\footnote{The $3\sigma$ uncertainties assigned to the charged lepton mass ratios are taken to be $0.3\%$ of their central values.}. Under these constraints, the three leptonic mixing angles, the two CP-violating phases, the neutrino masses, the effective Majorana mass $m_{\beta\beta}$ relevant for $0\nu\beta\beta$-decay, and the kinematical mass $m_{\beta}$ for beta decay are all confined to narrow regions. The corresponding predictions are summarized in table~\ref{tab:three_models_res}.

\begin{table}[t!]
\centering
\renewcommand{\arraystretch}{1.2}
\begin{tabular}{|c|c|c|c|c|c|c|c|}
\hline \hline
\texttt{Models}  &  $C^{(-4,2,-1)}_{(0,0,3)}-S^{(-2,-4)}$   & $C^{(-4,-3,-1)}_{(0,1,3)}-S^{(-2,4)}$  & $C^{(-4,0,1)}_{(0,0,1)}-S^{(-2,4)}$  \\ \hline

$\sin^2\theta_{13}$ & $[0.02064,0.02336]$ &  $[0.02064,0.02275]$ & $[0.02064,0.02418]$\\ 

$\sin^2\theta_{12}$ & $[0.2929,0.3295]$ &  $[0.3172,0.3295]$ & $[0.2893,0.3276]$ \\ 

$\sin^2\theta_{23}$  & $[0.458,0.471]$  & $[0.518,0.536]$ & $[0.542,0.556]$ \\ 

$\delta_{CP}/\pi$  & $[0.910, 0.916] \cup[1.084, 1.090]$ & $[1.707, 1.760]$ & $[1.552, 1.574]$ \\ 

$\alpha_{21}/\pi$  & $[0.689, 0.713]\cup[1.287, 1.311]$ & $[0.983, 0.985]$ & $[0.198, 0.263]$ \\   

$m_{2}/\text{meV}$  & $[8.503,8.843]$ & $[8.504,8.843]$ & $[8.504,8.843]$ \\   

$m_{3}/\text{meV}$  & $[49.50,50.75]$ &  $[49.50,50.75]$ & $[49.50,50.75]$ \\    

$\sum_{i}m_{i}/\text{meV}$  & $[58.00,59.60]$ & $[58.00,59.60]$ & $[58.00,59.59]$ \\    

$m_{\beta\beta}/\text{meV}$  & $[2.635,2.994]$ & $[2.918,3.293]$ & $[2.187,2.387]$ \\  

$m_{\beta}/\text{meV}$  & $[8.421,9.251]$ & $[8.570,9.085]$ & $[8.527,9.252]$ \\   \hline  \hline

\end{tabular}
\caption{\label{tab:three_models_res} The prediction ranges for the lepton mixing parameters, neutrino masses, the effective Majorana mass for $0\nu\beta\beta$-decay and the kinematical mass $m_{\beta}$ for beta decay in the three benchmark models. }
\end{table}

\begin{figure}[t!]
\centering
\includegraphics[width=5.0in]{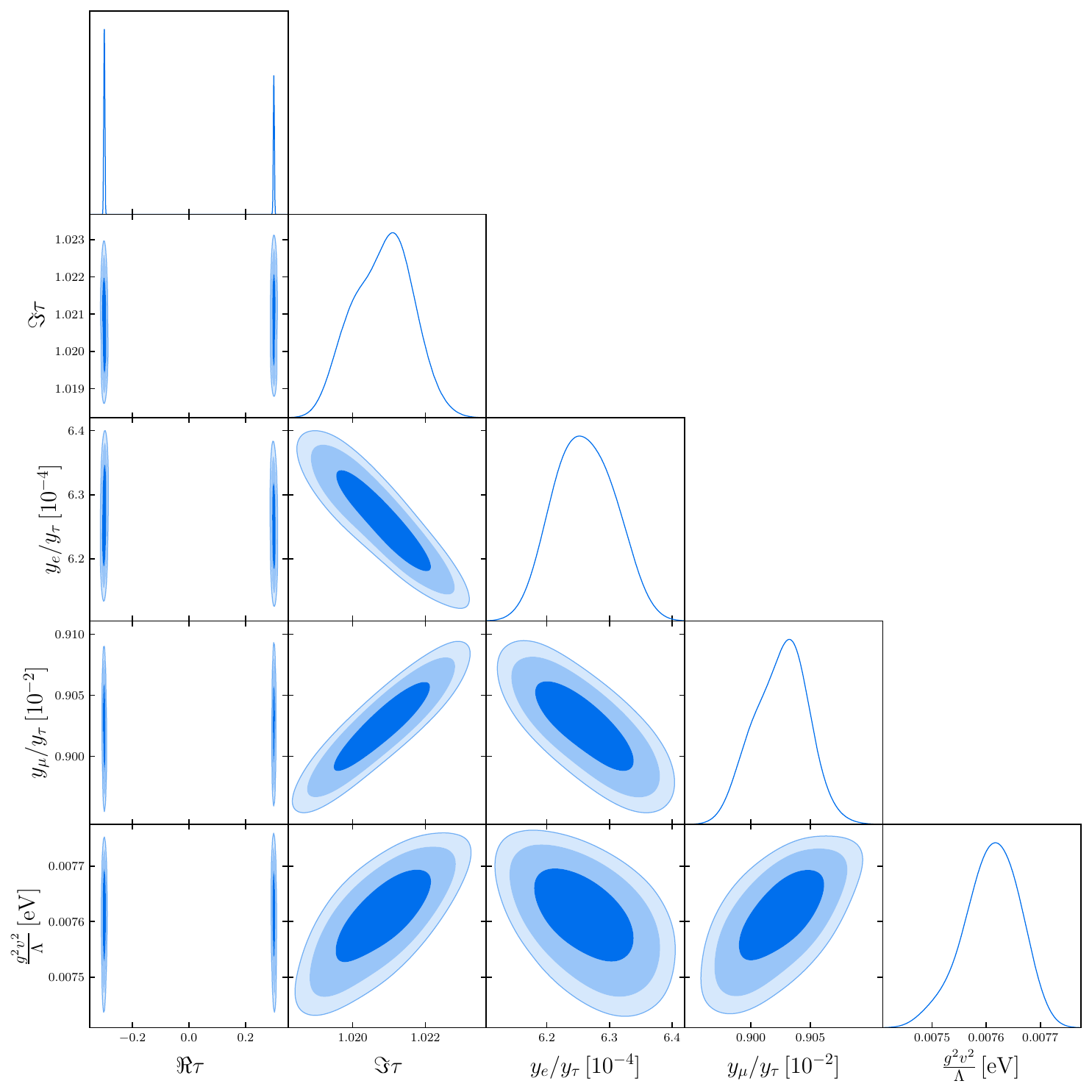}
\caption{\label{fig:input_model1}Allowed regions for the lepton input parameters for case $C^{(-4,2,-1)}_{(0,0,3)}-S^{(-2,-4)}$. Different color shadings correspond to the $1\sigma$, $2\sigma$, and $3\sigma$ confidence levels. }
\label{fig:lepton_input_1}
\end{figure}

\begin{figure}[t!]
\centering
\includegraphics[width=5.0in]{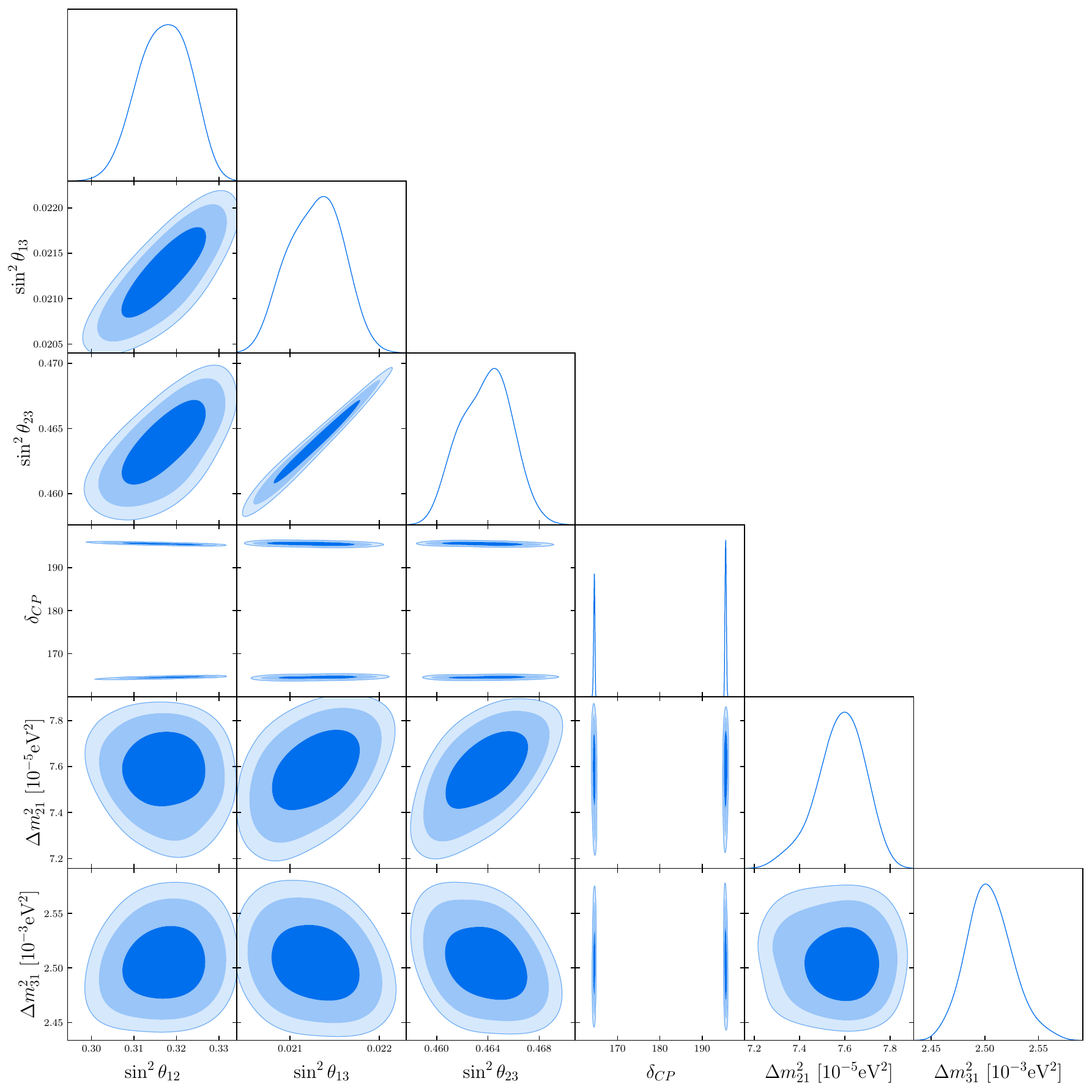}
\caption{\label{fig:output_model1}Allowed regions for the lepton observables for case $C^{(-4,2,-1)}_{(0,0,3)}-S^{(-2,-4)}$. Different color shadings correspond to the $1\sigma$, $2\sigma$, and $3\sigma$ confidence levels. }
\label{fig:lepton_obs_1}
\end{figure}

One finds that the three representative models are mainly distinguished by their predictions for the atmospheric mixing angle $\theta_{23}$ and the CP-violating phases $\delta_{CP}$ and $\alpha_{21}$, all of which are confined to relatively narrow intervals. Specifically, the atmospheric mixing angle $\theta_{23}$ lies in the first octant, $\sin^{2}\theta_{23} \in [0.458,\,0.471]$, for the model $C^{(-4,2,-1)}_{(0,0,3)}-S^{(-2,-4)}$, while it is located in the second octant for the other two models, with $\sin^{2}\theta_{23} \in [0.518,\,0.536]$ for $C^{(-4,-3,-1)}_{(0,1,3)}-S^{(-2,4)}$ and $\sin^{2}\theta_{23} \in [0.542,\,0.556]$ for $C^{(-4,0,1)}_{(0,0,1)}-S^{(-2,4)}$. The allowed regions for the Dirac CP-violating phase $\delta_{CP}$ are $[0.910\pi,\,0.916\pi] \cup [1.084\pi,\,1.090\pi]$ for $C^{(-4,2,-1)}_{(0,0,3)}-S^{(-2,-4)}$, $[1.707\pi,\,1.760\pi]$ for $C^{(-4,-3,-1)}_{(0,1,3)}-S^{(-2,4)}$, and $[1.552\pi,\,1.574\pi]$ for $C^{(-4,0,1)}_{(0,0,1)}-S^{(-2,4)}$. The forthcoming long-baseline neutrino experiments DUNE~\cite{DUNE:2020ypp} and T2HK~\cite{Hyper-KamiokandeProto-:2015xww,Hyper-Kamiokande:2018ofw} are designed to achieve precision measurements of the CP-violating phase $\delta_{CP}$ and to resolve the octant of the atmospheric mixing angle $\theta_{23}$. Their sensitivities to $\sin^2\theta_{23}$ and $\delta_{CP}$ exhibit a strong dependence on the true parameter values. For $\theta_{23}$ around $40^{\circ}$ or $50^{\circ}$, the expected precision on $\theta_{23}$ is approximately $0.2^{\circ}$ ($0.13^{\circ}$) for DUNE (T2HK), worsening to $\sim 2^{\circ}$ ($0.95^{\circ}$) at $\theta_{23}=45^{\circ}$. Combining DUNE and T2HK, $\delta_{CP}$ can be constrained with a $1\sigma$ uncertainty of $4.5^{\circ}$--$11^{\circ}$ after 10 years of running~\cite{Ballett:2016daj}. These high-precision measurements of $\theta_{23}$ and $\delta_{CP}$ could help distinguish between the three lepton flavor models. Furthermore, the allowed ranges of $\sin^{2}\theta_{13}$ and $\sin^{2}\theta_{12}$ span almost the entire $3\sigma$ intervals for the models $C^{(-4,2,-1)}_{(0,0,3)}-S^{(-2,-4)}$ and $C^{(-4,0,1)}_{(0,0,1)}-S^{(-2,4)}$. In contrast, for the model $C^{(-4,-3,-1)}_{(0,1,3)}-S^{(-2,4)}$, the predictions are more restricted, with $\sin^{2}\theta_{13}\in[0.02064,\,0.02275]$ and $\sin^{2}\theta_{12}\in[0.3172,\,0.3295]$. The running JUNO experiment is expected to improve the precision of $\Delta m^2_{21}$ and $\sin^{2}\theta_{12}$ to a world-leading level of $0.3\%$ and $0.5\%$, respectively, after six years of data taking~\cite{JUNO:2025gmd,JUNO:2022mxj}. The resulting high-precision determination of $\sin^{2}\theta_{12}$ will therefore provide a stringent test of these models.

In addition, we find that the neutrino mass observables $\sum_i m_i$ and $m_{\beta\beta}$ are also strongly constrained. The predicted sum of neutrino masses is nearly identical across the three models, lying in the range $[58.00~\text{meV},\,59.60~\text{meV}]$, which is consistent with the cosmological bound $\sum_i m_i < 120~\text{meV}$ from the Planck collaboration~\cite{Planck:2018vyg}. It is notable that the predicted neutrino mass sum lies within the projected sensitivity of upcoming cosmological surveys  such as Euclid+CMB-S4+LiteBIRD whose projected sensitivity is $\sum m_{i}<(44-76)~\text{meV}$~\cite{Euclid:2024imf}. The corresponding predictions for the effective Majorana mass $m_{\beta\beta}$ are $[2.635,\,2.994]~\text{meV}$, $[2.918,\,3.293]~\text{meV}$, and $[2.187,\,2.387]~\text{meV}$ for the three benchmark models, respectively. The most stringent current limit is provided by the KamLAND-Zen experiment with $m_{\beta\beta} < (28 - 122)~\text{meV}$ at 90\% C.L.~\cite{KamLAND-Zen:2024eml}. Future tonne-scale $0\nu\beta\beta$-decay experiments such as LEGEND-1000~\cite{LEGEND:2021bnm} and nEXO~\cite{nEXO:2021ujk} are expected to reach sensitivities of $m_{\beta\beta} \sim (9 - 21)~\text{meV}$ and $m_{\beta\beta} \sim (4.7 - 20.3)~\text{meV}$, respectively, for a ten-year exposure. For the kinematical mass $m_{\beta}$, the allowed ranges are $[8.421,\,9.251]~\text{meV}$, $[8.570,\,9.085]~\text{meV}$, and $[8.527,\,9.252]~\text{meV}$ for the three representative models. The current bound reported by the KATRIN experiment is $m_{\beta} < 0.45~\text{eV}$ at 90\% C.L.~\cite{KATRIN:2024cdt}, with future sensitivities expected to improve to $m_{\beta} < 0.2~\text{eV}$ for KATRIN~\cite{KATRIN:2021dfa} and $m_{\beta} < 0.04~\text{eV}$ for Project~8~\cite{Project8:2022wqh}. It follows that the predicted values of both $m_{\beta\beta}$ and $m_{\beta}$ in these models are below the reach of current and near-future experiments.

\begin{figure}[t!]
\centering
\includegraphics[width=5.0in]{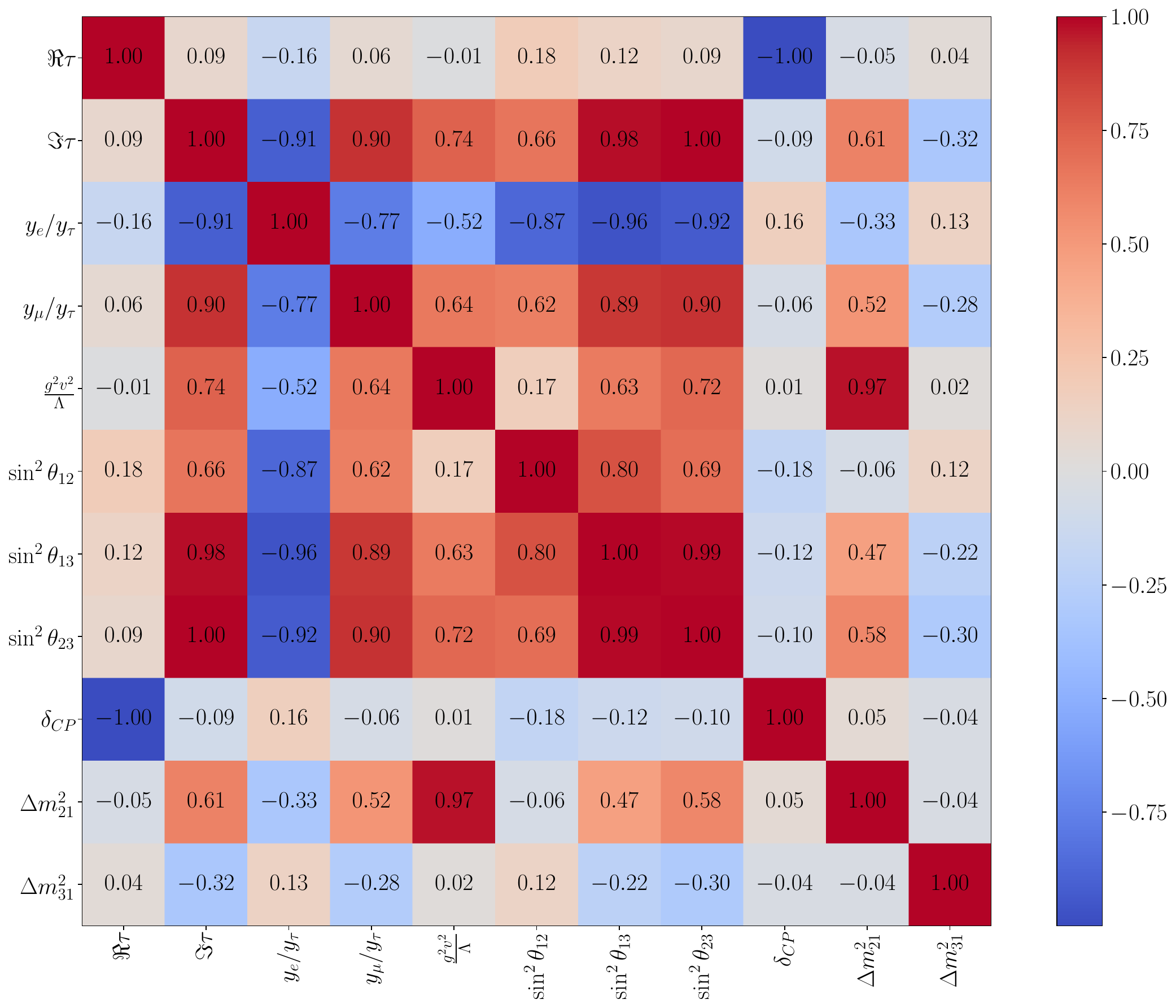}
\caption{Matrix of the correlations among the model parameters and lepton observables for the
lepton model $C^{(-4,2,-1)}_{(0,0,3)}-S^{(-2,-4)}$.  }
\label{fig:lepton_correalation_1}
\end{figure}

Furthermore, we investigate the correlations among the free parameters of the model and the predicted observables for the benchmark models. The correlations among the input parameters of the model 
$C^{(-4,2,-1)}_{(0,0,3)}-S^{(-2,-4)}$ are illustrated in figure~\ref{fig:lepton_input_1}. The ratio $y_{e}/y_{\tau}$ shows negative correlations with $\Im \tau$, $y_{\mu}/y_{\tau}$, and $\frac{g^{2}v^{2}}{\Lambda}$, whereas $\Im \tau$ is positively associated with $y_{\mu}/y_{\tau}$ and $\frac{g^{2}v^{2}}{\Lambda}$. A positive relation is also observed between $y_{\mu}/y_{\tau}$ and $\frac{g^{2}v^{2}}{\Lambda}$. It is notable that $\Re \tau$ is confined to two narrow regions around $\pm 0.3$. The correlations among the lepton mixing parameters are presented in figure~\ref{fig:lepton_obs_1}. The three mixing angles $\sin^{2}\theta_{12}$, $\sin^{2}\theta_{13}$, and $\sin^{2}\theta_{23}$ exhibit a common increasing trend, with a particularly strong interplay between $\sin^{2}\theta_{13}$ and $\sin^{2}\theta_{23}$. The Dirac CP-violating phase $\delta_{CP}$ is restricted to two separated regions around $0.91\pi$ and $1.08\pi$, in agreement with table~\ref{tab:three_models_res}. This feature originates from the fact that all couplings are real and $\Re \tau$ serves as the sole source of CP violation. Since $\Re \tau$ is localized around $\pm 0.3$, flipping its sign induces the transformation $\delta_{CP} \to -\delta_{CP}$. To further quantify these relations, we compute the correlation matrix for both input parameters and observables, as shown in figure~\ref{fig:lepton_correalation_1}. One finds that $\Re \tau$ is strongly anti-correlated with $\delta_{CP}$. In contrast, $\Im \tau$ and $y_{\mu}/y_{\tau}$ tend to increase together with the mixing angles and $\Delta m_{21}^{2}$, while $y_{e}/y_{\tau}$ follows the opposite trend. The overall mass scale $\frac{g^{2}v^{2}}{\Lambda}$ is most strongly linked to $\Delta m_{21}^{2}$, and also shows a clear association with $\sin^{2}\theta_{13}$ and $\sin^{2}\theta_{23}$.

\begin{figure}[t!]
\centering
\includegraphics[width=5.0in]{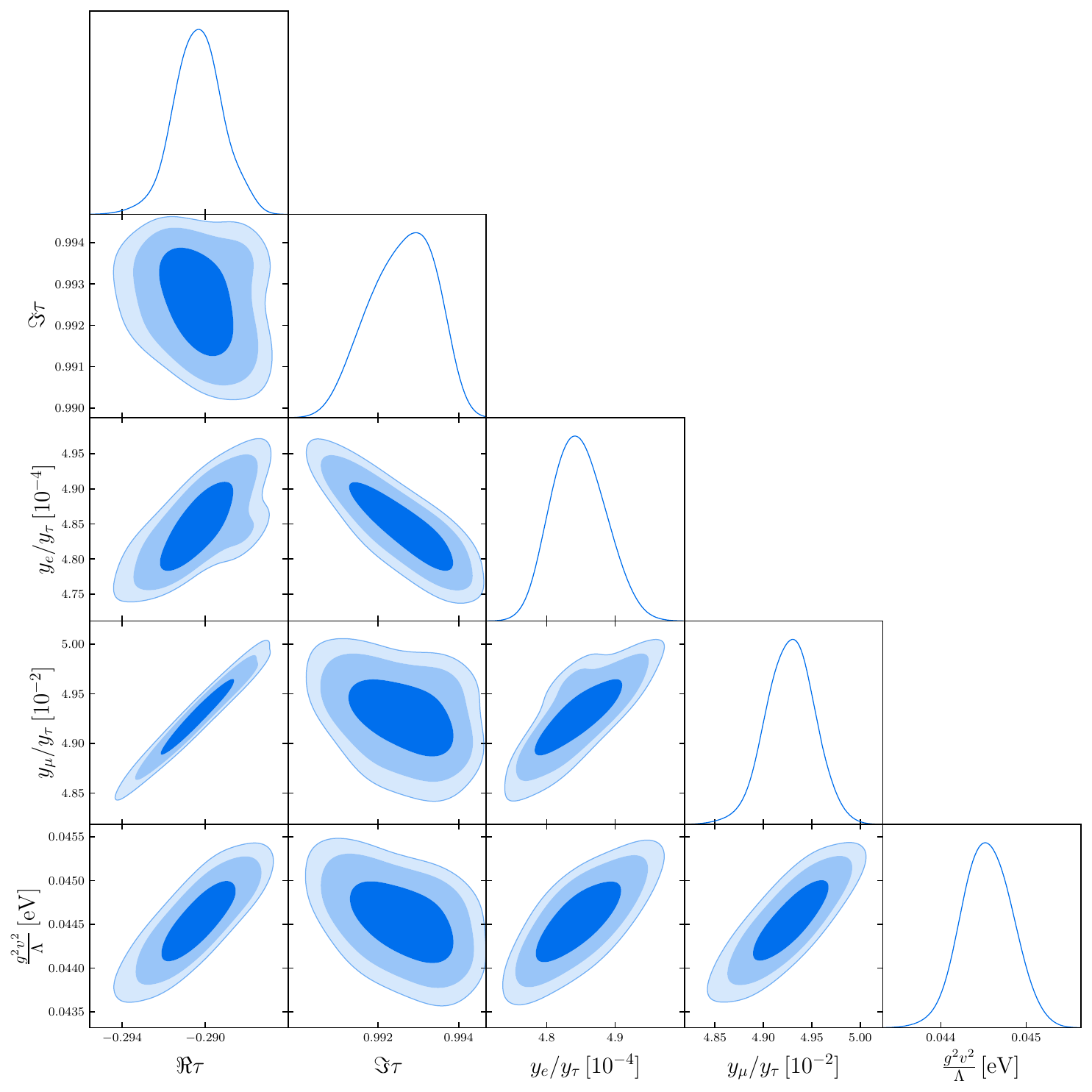}
\caption{Allowed regions for the lepton input parameters for case $C^{(-4,-3,-1)}_{(0,1,3)}-S^{(-2,4)}$. Different color shadings correspond to the $1\sigma$, $2\sigma$, and $3\sigma$ confidence levels. }
\label{fig:lepton_input_2}
\end{figure}

\begin{figure}[t!]
\centering
\includegraphics[width=5.0in]{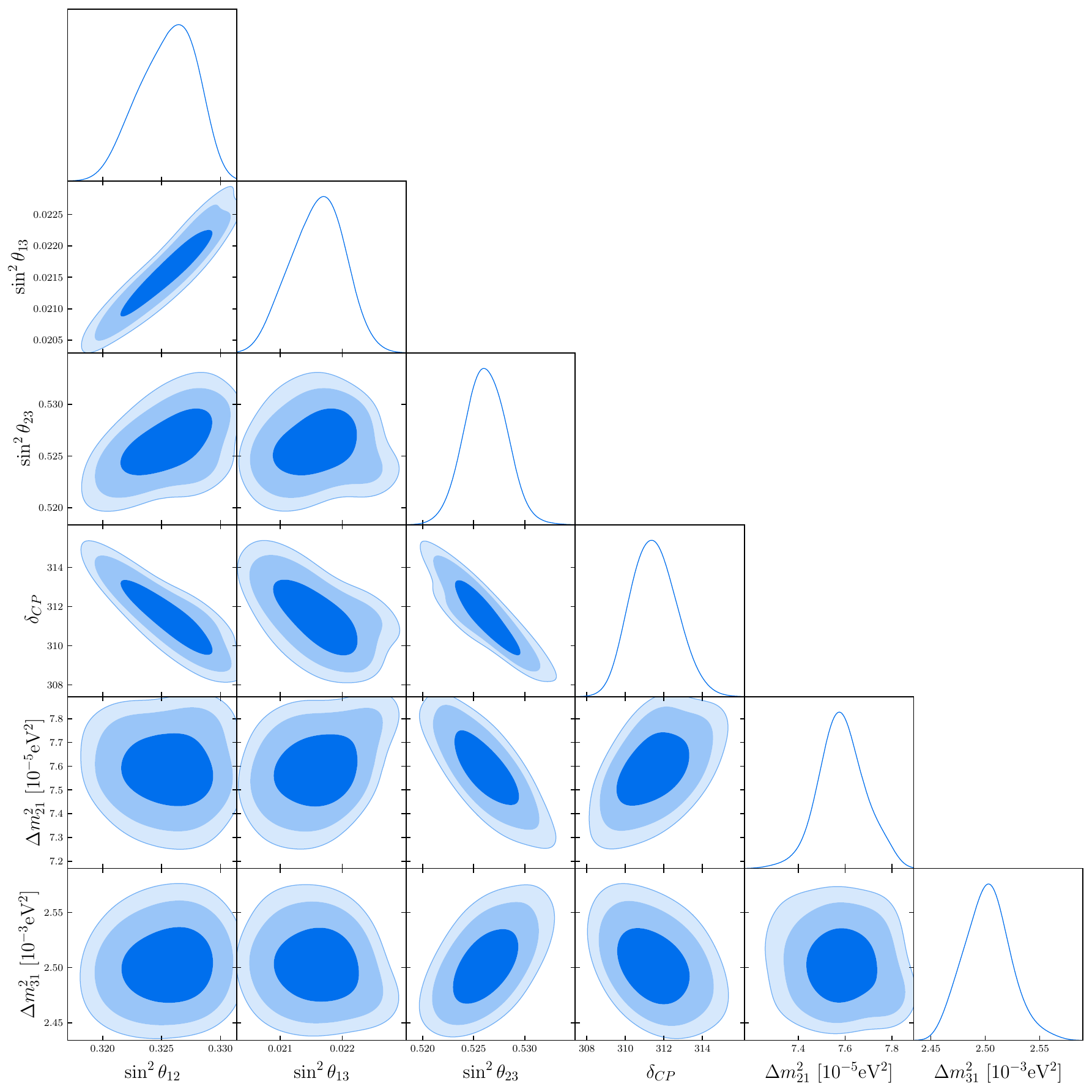}
\caption{Allowed regions for the lepton observables for case $C^{(-4,-3,-1)}_{(0,1,3)}-S^{(-2,4)}$. Different color shadings correspond to the $1\sigma$, $2\sigma$, and $3\sigma$ confidence levels. }
\label{fig:lepton_obs_2}
\end{figure}

\begin{figure}[t!]
\centering
\includegraphics[width=5.0in]{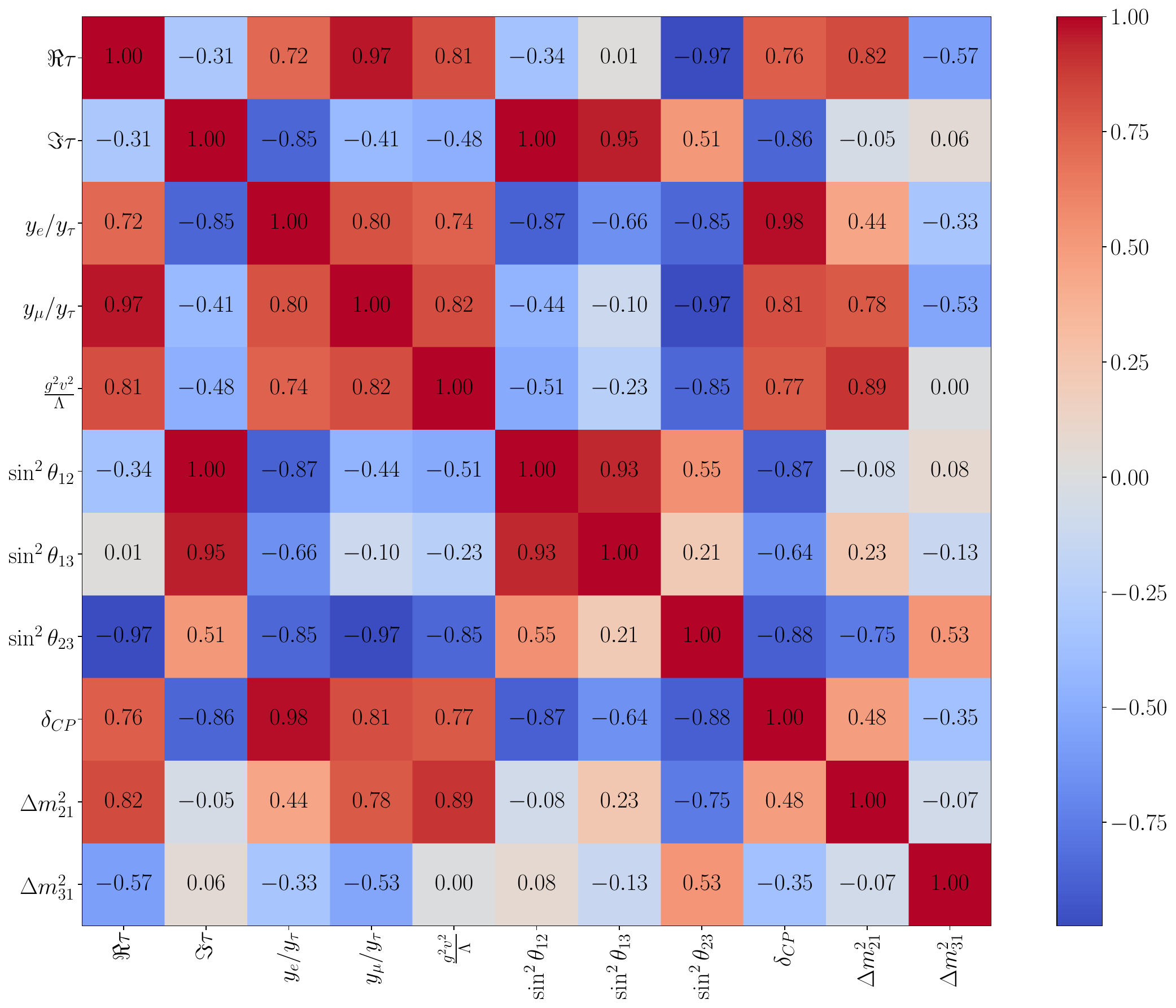}
\caption{Matrix of the correlations among the input parameters and lepton observables for the lepton model $C^{(-4,-3,-1)}_{(0,1,3)}-S^{(-2,4)}$. }
\label{fig:lepton_correalation_2}
\end{figure}

Figure~\ref{fig:lepton_input_2} shows the relations among the input parameters of the model $C^{(-4,-3,-1)}_{(0,1,3)}-S^{(-2,4)}$. The parameters $\Re \tau$, $y_{e}/y_{\tau}$, $y_{\mu}/y_{\tau}$, and $\frac{g^{2}v^{2}}{\Lambda}$ follow a common increasing trend, whereas $\Im \tau$ decreases with respect to $y_{e}/y_{\tau}$. The corresponding behavior of the neutrino observables is illustrated in figure~\ref{fig:lepton_obs_2}. The mixing angles $\sin^{2}\theta_{12}$ and $\sin^{2}\theta_{13}$ evolve coherently, while $\delta_{CP}$ decreases as any of the three mixing angles grows. In addition, $\sin^{2}\theta_{23}$ exhibits an inverse relation with $\Delta m_{21}^{2}$. A more global picture is provided by the correlation matrix in 
figure~\ref{fig:lepton_correalation_2}. The set $\{\Re \tau,\, y_{e}/y_{\tau},\, y_{\mu}/y_{\tau},\, \frac{g^{2}v^{2}}{\Lambda}\}$ tends to vary oppositely to $\sin^{2}\theta_{23}$, but aligns with $\delta_{CP}$ and $\Delta m_{21}^{2}$. Moreover, $y_{e}/y_{\tau}$ decreases with $\sin^{2}\theta_{12}$, $\sin^{2}\theta_{13}$ and $\sin^{2}\theta_{23}$. The parameter $\Im \tau$ shows a strong association with $\sin^{2}\theta_{12}$ and $\sin^{2}\theta_{13}$, while displaying an opposite tendency with respect to $\delta_{CP}$.

\begin{figure}[t!]
\centering
\includegraphics[width=5.0in]{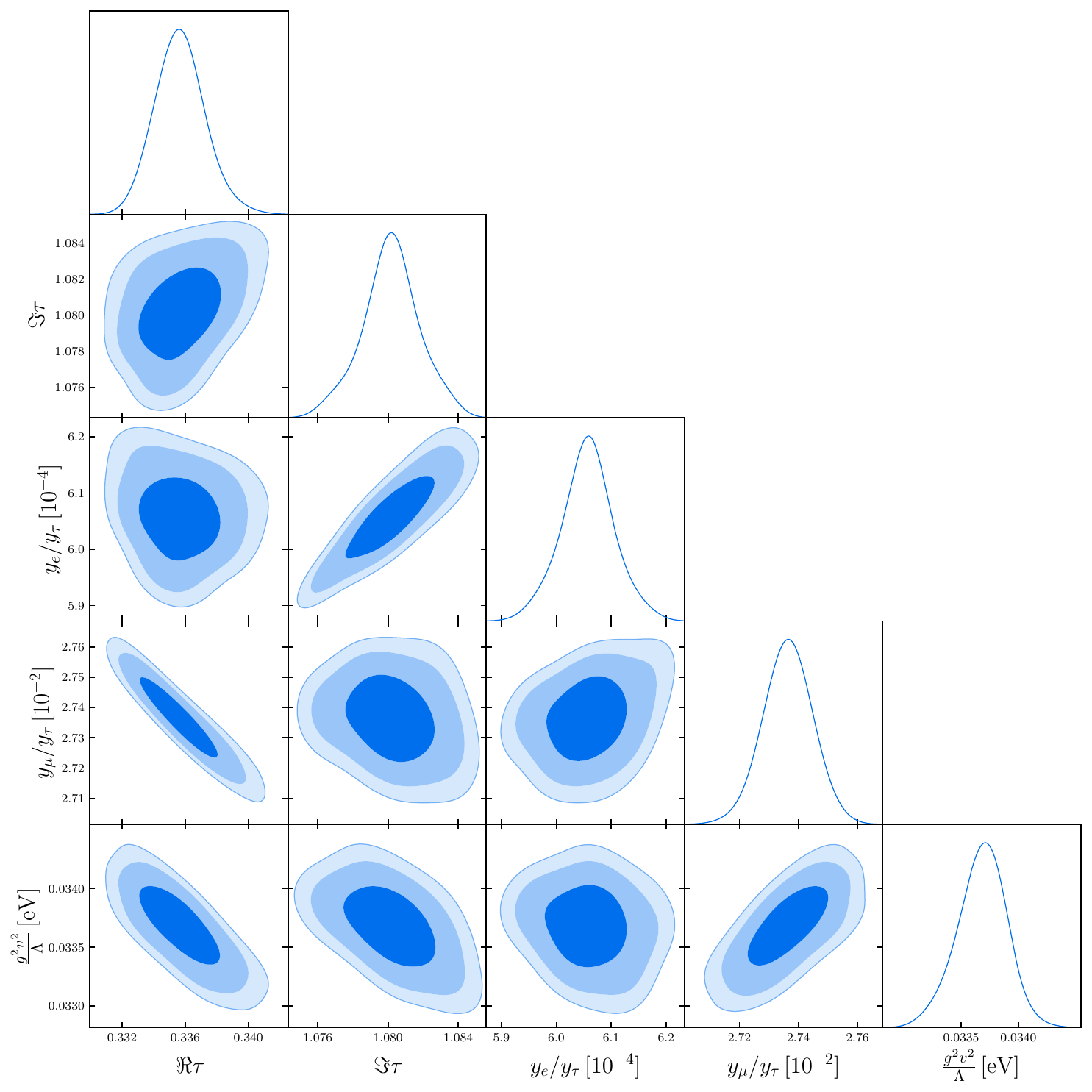}
\caption{Allowed regions for the lepton input parameters for case $C^{(-4,0,1)}_{(0,0,1)}-S^{(-2,4)}$. Different color shadings correspond to the $1\sigma$, $2\sigma$, and $3\sigma$ confidence levels. }
\label{fig:lepton_input_3}
\end{figure}

\begin{figure}[t!]
\centering
\includegraphics[width=5.0in]{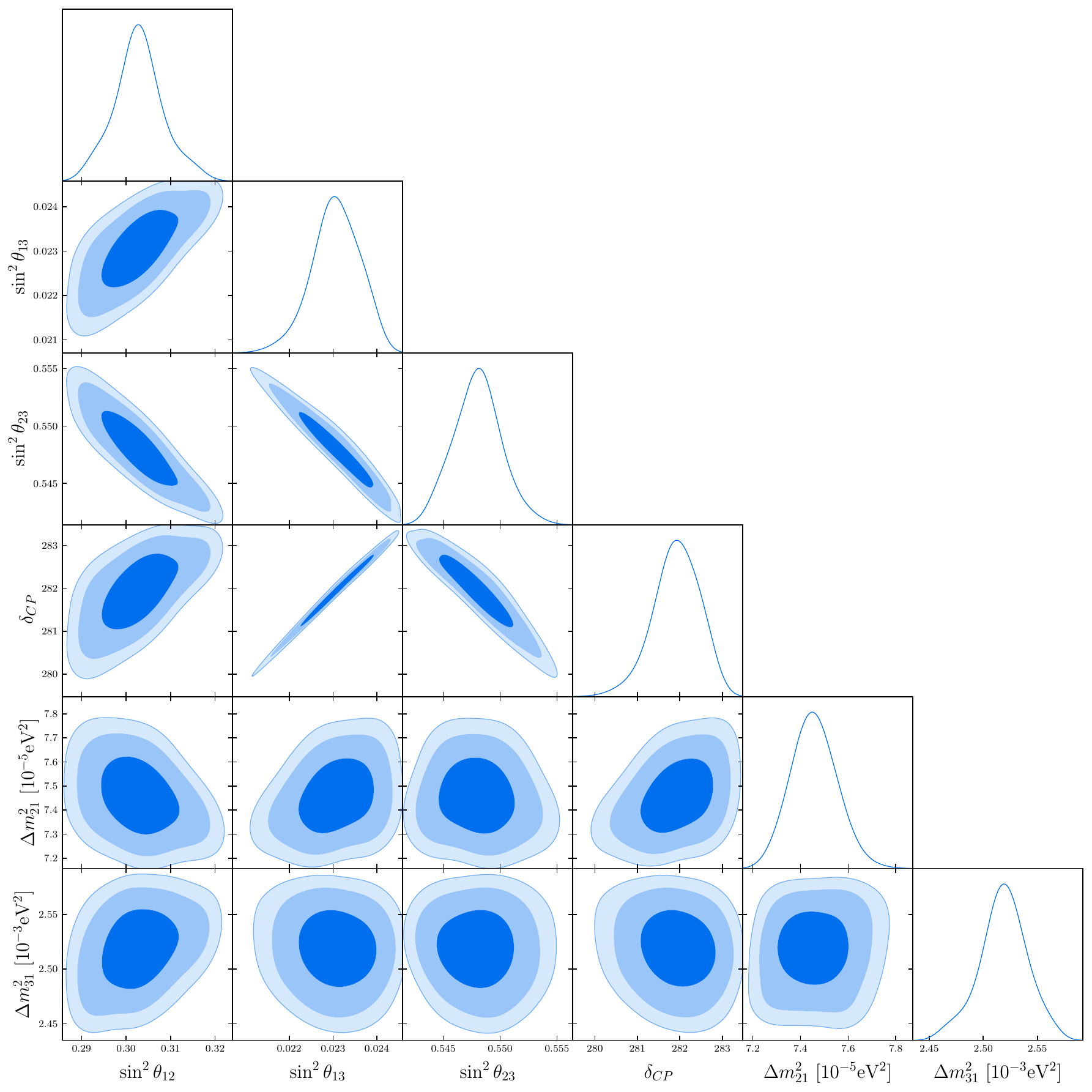}
\caption{Allowed regions for the lepton observables for case $C^{(-4,0,1)}_{(0,0,1)}-S^{(-2,4)}$. Different color shadings correspond to the $1\sigma$, $2\sigma$, and $3\sigma$ confidence levels. }
\label{fig:lepton_obs_3}
\end{figure}

\begin{figure}[t!]
\centering
\includegraphics[width=5.0in]{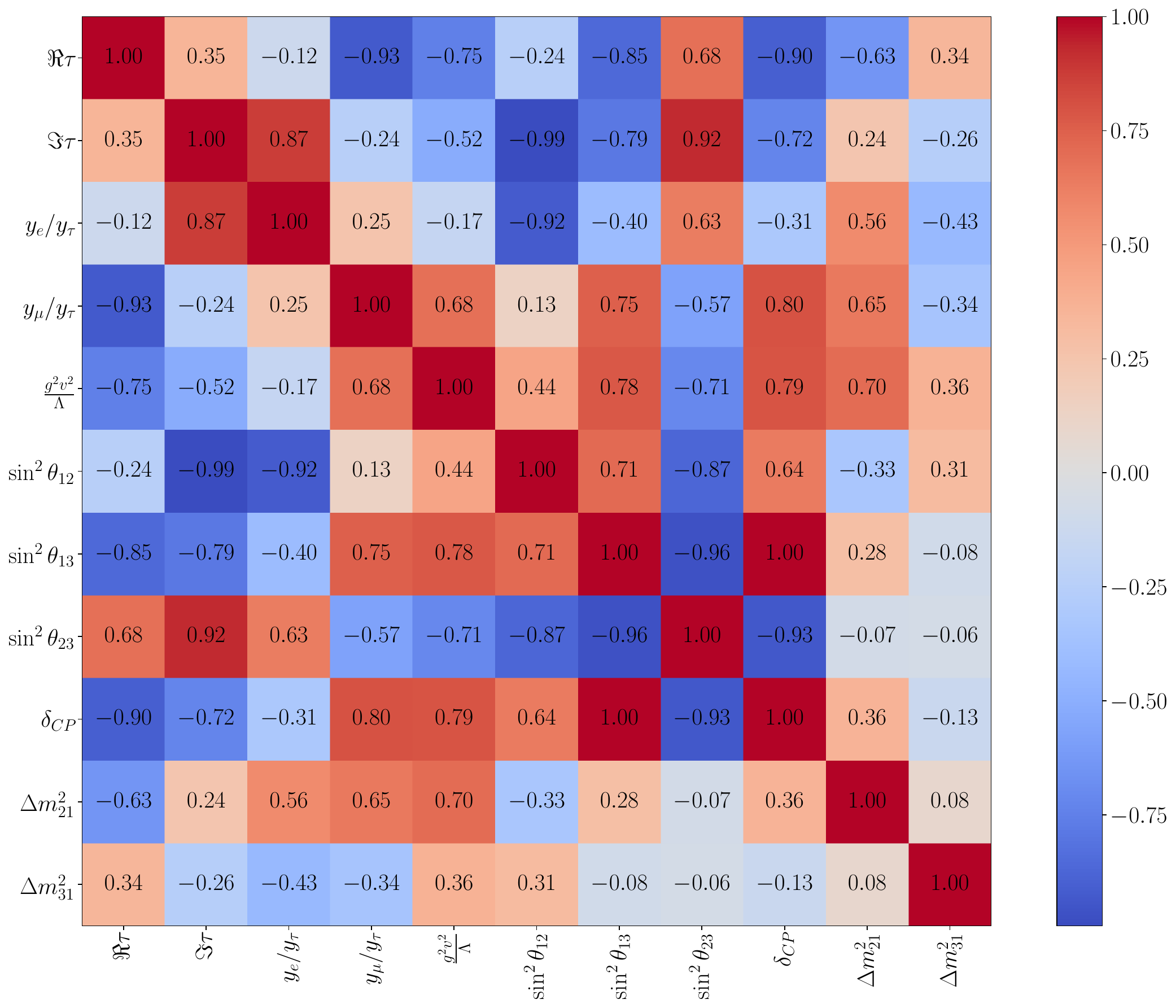}
\caption{Matrix of the correlations among the input parameters and lepton observables for the lepton model $C^{(-4,0,1)}_{(0,0,1)}-S^{(-2,4)}$. }
\label{fig:lepton_correalation_3}
\end{figure}

Figure~\ref{fig:lepton_input_3} summarizes the correlations among 
the input parameters of the model 
$C^{(-4,0,1)}_{(0,0,1)}-S^{(-2,4)}$. The $\Re \tau$ is found to be decreased with $y_{\mu}/y_{\tau}$ and $\frac{g^{2}v^{2}}{\Lambda}$, while $\Im \tau$ 
is positively correlated with $y_{e}/y_{\tau}$. Besides, there is positive correlation between $y_{\mu}/y_{\tau}$ and 
$\frac{g^{2}v^{2}}{\Lambda}$. The correlations among the neutrino observables are shown in figure~\ref{fig:lepton_obs_3}. The angle $\sin^{2}\theta_{12}$ 
increases with $\sin^{2}\theta_{13}$ and $\delta_{CP}$, but 
decreases with $\sin^{2}\theta_{23}$. A similar trend is observed 
for $\sin^{2}\theta_{13}$, which is strongly aligned with 
$\delta_{CP}$ while being anticorrelated with 
$\sin^{2}\theta_{23}$. Accordingly, $\sin^{2}\theta_{23}$ and 
$\delta_{CP}$ display a clear anticorrelation. The correlation matrix of this model is presented in figure~\ref{fig:lepton_correalation_3}. The parameter $\Re \tau$ is positively associated with $\sin^{2}\theta_{23}$, but shows the 
opposite behavior with $\sin^{2}\theta_{13}$, $\delta_{CP}$, and 
$\Delta m_{21}^{2}$. The imaginary part $\Im \tau$, suppresses 
$\sin^{2}\theta_{12}$, $\sin^{2}\theta_{13}$, and $\delta_{CP}$, 
while enhancing $\sin^{2}\theta_{23}$. The parameter $y_{e}/y_{\tau}$ is negative correlated with $\sin^{2}\theta_{12}$ but positive with $\sin^{2}\theta_{23}$. Meanwhile, $y_{\mu}/y_{\tau}$ and $\frac{g^{2}v^{2}}{\Lambda}$ increase with $\sin^{2}\theta_{13}$, $\delta_{CP}$ and $\Delta m_{21}^{2}$, but decrease with $\sin^{2}\theta_{23}$.

\section{\label{sec:leptogenesis}Prediction for leptogenesis}

The origin of the baryon asymmetry of the Universe remains one of the fundamental unresolved problems in particle physics and cosmology. It has been precisely determined from cosmological observations, with Big Bang nucleosynthesis and measurements of the cosmic microwave background yielding consistent results. The observed baryon asymmetry, normalized to the entropy density, is given by~\cite{Planck:2018vyg}
\begin{equation}
\label{eq:YB observation}
Y_{B}=(8.70300\pm 0.11288)\times10^{-11} \quad (95\%\,\, \text{CL})\,.
\end{equation}

The seesaw mechanism not only accounts for the observed lightness of neutrino masses but also naturally leads to the leptogenesis paradigm, which offers a compelling resolution to the cosmological baryon asymmetry puzzle~\cite{Fukugita:1986hr}. In this mechanism, an initial lepton asymmetry is produced  by the CP-violating decays of heavy RH neutrinos, and subsequently a fraction of this lepton excess is transformed into the observed baryon abundance through the sphaleron processes which anomalously break $B+L$~\cite{Klinkhamer:1984di, Arnold:1987mh, Kuzmin:1985mm, Rubakov:1996vz}.

In our setup, the light neutrino masses are generated through the minimal type-I seesaw mechanism involving two RH neutrinos. Remarkably, both the Dirac neutrino mass matrix $M_D$ and the heavy Majorana mass matrix $M_N$ are uniquely fixed by the modulus $\tau$, up to the overall normalization parameters $g$ and $\Lambda$, as summarized in table~\ref{tab:3Ms_mass}. In what follows, we explore the implications of these benchmark models for the generation of the baryon asymmetry via leptogenesis. Depending on whether charged lepton flavor effects are dynamically resolved, leptogenesis can be treated in either the flavored or unflavored regime. For $M_{1}\lesssim10^{12}\,\mathrm{GeV}$, flavor effects play an important role in determining the final baryon asymmetry~\cite{Abada:2006ea,Abada:2006fw,Nardi:2006fx,Antusch:2006cw}. However, none of the viable models identified in this work can reproduce the observed baryon asymmetry $Y_{B}$ in this mass range. We therefore restrict our analysis to the unflavored regime, valid for $M_{1}\gtrsim10^{12}\,\mathrm{GeV}$.  In this temperature range, the processes induced by all charged lepton Yukawa interactions remain out of equilibrium, rendering the individual lepton flavors effectively indistinguishable, and the lepton sector behaves effectively as a single coherent degree of freedom. Under these conditions, the baryon asymmetry is given by~\cite{Davidson:2008bu}
\begin{equation}
\label{eq:YB SM}
Y_{B}=\frac{12}{37}  Y_{\Delta}\,,
\end{equation}
where $Y_{\Delta}\equiv Y_{B-L}$ denotes the $B-L$ asymmetry, which remains conserved under sphaleron processes and all other SM interactions. Here  particle number densities are normalized to the entropy density.

For the 36 viable minimal models identified in section~\ref{sec:num_analysis}, the RH neutrino mass matrix only depends on $\tau$ up to the overall scale $\Lambda$, as shown in Eq.~\eqref{eq:nu_masses_MN}. Hence  the mass ratio $M_{2}/M_{1}$ of the two RH neutrinos is entirely determined by the complex modulus $\tau$. When evaluated at the corresponding best-fit values of $\tau$, this ratio is consistently found to be close to 5 in all cases. This implies a hierarchical mass spectrum for the RH neutrinos, with $M_{2} \gg M_{1}$. In this hierarchical limit, leptogenesis is dominated by the lightest RH neutrino $N_{1}$, while the contribution from the heavier state $N_{2}$ is strongly washed out. The phases of the coupling constants can be removed by field redefinition in these models, thus CP violations have a unique origin from the real part of $\tau$. The CP asymmetry generated in the out of equilibrium decays of $N_{1}$ decays is approximately given by~\cite{Covi:1996wh,Buchmuller:2004nz,Buchmuller:2005eh,Davidson:2008bu}:
\begin{equation}\label{eq:CP_epsilon1}
\varepsilon_{1}\approx-\frac{3}{16 \pi} \frac{\Im\left[(\lambda \lambda^{\dagger})^2_{12}\right]}{{\left(\lambda \lambda^{\dagger}\right)}_{11}} \frac{M_{1}}{M_{2}} \,,
\end{equation}
in the hierarchical limit $M_{2}\gg M_{1}$. Here, $\lambda$ denotes the neutrino Yukawa coupling matrix, expressed in the basis where both the charged lepton Yukawa couplings and the Majorana mass matrix $M_{N}$ of the RH neutrinos are diagonal and real. We adopt the convention in which the Yukawa interaction term of neutrinos is written as $N^{c}\lambda LH$. The baryon asymmetry generated through leptogenesis can be obtained by solving the following Boltzmann equations for the asymmetry $Y_{\Delta}$ and the  abundance $Y_{N_1}$ of the lightest RH neutrino~\cite{Davidson:2008bu} 
\begin{eqnarray}
\frac{\text{d} Y_{N_{1}}}{\text{d} z}&=& K\frac{z  f_{1}(z)K_{1}(z)}{K_{2}(z)}\left(Y_{N_{1}}^{\text{eq}}-Y_{N_{1}}\right)\,,\nonumber\\
\label{eq:Bol_Eq2} \frac{\text{d} Y_{\Delta}}{\text{d} z}&=&K  \frac{z  K_{1}(z)}{K_{2}(z)}\left[\varepsilon_{1}  f_{1}(z)\left(Y_{N_{1}}^{\text{eq}}-Y_{N_{1}}\right)- f_{2}(z) Y_{N_{1}}^{\text{eq}} \frac{Y_{\Delta}}{Y_{\ell}^{\text{eq}}}\right]\,,
\end{eqnarray}
where the dimensionless variable $z= M_1/T$ is inversely proportional to the temperature of the Universe. The functions $K_{1}(z)$ and $K_{2}(z)$ are the modified Bessel functions of the second kind. The quantities $Y_{N_{1}}^{\text{eq}}$ and $Y_{\ell}^{\text{eq}}$ denote the equilibrium abundances of the RH neutrino $N_1$ and the lepton asymmetry normalized to the entropy density, respectively. They are given by 
\begin{equation}\label{eq:Yleq_Ynieq}
Y_{N_{1}}^{\text{eq}}\simeq\frac{45z^{2}K_{2}(z)}{2\pi^4g_{*}},\qquad \qquad  Y_{\ell}^{\text{eq}}\simeq\frac{45}{\pi^4g_{*}}\,,
\end{equation}
with the effective number of relativistic degrees of freedom $g_{*}=106.75$ in the SM. The efficiency of washout processes is described by the parameter 
\begin{equation}
  \label{eq:Ka}
  K=\frac{\widetilde{m}_{1}}{m^{*}_{\text{SM}}}\,,
\end{equation}
where $m^{*}_{\text{SM}}\simeq1.08\times10^{-3}$ eV in the SM. The effective neutrino mass $\widetilde{m}_{1}$, which depends on the total decay rate of $N_{1}$, is given by
\begin{equation}\label{eq:tildem1a}
\widetilde{m}_{1}\equiv\frac{(\lambda \lambda^{\dagger})_{11}v^2}{M_{1}}\,,
\end{equation}
where the parameter $v=174$ GeV represents the VEV of the Higgs field. It is straightforward to verify that the washout parameter $K\approx48$ for all 36 viable models at the best fit points listed in table~\ref{tab:best_fit_noGCP_input}. Consequently, the RH neutrino $N_1$ is nearly in thermal equilibrium, and the leptogenesis occurs in the strong washout regime.   

In the Boltzmann equations of Eq.~\eqref{eq:Bol_Eq2}, the functions $f_1(z)$ and $f_2(z)$ account for the effects of $\Delta L=1$ scatterings induced by neutrino and top-quark Yukawa interactions on RH neutrino production and lepton asymmetry washout, respectively~\cite{Giudice:2003jh,Buchmuller:2004nz}.  In the unflavored and strong washout regime, one has $f_2(z)\approx f_1(z)/2$, and their analytic approximations take the form~\cite{Giudice:2003jh,Buchmuller:2004nz}
\begin{equation}
\label{eq:f1f2}
f_{1}(z)\approx 2f_{2}(z)\approx\left[\frac{9}{8\pi^2z }+\frac{z}{t} \ln \left(1+\frac{t}{z}\right)\right] \frac{K_{2}(z)}{K_{1}(z)}\,,
\end{equation}
where 
\begin{equation}\label{eq:a-Ks-over-K}
t=\frac{8\pi^2}{9\ln(M_{1}/M_{h})}\,,
\end{equation}
and $M_{h}=125$ GeV is the Higgs boson mass.

\subsection{Numerical analysis}

The 36 phenomenologically viable models listed in tables~\ref{tab:best_fit_noGCP_input} and \ref{tab:best_fit_noGCP_output} naturally fall into three distinct categories, each comprising twelve models that have the same set of neutrino mass matrices and  yield nearly identical predictions for the lepton mixing observables. This implies that all models belonging to a given category yield the same baryon asymmetry prediction. Hence it suffices to examine the three representative models $C^{(-4,2,-1)}_{(0,0,3)}-S^{(-2,-4)}$, $C^{(-4,-3,-1)}_{(0,1,3)}-S^{(-2,4)}$  and $C^{(-4,0,1)}_{(0,0,1)}-S^{(-2,4)}$ studied in section~\ref{sec:example_model}. In the original basis, the charged lepton mass matrix and the Majorana mass matrix of the RH neutrinos are both non-diagonal for the each of the three benchmark models. They can be diagonalized by the unitary transformations $U_{l}$ and $U_{N}$ respectively,
\begin{equation}
U^{\dagger}_{l}M^{\dagger}_{l}M_{l}U_{l}=\text{diag}(m^{2}_{e},m^{2}_{\mu},m^{2}_{\tau}), \qquad U^{T}_{N}M_{N}U_{N}=\text{diag}(M_{1},M_{2})\,,
\end{equation}
where $m_{e,\mu,\tau}$ are the charged lepton masses and $M_{1,2}$ are the masses of the heavy Majorana neutrinos. Under this change of basis, the Dirac neutrino mass matrix $M_{D}$ is correspondingly transformed as  
\begin{equation}
M^{\prime}_{D}=U^{T}_{N}M_{D}U_{l}\,.
\end{equation}
Consequently, the Dirac Yukawa coupling matrix entering Eq.~\eqref{eq:CP_epsilon1} is given by $\lambda=M^{\prime}_{D}/v$.

\begin{figure}[t!]
\centering
\includegraphics[width=0.6\textwidth]{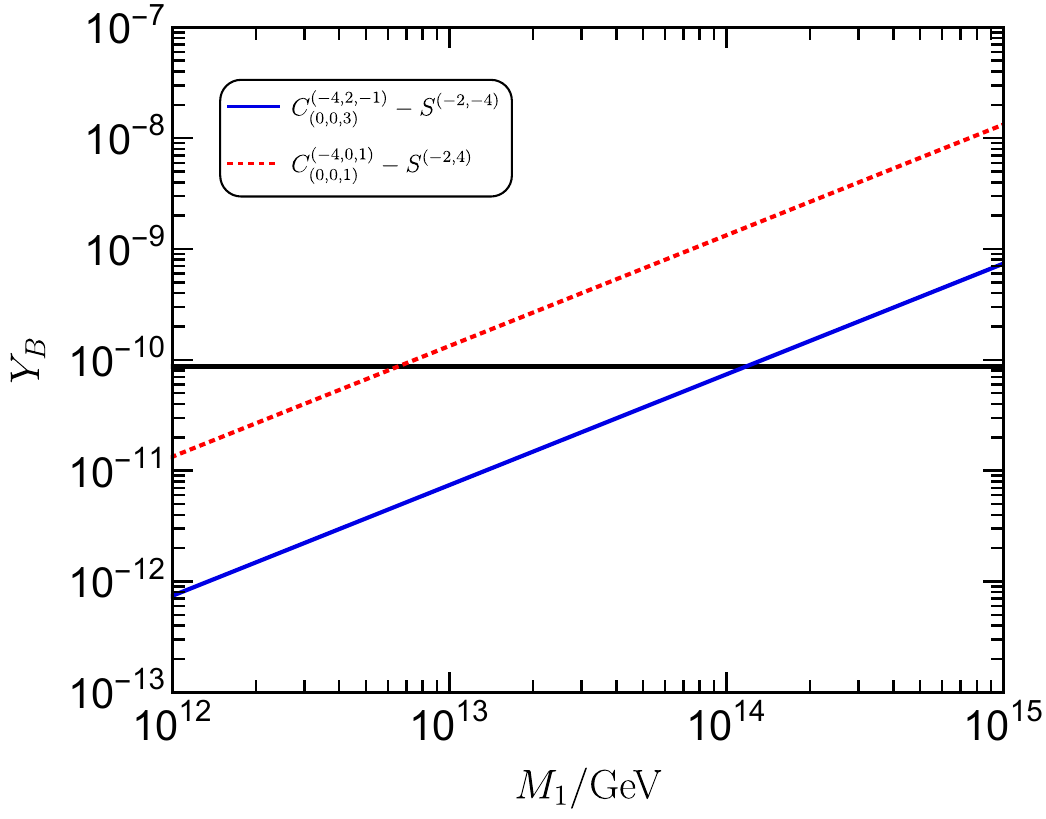}
\caption{\label{fig:YB-M1}  Correlation between the baryon asymmetry $Y_B$ and the RH neutrino mass $M_1$ for the representative models $C^{(-4,2,-1)}_{(0,0,3)}-S^{(-2,-4)}$ and $C^{(-4,0,1)}_{(0,0,1)}-S^{(-2,4)}$. The black horizontal line represents the observed baryon asymmetry $Y_B=8.703\times 10^{-11}$. }
\end{figure}

\begin{figure}[t!]
\centering
\includegraphics[width=3.1in]{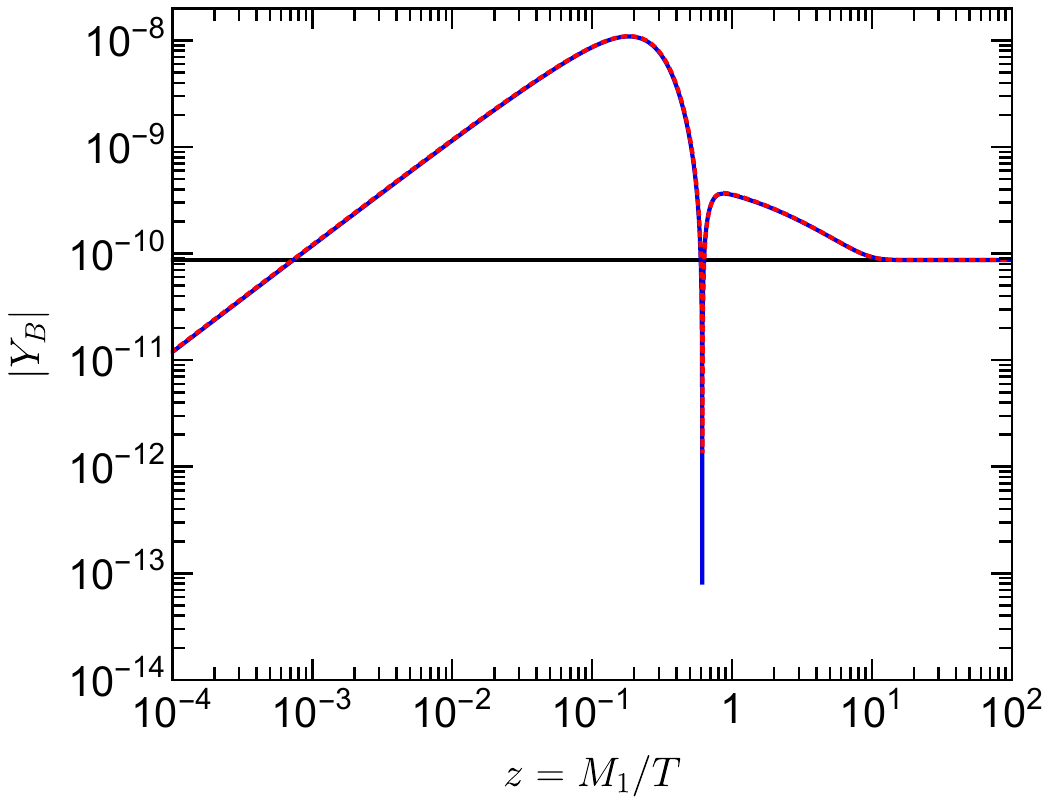}
\includegraphics[width=3.1in]{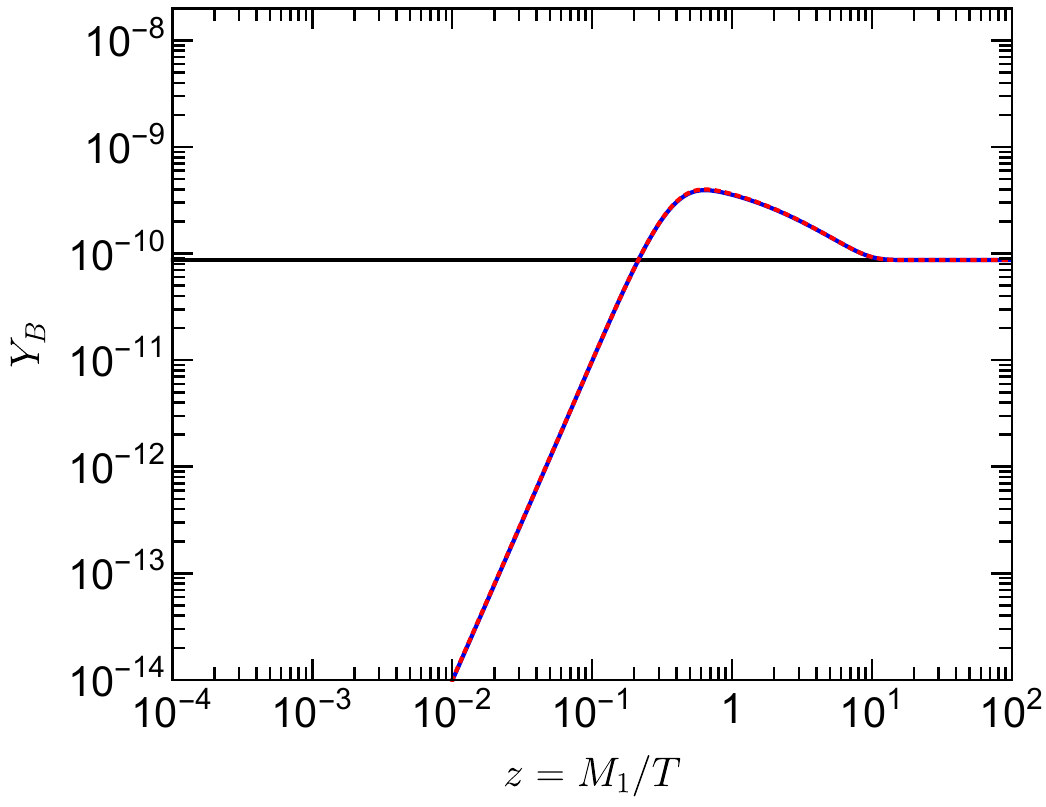}
\caption{\label{fig:YB-T} The evolution of $Y_B$ with $M_1/T$ for the representative models $C^{(-4,2,-1)}_{(0,0,3)}-S^{(-2,-4)}$ and $C^{(-4,0,1)}_{(0,0,1)}-S^{(-2,4)}$ denoted by blue solid and red dashed lines respectively, the left and right panels are for vanishing and thermal initial RH neutrino abundances respectively. The black horizontal line represents the observed baryon asymmetry $Y_B=8.703\times 10^{-11}$.  }
\end{figure}

For the three benchmark models, all input parameters are fixed at their best-fit values given in table~\ref{tab:best_fit_noGCP_input}, ensuring consistency with current neutrino oscillation data.  Upon applying the seesaw mechanism, the light neutrino mass matrix depends only on the effective combination $g^2v^2/\Lambda$ which controls the overall neutrino mass scale. The masses of the RH neutrinos are crucial for leptogenesis. Accordingly, the heavy Majorana mass scale $\Lambda$ is treated as a free parameter varied over the range $10^{12}$ GeV to $10^{16}$ GeV.  Then we numerically solve the Boltzmann equations in Eq.~\eqref{eq:Bol_Eq2} to determine the correlation between the baryon asymmetry $Y_{B}$ and the lightest RH neutrino mass $M_{1}$. The results indicate that only the models $C^{(-4,2,-1)}_{(0,0,3)}-S^{(-2,-4)}$ and $C^{(-4,0,1)}_{(0,0,1)}-S^{(-2,4)}$ can successfully reproduce the observed baryon asymmetry, as illustrated in figure~\ref{fig:YB-M1}. Notice that the baryon asymmetry $Y_B$ is predicted to be negative in the remaining model  $C^{(-4,-3,-1)}_{(0,1,3)}-S^{(-2,4)}$. Setting $\Lambda = 1.5411\times10^{15}\,\text{GeV}$ and $1.6388\times10^{13}\,\text{GeV}$ in the two viable models respectively, solving the Boltzmann equations of Eq.~\eqref{eq:Bol_Eq2} yields $Y_B = 8.703\times10^{-11}$ in agreement with observational data. The corresponding RH neutrino masses are as follows: 
\begin{eqnarray}
\nonumber\hskip-0.4in C^{(-4,2,-1)}_{(0,0,3)}-S^{(-2,-4)}&: &M_{1}\approx1.175\times 10^{14}\text{GeV} \,, ~ M_{2}\approx6.307\times 10^{14}\text{GeV}, ~ M^{2}_{2}/M^{2}_{1}\approx28.80\,, \\
\hskip-0.4in C^{(-4,0,1)}_{(0,0,1)}-S^{(-2,4)}&: &M_{1}\approx6.521\times 10^{12}\text{GeV} \,, ~M_{2}\approx3.243\times 10^{13}\text{GeV}, ~ M^{2}_{2}/M^{2}_{1}\approx24.74\,, ~~~
\end{eqnarray}
which clearly satisfy $M^{2}_{2}/M^{2}_{1}\gg1$, consistent with the standard $N_{1}$-dominant leptogenesis. Accordingly the two viable models predict nearly identical values of the CP asymmetry $\varepsilon_{1}$ and the washout parameter $K$: 
\begin{eqnarray}
 \nonumber &&\varepsilon_{1}\approx-1.558\times 10^{-5}, \qquad K \approx 47.84 \qquad \text{for} \qquad C^{(-4,2,-1)}_{(0,0,3)}-S^{(-2,-4)} \,, \\
 &&\varepsilon_{1}\approx-1.568\times 10^{-5}, \qquad K \approx 48.20 \qquad \text{for} \qquad C^{(-4,0,1)}_{(0,0,1)}-S^{(-2,4)} \,.
\end{eqnarray}
Since the Boltzmann equations in Eq.~\eqref{eq:Bol_Eq2} depend on the model only solely through $\varepsilon_{1}$ and $K$, the two models lead to nearly identical predictions for the baryon asymmetry $Y_{B}$. Consequently the evolution of $Y_{B}$ in both cases follows almost the same trajectory, resulting in nearly overlapping curves in figure~\ref{fig:YB-T}. 

In figure~\ref{fig:YB-T}, we present the evolution of the baryon asymmetry as a function of $z=M_1/T$ for the two viable benchmark models. The left panel corresponds to a vanishing initial abundance of RH neutrinos, while the right panel assumes a thermal initial abundance. In the left panel, the baryon asymmetry $Y_B$ changes sign from negative to positive around $M_1/T \approx 0.6$ as the temperature decreases. Since the vertical axis uses a logarithmic scale, negative values cannot be plotted directly. Therefore the absolute value $|Y_B|$ is plotted. This treatment preserves the magnitude of the asymmetry but omits the sign information. In contrast, for the thermal initial abundance of RH neutrinos in right panel, $Y_B$ remains positive throughout the relevant temperature range, so no absolute value is needed and $Y_B$ is explicitly plotted. Comparing the left and right panels, we observe that the final $Y_B$ at low temperatures is independent of the initial condition of the RH neutrinos. This behavior is characteristic of the strong washout regime, where the generated asymmetry is predominantly determined by the decay and inverse decay processes, and any memory of the initial abundance is erased.

\section{\label{sec:conclusion} Conclusion }

The non-holomorphic extension of modular flavor symmetry removes the necessity of supersymmetry and grants access to negative weight modular multiplets, thereby substantially enriching the landscape of viable flavor model constructions. In this work, we have systematically investigated minimal lepton models based on non-holomorphic modular $S^{\prime}_{4}$ symmetry, within the framework of minimal seesaw with 2RHNs. The Yukawa couplings are polyharmonic Maa{\ss} forms of level $N=4$. 

We have assigned the three generations of LH lepton doublets to an $S^{\prime}_4$ triplet, the RH charged leptons to singlets, and the RH neutrinos to a doublet. Focusing on the polyharmonic Maa{\ss} forms of weights $k\in[-4,6]$ at level $N=4$, we have identified a total of 7280 distinct lepton models which are all governed by only six real input parameters: $\Re\tau$, $\Im\tau$, three real couplings in the charged lepton sector $y_{e}$, $y_{\mu}$ and $y_{\tau}$, and the overall neutrino mass scale $g^{2}v^{2}/\Lambda$. Notice that the phases of $y_{e}$, $y_{\mu}$ and $y_{\tau}$ can be absorbed by the charged lepton fields, so that they can be taken to be real and the generalized CP symmetry is not imposed.

Through an exhaustive numerical scan the parameter space and a statistical evaluation employing the marginalized $\chi^2$ projections from the \texttt{NuFIT}-v6.1 global analysis, we find 36 phenomenologically viable models that accommodate all experimental constraints on lepton masses and mixing parameters for the normal mass ordering, while none of them can accommodate the experimental data when neutrino mass spectrum is inverted ordering. The best-fit values of the input parameters for the 36 viable models, together with the corresponding predictions for the lepton observables, are summarized in tables~\ref{tab:best_fit_noGCP_input} and~\ref{tab:best_fit_noGCP_output}. We find that these models can be grouped into three distinct categories, each containing twelve models with nearly degenerate predictions for the mixing observables. Within each category, the assignments of representations and modular weights for the lepton fields are nearly identical, differing only in the assignments of $E^{c}_{1}$. To further explore the phenomenological implications of these models, we perform a detailed statistical analysis of three benchmark models, $C^{(-4,2,-1)}_{(0,0,3)}-S^{(-2,-4)}$, $C^{(-4,-3,-1)}_{(0,1,3)}-S^{(-2,4)}$, and $C^{(-4,0,1)}_{(0,0,1)}-S^{(-2,4)}$, which are representative of the three categories of viable models. The predicted allowed regions of neutrino masses and mixing parameters are summarized in table~\ref{tab:three_models_res}. We find that these benchmark models are primarily distinguished by their predictions for the atmospheric mixing angle $\theta_{23}$ and the CP-violating phase $\delta_{CP}$, both of which are confined to relatively narrow intervals. In particular, $\sin^{2}\theta_{23}$ lies in the first octant $[0.458,0.471]$ for $C^{(-4,2,-1)}_{(0,0,3)}-S^{(-2,-4)}$, while it resides in the second octant, $[0.518,0.536]$ and $[0.542,0.556]$, for the other two benchmark models. Meanwhile, $\delta_{CP}$ is restricted to two separate narrow regions around $0.91\pi$ and $1.08\pi$ for $C^{(-4,2,-1)}_{(0,0,3)}-S^{(-2,-4)}$, and around $1.73\pi$ and $1.56\pi$ for $C^{(-4,-3,-1)}_{(0,1,3)}-S^{(-2,4)}$ and $C^{(-4,0,1)}_{(0,0,1)}-S^{(-2,4)}$, respectively. Future high-precision measurements of $\theta_{12}$, $\theta_{23}$, and $\delta_{CP}$ at JUNO, DUNE and T2HK will help to distinguish among these models. In addition, the predicted sum of neutrino masses for all three models lies in the range $[58.00~\text{meV},59.60~\text{meV}]$, which is consistent with the current cosmological bound $\sum_i m_i < 120~\text{meV}$ from Planck~\cite{Planck:2018vyg} and it is within the projected sensitivity of upcoming cosmological surveys such as Euclid+CMB-S4+LiteBIRD~\cite{Euclid:2024imf}. However, the predicted values of both $m_{\beta\beta}$ and $m_{\beta}$ are below the reach of current and near-future experiments.

Furthermore, the correlations among the input parameters and observables of the three benchmark models are shown in figures~\ref{fig:lepton_input_1}--\ref{fig:lepton_correalation_3}. We observe rich and structured correlations between the model input parameters and neutrino observables, with several robust features shared across different models. The modular parameter $\tau$ plays a central role: its real part $\Re \tau$ governs CP violation and exhibits strong (anti-)correlations with $\delta_{CP}$, while its imaginary part $\Im \tau$ is closely related to the mixing angles. The ratios $y_e/y_{\tau}$ and $y_{\mu}/y_{\tau}$, together with the overall scale $\frac{g^{2}v^{2}}{\Lambda}$, show clear and model-dependent correlations with both mixing angles and mass-squared differences. The mixing angles are strongly correlated with each other and also exhibit correlations with $\delta_{CP}$. Overall, these results demonstrate that the considered models not only accommodate current experimental data but also exhibit strong predictive power through their distinctive correlation patterns. Consequently they can be fully test at future neutrino facilities.

Moreover, we show that the non-holomorphic $S^{\prime}_{4}$ modular framework provides a natural setting for successful thermal leptogenesis. In all viable models, the RH neutrinos exhibit a hierarchical mass spectrum, such that leptogenesis is dominated by the out of equilibrium decays of the lightest RH neutrino $N_1$. Moreover, all viable models lie in the strong washout regime. In the unflavored regime with $M_{1}> 10^{12}$ GeV, numerically solving the Boltzmann equations for the three benchmark models shows that only the models $C^{(-4,2,-1)}_{(0,0,3)}-S^{(-2,-4)}$ and $C^{(-4,0,1)}_{(0,0,1)}-S^{(-2,4)}$ can reproduce the observed baryon asymmetry $Y_B = 8.703\times10^{-11}$ at the best-fit points of the free parameter which ensures excellent description of lepton mixing parameters. The corresponding masses of the lightest RH neutrino are $1.175\times 10^{14}\,\text{GeV}$ and $6.521\times 10^{12}\,\text{GeV}$, respectively. This framework unifies the origin of the low energy neutrino flavor structure with the high scale mechanism of baryogenesis within a common parameter space. In particular, $\Re \tau$ acts as the unique source of CP violation, simultaneously controlling the leptonic CP phase and the CP asymmetry required for leptogenesis.

In summary, non-holomorphic modular $S^{\prime}_{4}$ symmetry provides a unified and highly constrained framework in which neutrino masses, lepton mixing, CP violation, and baryogenesis are intrinsically linked. The predictive nature of these minimal constructions, together with their testable implications for upcoming experiments, highlights their potential as a compelling avenue for physics beyond the SM.

\section*{Acknowledgements}

This work is supported by the National Natural Science Foundation of China under Grant Nos. 12547106, 12247103, 12505133, 12375104, and Guizhou Provincial Major Scientific and Technological Program XKBF (2025)010.

\newpage        
        
\section*{Appendix}
        
\begin{appendix}
                
\section{\label{sec:S4p_group_theory}Finite modular group $S^{\prime}_{4}$}

\begin{table}[t!]
\begin{center}
\renewcommand{\tabcolsep}{2.8mm}
\renewcommand{\arraystretch}{1.2}
\begin{tabular}{|c|c|c|c|c|c|}\hline\hline
$\bm{r}$ & $\rho_{\bm{r}}(S)$ & $\rho_{\bm{r}}(T)$ \\ \hline
$\bm{1_{p}}$ & $i^{p}$ & $(-i)^{p}$ \\ \hline
& &  \\[-0.16in]
                                        
$\bm{2_{a}}$ & $\dfrac{i^a}{2}\begin{pmatrix}
-1 ~&\sqrt{3} \\
\sqrt{3} ~& 1 \\
\end{pmatrix}$ & $(-i)^{a}\begin{pmatrix}
1 ~& 0 \\
0 ~& -1 \\
\end{pmatrix}$  \\ [0.18in]  \hline
& & \\[-0.16in]
                                        
$\bm{3_{p}}$ & $\dfrac{i^{p}}{2}\begin{pmatrix}
0 ~&\sqrt{2} ~&\sqrt{2} \\
\sqrt{2} ~& -1 ~& 1 \\
\sqrt{2} ~& 1 ~& -1
\end{pmatrix}$ & $(-i)^{p}\begin{pmatrix}
1 ~& 0 ~& 0 \\
0 ~& i ~& 0 \\
0 ~& 0 ~& -i
\end{pmatrix}$ \\ [0.18in] \hline \hline
\end{tabular}
\caption{\label{tab:S4p_rep_matrices}The representation matrices of the generators $S$ and $T$ in different irreducible representations of $S^{\prime}_{4}$, where $a=0,1$ and $p=0,1,2,3$. }
\end{center}
\end{table}

The homogeneous finite modular group $\Gamma^{\prime}_4\cong S^{\prime}_4$ is the double cover of $\Gamma_4\cong S_4$, which is the permutation group associated with the symmetries of a cube. It is generated by two elements $S$ and $T$ which satisfy the defining relations~\cite{Ding:2023htn}
\begin{equation}
S^4=(ST)^3=T^4=1, \qquad S^{2}T=TS^{2}\,.
\end{equation}
The group $S^{\prime}_{4}$ admits ten irreducible representations: four singlet representations $\bm{1_{p}}$, two doublet representations $\bm{2_{a}}$, and four triplet representations $\bm{3_{p}}$, where $a=0,1$ and $p=0,1,2,3$. In the notation of Ref.~\cite{Ding:2023htn}, these representations correspond to $\bm{1_{0,1,2,3}}=\bm{1}, \bm{\widehat{1}}, \bm{1^{\prime}}, \bm{\widehat{1}^{\prime}}$, $\bm{2_{0,1}}=\bm{2},\bm{\widehat{2}}$, and $\bm{3_{0,1,2,3}}=\bm{3},\bm{\widehat{3}},\bm{3^{\prime}},\bm{\widehat{3}^{\prime}}$, respectively. The explicit representation matrices for the generators $S$ and $T$ are taken from Ref.~\cite{Ding:2023htn} and listed in table~\ref{tab:S4p_rep_matrices}.  The Kronecker products between different irreducible representations of $S^{\prime}_4$ read as follows:            
\begin{eqnarray}
\nonumber&&\bm{1_{p}} \otimes \bm{1_{q}}=\bm{1_{[p+q]}}, \qquad \bm{1_{p}} \otimes \bm{2_{a}}=\bm{2_{\langle p+a\rangle}}, \qquad \bm{1_{p}} \otimes \bm{3_{q}}=\bm{3_{[p+q]}}  ,\\
\nonumber&&\bm{2_{a}} \otimes \bm{2_{b}}=\bm{1_{a+b}} \oplus \bm{1_{[a+b+2]}} \oplus \bm{2_{\langle a+b\rangle}},\qquad \bm{2_{a}} \otimes \bm{3_{p}}=\bm{3_{[a+p]}} \oplus \bm{3_{[a+p+2]}} ,\\
\label{eq:Kronecker_products}   &&\bm{3_{p}} \otimes \bm{3_{q}}=\bm{1_{[p+q]}} \oplus \bm{2_{\langle p+q\rangle}} \oplus \bm{3_{[p+q]}} \oplus \bm{3_{[p+q+2]}}\,.
\end{eqnarray}  
where $p,q=0,1,2,3$ and $a,b=0,1$. Here, the notation $[x]$ denotes $x\,(\text{mod}\,4)$, while $\langle x\rangle$ stands for $x\,(\text{mod}\,2)$.

In what follows, we present the CG coefficients of $S^{\prime}_{4}$  in the tensor product form $\alpha\otimes\beta$. The components of the vectors are labeled as $\alpha_i$ and $\beta_i$,  referring respectively to the entries of the basis vectors $\alpha$ and $\beta$. For notational convenience, we also introduce
\begin{equation}\label{eq:def_P2}
 P_{2}=\begin{pmatrix}
0 & 1\\-1 & 0
\end{pmatrix}\,,
\end{equation}
which will be used to simplify the expressions of the CG coefficients. Then the results of the CG coefficients are summarized in table~\ref{tab:S4p_CG}.

\begin{table}[t!]
\centering
\renewcommand{\tabcolsep}{2.1mm}
\renewcommand{\arraystretch}{1.2}
\begin{tabular}{|c|c|c|c|c|c|c|c|}\hline\hline
\multicolumn{2}{|c}{$\bm{1_{p}} \otimes \bm{1_{q}}=\bm{1_{[p+q]}}$}  & \multicolumn{2}{|c}{$\bm{1_{p}} \otimes \bm{2_{a}}=\bm{2_{\langle p+a\rangle}}$}  & \multicolumn{2}{|c|}{$\bm{1_{p}} \otimes \bm{3_{q}}=\bm{3_{[p+q]}}$} \\ \hline
\multicolumn{2}{|c}{$\bm{1_{[p+q]}}:\alpha\beta$} & 
\multicolumn{2}{|c}{$\bm{2_{\langle p+a\rangle}}:\alpha P^{\left\lfloor \frac{p+a}{2} \right\rfloor}_{2}\begin{pmatrix}
\beta_1\\
\beta_2
\end{pmatrix}$} & 
\multicolumn{2}{|c|}{$\bm{3_{[p+q]}}:\alpha\begin{pmatrix}
\beta_1\\
\beta_2\\
\beta_3
\end{pmatrix}$} \\ \hline\hline

\multicolumn{3}{|c}{$\bm{2_{a}} \otimes \bm{2_{b}}=\bm{1_{a+b}} \oplus \bm{1_{[a+b+2]}} \oplus \bm{2_{\langle a+b\rangle}}$} &  \multicolumn{3}{|c|}{$\bm{2_{a}} \otimes \bm{3_{p}}=\bm{3_{[a+p]}} \oplus \bm{3_{[a+p+2]}}$} \\ \hline
\multicolumn{3}{|c}{$ \begin{array}{l}
\bm{1_{a+b}} : \alpha_1\beta_1+\alpha_2\beta_2\\[1ex]
\bm{1_{[a+b+2]}} : \alpha_1\beta_2-\alpha_2\beta_1\\[1ex]
\bm{2_{\langle a+b\rangle}} :  P^{\left\lfloor \frac{a+b}{2} \right\rfloor}_{2}\begin{pmatrix}
-\alpha_1\beta_1+\alpha_2\beta_2\\
\alpha_1\beta_2+\alpha_2\beta_1
\end{pmatrix}
\end{array}$} & \multicolumn{3}{|c|}{$\begin{array}{l}
\bm{3_{[a+p]}} : \begin{pmatrix}
2\alpha_1\beta_1\\
-\alpha_1\beta_2+\sqrt{3}\alpha_2\beta_3\\
-\alpha_1\beta_3+\sqrt{3}\alpha_2\beta_2
\end{pmatrix} \\ [1ex]          
\bm{3_{[a+p+2]}} : \begin{pmatrix}
-2\alpha_2\beta_1\\
\sqrt{3}\alpha_1\beta_3+\alpha_2\beta_2\\
\sqrt{3}\alpha_1\beta_2+\alpha_2\beta_3
\end{pmatrix}
\end{array}$} \\ \hline\hline 

\multicolumn{6}{|c|}{$\bm{3_{p}}\otimes\bm{3_{q}}=\bm{1_{[p+q]}}\oplus\bm{2_{\langle p+q\rangle}}\oplus\bm{3_{[p+q]}}\oplus\bm{3_{[p+q+2]}}$} \\ \hline

\multicolumn{6}{|c|}{$\begin{array}{l}
\bm{1_{[p+q]}} : \alpha_1\beta_1+\alpha_2\beta_3+\alpha_3\beta_2\\[3ex]
\bm{2_{\langle p+q\rangle}} :  P^{\left\lfloor \frac{p+q}{2} \right\rfloor}_{2}\begin{pmatrix}
2\alpha_1\beta_1-\alpha_2\beta_3-\alpha_3\beta_2\\
\sqrt{3}(\alpha_2\beta_2+\alpha_3\beta_3)
\end{pmatrix}\\[3ex]
\bm{3_{[p+q]}} :\begin{pmatrix}
\alpha_2\beta_3-\alpha_3\beta_2\\
\alpha_1\beta_2-\alpha_2\beta_1\\
-\alpha_1\beta_3+\alpha_3\beta_1
\end{pmatrix}\\[5ex]
\bm{3_{[p+q+2]}} : \begin{pmatrix}
\alpha_2\beta_2-\alpha_3\beta_3\\
-\alpha_1\beta_3-\alpha_3\beta_1\\
\alpha_1\beta_2+\alpha_2\beta_1
\end{pmatrix}
\end{array}$} \\\hline\hline    
\end{tabular} \\[20pt] 
\caption{\label{tab:S4p_CG}Tensor products and the corresponding CG coefficients for the the finite modular group $S^{\prime}_4$. Here, we use $\alpha_i(\beta_i)$ to denote the component of the basis vector $\alpha (\beta)$. The matrix  $P_{2}$ is defined in Eq.~\eqref{eq:def_P2} and the symbol $\left\lfloor x \right\rfloor$ denotes the floor function.} 
\end{table}     
        
\end{appendix}  

\vskip 2cm


\begin{thebibliography}{10}

\bibitem{ParticleDataGroup:2026aaa}
{\bfseries Particle Data Group} Collaboration, F.~Takahashi {\em et~al.},
  ``{Review of Particle Physics},''
  \href{http://dx.doi.org/10.1142/S0217751X26300115}{{\em Int. J. Mod. Phys. A}
  {\bfseries 41} (2026) 2630011}.

\bibitem{Altarelli:2010gt}
G.~Altarelli and F.~Feruglio, ``{Discrete Flavor Symmetries and Models of
  Neutrino Mixing},'' \href{http://dx.doi.org/10.1103/RevModPhys.82.2701}{{\em
  Rev. Mod. Phys.} {\bfseries 82} (2010) 2701--2729},
  \href{http://arxiv.org/abs/1002.0211}{{\ttfamily arXiv:1002.0211 [hep-ph]}}.

\bibitem{Ishimori:2010au}
H.~Ishimori, T.~Kobayashi, H.~Ohki, Y.~Shimizu, H.~Okada, and M.~Tanimoto,
  ``{Non-Abelian Discrete Symmetries in Particle Physics},''
  \href{http://dx.doi.org/10.1143/PTPS.183.1}{{\em Prog. Theor. Phys. Suppl.}
  {\bfseries 183} (2010) 1--163},
  \href{http://arxiv.org/abs/1003.3552}{{\ttfamily arXiv:1003.3552 [hep-th]}}.

\bibitem{King:2013eh}
S.~F. King and C.~Luhn, ``{Neutrino Mass and Mixing with Discrete Symmetry},''
  \href{http://dx.doi.org/10.1088/0034-4885/76/5/056201}{{\em Rept. Prog.
  Phys.} {\bfseries 76} (2013) 056201},
  \href{http://arxiv.org/abs/1301.1340}{{\ttfamily arXiv:1301.1340 [hep-ph]}}.

\bibitem{King:2014nza}
S.~F. King, A.~Merle, S.~Morisi, Y.~Shimizu, and M.~Tanimoto, ``{Neutrino Mass
  and Mixing: from Theory to Experiment},''
  \href{http://dx.doi.org/10.1088/1367-2630/16/4/045018}{{\em New J. Phys.}
  {\bfseries 16} (2014) 045018},
  \href{http://arxiv.org/abs/1402.4271}{{\ttfamily arXiv:1402.4271 [hep-ph]}}.

\bibitem{King:2017guk}
S.~F. King, ``{Unified Models of Neutrinos, Flavour and CP Violation},''
  \href{http://dx.doi.org/10.1016/j.ppnp.2017.01.003}{{\em Prog. Part. Nucl.
  Phys.} {\bfseries 94} (2017) 217--256},
  \href{http://arxiv.org/abs/1701.04413}{{\ttfamily arXiv:1701.04413
  [hep-ph]}}.

\bibitem{Petcov:2017ggy}
S.~T. Petcov, ``{Discrete Flavour Symmetries, Neutrino Mixing and Leptonic CP
  Violation},'' \href{http://dx.doi.org/10.1140/epjc/s10052-018-6158-5}{{\em
  Eur. Phys. J. C} {\bfseries 78} no.~9, (2018) 709},
  \href{http://arxiv.org/abs/1711.10806}{{\ttfamily arXiv:1711.10806
  [hep-ph]}}.

\bibitem{Xing:2020ijf}
Z.-z. Xing, ``{Flavor structures of charged fermions and massive neutrinos},''
  \href{http://dx.doi.org/10.1016/j.physrep.2020.02.001}{{\em Phys. Rept.}
  {\bfseries 854} (2020) 1--147},
  \href{http://arxiv.org/abs/1909.09610}{{\ttfamily arXiv:1909.09610
  [hep-ph]}}.

\bibitem{Feruglio:2019ybq}
F.~Feruglio and A.~Romanino, ``{Lepton flavor symmetries},''
  \href{http://dx.doi.org/10.1103/RevModPhys.93.015007}{{\em Rev. Mod. Phys.}
  {\bfseries 93} no.~1, (2021) 015007},
  \href{http://arxiv.org/abs/1912.06028}{{\ttfamily arXiv:1912.06028
  [hep-ph]}}.

\bibitem{Almumin:2022rml}
Y.~Almumin, M.-C. Chen, M.~Cheng, V.~Knapp-Perez, Y.~Li, A.~Mondol,
  S.~Ramos-Sanchez, M.~Ratz, and S.~Shukla, ``{Neutrino Flavor Model Building
  and the Origins of Flavor and CP Violation},''
  \href{http://dx.doi.org/10.3390/universe9120512}{{\em Universe} {\bfseries 9}
  no.~12, (2023) 512}, \href{http://arxiv.org/abs/2204.08668}{{\ttfamily
  arXiv:2204.08668 [hep-ph]}}.

\bibitem{Ding:2024ozt}
G.-J. Ding and J.~W.~F. Valle, ``{The symmetry approach to quark and lepton
  masses and mixing},''
  \href{http://dx.doi.org/10.1016/j.physrep.2024.12.005}{{\em Phys. Rept.}
  {\bfseries 1109} (2025) 1--105},
  \href{http://arxiv.org/abs/2402.16963}{{\ttfamily arXiv:2402.16963
  [hep-ph]}}.

\bibitem{Feruglio:2017spp}
F.~Feruglio, {\em {Are neutrino masses modular forms?}},
  \href{http://dx.doi.org/10.1142/9789813238053_0012}{pp.~227--266}.
\newblock 2019.
\newblock \href{http://arxiv.org/abs/1706.08749}{{\ttfamily arXiv:1706.08749
  [hep-ph]}}.

\bibitem{Ding:2023htn}
G.-J. Ding and S.~F. King, ``{Neutrino mass and mixing with modular
  symmetry},'' \href{http://dx.doi.org/10.1088/1361-6633/ad52a3}{{\em Rept.
  Prog. Phys.} {\bfseries 87} no.~8, (2024) 084201},
  \href{http://arxiv.org/abs/2311.09282}{{\ttfamily arXiv:2311.09282
  [hep-ph]}}.

\bibitem{Kobayashi:2023zzc}
T.~Kobayashi and M.~Tanimoto, ``{Modular flavor symmetric models},''
\newblock 7, 2023.
\newblock \href{http://arxiv.org/abs/2307.03384}{{\ttfamily arXiv:2307.03384
  [hep-ph]}}.

\bibitem{Liu:2019khw}
X.-G. Liu and G.-J. Ding, ``{Neutrino Masses and Mixing from Double Covering of
  Finite Modular Groups},''
  \href{http://dx.doi.org/10.1007/JHEP08(2019)134}{{\em JHEP} {\bfseries 08}
  (2019) 134}, \href{http://arxiv.org/abs/1907.01488}{{\ttfamily
  arXiv:1907.01488 [hep-ph]}}.

\bibitem{Lauer:1989ax}
J.~Lauer, J.~Mas, and H.~P. Nilles, ``{Duality and the Role of Nonperturbative
  Effects on the World Sheet},''
  \href{http://dx.doi.org/10.1016/0370-2693(89)91190-8}{{\em Phys. Lett. B}
  {\bfseries 226} (1989) 251--256}.

\bibitem{Ferrara:1989bc}
S.~Ferrara, D.~Lust, A.~D. Shapere, and S.~Theisen, ``{Modular Invariance in
  Supersymmetric Field Theories},''
  \href{http://dx.doi.org/10.1016/0370-2693(89)90583-2}{{\em Phys. Lett. B}
  {\bfseries 225} (1989) 363}.

\bibitem{Ferrara:1989qb}
S.~Ferrara, .~D. Lust, and S.~Theisen, ``{Target Space Modular Invariance and
  Low-Energy Couplings in Orbifold Compactifications},''
  \href{http://dx.doi.org/10.1016/0370-2693(89)90631-X}{{\em Phys. Lett. B}
  {\bfseries 233} (1989) 147--152}.

\bibitem{Qu:2024rns}
B.-Y. Qu and G.-J. Ding, ``{Non-holomorphic modular flavor symmetry},''
  \href{http://dx.doi.org/10.1007/JHEP08(2024)136}{{\em JHEP} {\bfseries 08}
  (2024) 136}, \href{http://arxiv.org/abs/2406.02527}{{\ttfamily
  arXiv:2406.02527 [hep-ph]}}.

\bibitem{Qu:2025ddz}
B.-Y. Qu, J.-N. Lu, and G.-J. Ding, ``{Non-holomorphic modular flavor symmetry
  and odd weight polyharmonic Maa{\ss} form},''
  \href{http://arxiv.org/abs/2506.19822}{{\ttfamily arXiv:2506.19822
  [hep-ph]}}.

\bibitem{Green:1997tv}
M.~B. Green and M.~Gutperle, ``{Effects of D instantons},''
  \href{http://dx.doi.org/10.1016/S0550-3213(97)00269-1}{{\em Nucl. Phys. B}
  {\bfseries 498} (1997) 195--227},
  \href{http://arxiv.org/abs/hep-th/9701093}{{\ttfamily arXiv:hep-th/9701093}}.

\bibitem{Green:1997me}
M.~B. Green, M.~Gutperle, and H.-h. Kwon, ``{Sixteen fermion and related terms
  in M theory on T**2},''
  \href{http://dx.doi.org/10.1016/S0370-2693(97)01551-7}{{\em Phys. Lett. B}
  {\bfseries 421} (1998) 149--161},
  \href{http://arxiv.org/abs/hep-th/9710151}{{\ttfamily arXiv:hep-th/9710151}}.

\bibitem{Pioline:1998mn}
B.~Pioline, ``{A Note on nonperturbative R**4 couplings},''
  \href{http://dx.doi.org/10.1016/S0370-2693(98)00554-1}{{\em Phys. Lett. B}
  {\bfseries 431} (1998) 73--76},
  \href{http://arxiv.org/abs/hep-th/9804023}{{\ttfamily arXiv:hep-th/9804023}}.

\bibitem{Green:1998by}
M.~B. Green and S.~Sethi, ``{Supersymmetry constraints on type IIB
  supergravity},'' \href{http://dx.doi.org/10.1103/PhysRevD.59.046006}{{\em
  Phys. Rev. D} {\bfseries 59} (1999) 046006},
  \href{http://arxiv.org/abs/hep-th/9808061}{{\ttfamily arXiv:hep-th/9808061}}.

\bibitem{deHaro:2002vk}
S.~de~Haro, A.~Sinkovics, and K.~Skenderis, ``{On a supersymmetric completion
  of the R4 term in 2B supergravity},''
  \href{http://dx.doi.org/10.1103/PhysRevD.67.084010}{{\em Phys. Rev. D}
  {\bfseries 67} (2003) 084010},
  \href{http://arxiv.org/abs/hep-th/0210080}{{\ttfamily arXiv:hep-th/0210080}}.

\bibitem{Green:2010wi}
M.~B. Green, J.~G. Russo, and P.~Vanhove, ``{Automorphic properties of low
  energy string amplitudes in various dimensions},''
  \href{http://dx.doi.org/10.1103/PhysRevD.81.086008}{{\em Phys. Rev. D}
  {\bfseries 81} (2010) 086008},
  \href{http://arxiv.org/abs/1001.2535}{{\ttfamily arXiv:1001.2535 [hep-th]}}.

\bibitem{Basu:2011he}
A.~Basu, ``{Supersymmetry constraints on the $R^4$ multiplet in type IIB on
  $T^2$},'' \href{http://dx.doi.org/10.1088/0264-9381/28/22/225018}{{\em Class.
  Quant. Grav.} {\bfseries 28} (2011) 225018},
  \href{http://arxiv.org/abs/1107.3353}{{\ttfamily arXiv:1107.3353 [hep-th]}}.

\bibitem{Peeters:2000qj}
K.~Peeters, P.~Vanhove, and A.~Westerberg, ``{Supersymmetric higher derivative
  actions in ten-dimensions and eleven-dimensions, the associated superalgebras
  and their formulation in superspace},''
  \href{http://dx.doi.org/10.1088/0264-9381/18/5/307}{{\em Class. Quant. Grav.}
  {\bfseries 18} (2001) 843--890},
  \href{http://arxiv.org/abs/hep-th/0010167}{{\ttfamily arXiv:hep-th/0010167}}.

\bibitem{Sinha:2002zr}
A.~Sinha, ``{The G(hat)**4 lambda**16 term in IIB supergravity},''
  \href{http://dx.doi.org/10.1088/1126-6708/2002/08/017}{{\em JHEP} {\bfseries
  08} (2002) 017}, \href{http://arxiv.org/abs/hep-th/0207070}{{\ttfamily
  arXiv:hep-th/0207070}}.

\bibitem{Okada:2025jjo}
H.~Okada and Y.~Orikasa, ``{A radiative seesaw in a non-holomorphic modular
  $S_3$ flavor symmetry},'' \href{http://arxiv.org/abs/2501.15748}{{\ttfamily
  arXiv:2501.15748 [hep-ph]}}.

\bibitem{Kumar:2024uxn}
B.~Kumar and M.~K. Das, ``{Study of neutrino phenomenology and
  0{\ensuremath{\nu}}{\ensuremath{\beta}}{\ensuremath{\beta}} decay using
  polyharmonic Maa{\ensuremath{\beta}} forms},''
  \href{http://dx.doi.org/10.1142/S0217751X25500903}{{\em Int. J. Mod. Phys. A}
  {\bfseries 40} no.~23, (2025) 2550090},
  \href{http://arxiv.org/abs/2405.10586}{{\ttfamily arXiv:2405.10586
  [hep-ph]}}.

\bibitem{Nomura:2024atp}
T.~Nomura and H.~Okada, ``{Type-II seesaw of a non-holomorphic modular $A_4$
  symmetry},'' \href{http://arxiv.org/abs/2408.01143}{{\ttfamily
  arXiv:2408.01143 [hep-ph]}}.

\bibitem{Nomura:2024nwh}
T.~Nomura and H.~Okada, ``{Zee model in a non-holomorphic modular A4
  symmetry},'' \href{http://dx.doi.org/10.1016/j.physletb.2025.139618}{{\em
  Phys. Lett. B} {\bfseries 867} (2025) 139618},
  \href{http://arxiv.org/abs/2412.18095}{{\ttfamily arXiv:2412.18095
  [hep-ph]}}.

\bibitem{Kobayashi:2025hnc}
T.~Kobayashi, H.~Okada, and Y.~Orikasa, ``{Zee-Babu model in a non-holomorphic
  modular $A_4$ symmetry and modular stabilization},''
  \href{http://arxiv.org/abs/2502.12662}{{\ttfamily arXiv:2502.12662
  [hep-ph]}}.

\bibitem{Loualidi:2025tgw}
M.~A. Loualidi, M.~Miskaoui, and S.~Nasri, ``{Nonholomorphic A4 modular
  invariance for fermion masses and mixing in SU(5) GUT},''
  \href{http://dx.doi.org/10.1103/1py2-cmfx}{{\em Phys. Rev. D} {\bfseries 112}
  no.~1, (2025) 015008}, \href{http://arxiv.org/abs/2503.12594}{{\ttfamily
  arXiv:2503.12594 [hep-ph]}}.

\bibitem{Kumar:2025bfe}
B.~Kumar and M.~K. Das, ``{Leptogenesis,
  0{\ensuremath{\nu}}{\ensuremath{\beta}}{\ensuremath{\beta}} and lepton flavor
  violation in modular left-right asymmetric model with polyharmonic Maa{\ss}
  forms},'' \href{http://dx.doi.org/10.1007/JHEP09(2025)071}{{\em JHEP}
  {\bfseries 09} (2025) 071}, \href{http://arxiv.org/abs/2504.21701}{{\ttfamily
  arXiv:2504.21701 [hep-ph]}}.

\bibitem{Nomura:2025ovm}
T.~Nomura, H.~Okada, and X.-Y. Wang, ``{A radiative neutrino mass model with
  leptoquarks under non-holomorphic modular A$_{4}$ symmetry},''
  \href{http://dx.doi.org/10.1007/JHEP09(2025)163}{{\em JHEP} {\bfseries 09}
  (2025) 163}, \href{http://arxiv.org/abs/2504.21404}{{\ttfamily
  arXiv:2504.21404 [hep-ph]}}.

\bibitem{Nomura:2025raf}
T.~Nomura and H.~Okada, ``{Neutrino mass model at a three-loop level from a
  non-holomorphic modular $A_4$ symmetry},''
  \href{http://arxiv.org/abs/2506.02639}{{\ttfamily arXiv:2506.02639
  [hep-ph]}}.

\bibitem{Zhang:2025dsa}
X.~Zhang and Y.~Reyimuaji, ``{Inverse seesaw model in nonholomorphic modular A4
  flavor symmetry},'' \href{http://dx.doi.org/10.1103/17p3-bw5r}{{\em Phys.
  Rev. D} {\bfseries 112} no.~7, (2025) 075050},
  \href{http://arxiv.org/abs/2507.06945}{{\ttfamily arXiv:2507.06945
  [hep-ph]}}.

\bibitem{Priya:2025wdm}
Priya, L.~Singh, B.~C. Chauhan, and S.~Verma, ``{Type-III Seesaw in
  Non-Holomorphic Modular Symmetry and Leptogenesis},''
  \href{http://arxiv.org/abs/2508.05047}{{\ttfamily arXiv:2508.05047
  [hep-ph]}}.

\bibitem{Kumar:2025nut}
B.~Kumar and M.~K. Das, ``{Neutrino phenomenology and Dark matter in a
  left-right asymmetric model with non-holomorphic modular $A_{4}$ group},''
  \href{http://arxiv.org/abs/2509.01205}{{\ttfamily arXiv:2509.01205
  [hep-ph]}}.

\bibitem{Nanda:2025lem}
S.~K. Nanda, M.~Ricky~Devi, and S.~Patra, ``{Non-Holomorphic $A_4$ Modular
  Symmetry in Type-I Seesaw: Implications for Neutrino Masses and
  Leptogenesis},'' \href{http://arxiv.org/abs/2509.22108}{{\ttfamily
  arXiv:2509.22108 [hep-ph]}}.

\bibitem{Jangid:2025thp}
S.~Jangid and H.~Okada, ``{A radiative seesaw model in a non-invertible
  selection rule with the assistance of a non-holomorphic modular $A_4$
  symmetry},'' \href{http://arxiv.org/abs/2510.17292}{{\ttfamily
  arXiv:2510.17292 [hep-ph]}}.

\bibitem{Gao:2025jlw}
X.-Y. Gao and C.-C. Li, ``{Minimal lepton models with non-holomorphic modular A
  $_{4}$ symmetry*},'' \href{http://dx.doi.org/10.1088/1674-1137/ae3f0a}{{\em
  Chin. Phys.} {\bfseries 50} no.~5, (2026) 053109},
  \href{http://arxiv.org/abs/2512.07158}{{\ttfamily arXiv:2512.07158
  [hep-ph]}}.

\bibitem{Nasri:2026nbf}
S.~Nasri, L.~Singh, Tapender, and S.~Verma, ``{Dark-Portal Leptogenesis in a
  Non-Holomorphic Modular Scoto-Seesaw Model},''
  \href{http://arxiv.org/abs/2601.06435}{{\ttfamily arXiv:2601.06435
  [hep-ph]}}.

\bibitem{Tapender:2026ets}
Tapender and S.~Verma, ``{Tri-Resonant Leptogenesis in a Non-Holomorphic
  Modular A$_4$ Scotogenic Model},''
  \href{http://arxiv.org/abs/2602.17243}{{\ttfamily arXiv:2602.17243
  [hep-ph]}}.

\bibitem{Majhi:2026jdk}
R.~Majhi, M.~K. Behera, and R.~Mohanta, ``{A Predictive Non-Holomorphic Modular
  $A_4$ Linear Seesaw Framework Testable at DUNE},''
  \href{http://arxiv.org/abs/2602.23018}{{\ttfamily arXiv:2602.23018
  [hep-ph]}}.

\bibitem{Priya:2026ehe}
Priya, B.~C. Chauhan, D.~Kumar, and T.~Nomura, ``{Predictions of Modular
  Symmetry Fixed Points on Neutrino Masses, Mixing, and Leptogenesis},''
  \href{http://arxiv.org/abs/2604.04585}{{\ttfamily arXiv:2604.04585
  [hep-ph]}}.

\bibitem{Abbas:2026siw}
M.~Abbas, ``{Lepton masses and mixing in non-holomorphic modular $A_4$ with
  universal couplings},'' \href{http://arxiv.org/abs/2604.16130}{{\ttfamily
  arXiv:2604.16130 [hep-ph]}}.

\bibitem{Ding:2024inn}
G.-J. Ding, J.-N. Lu, S.~T. Petcov, and B.-Y. Qu, ``{Non-holomorphic modular
  S$_{4}$ lepton flavour models},''
  \href{http://dx.doi.org/10.1007/JHEP01(2025)191}{{\em JHEP} {\bfseries 01}
  (2025) 191}, \href{http://arxiv.org/abs/2408.15988}{{\ttfamily
  arXiv:2408.15988 [hep-ph]}}.

\bibitem{Li:2024svh}
C.-C. Li, J.-N. Lu, and G.-J. Ding, ``{Non-holomorphic modular A$_{5}$ symmetry
  for lepton masses and mixing},''
  \href{http://dx.doi.org/10.1007/JHEP12(2024)189}{{\em JHEP} {\bfseries 12}
  (2024) 189}, \href{http://arxiv.org/abs/2410.24103}{{\ttfamily
  arXiv:2410.24103 [hep-ph]}}.

\bibitem{Loualidi:2026pld}
M.~A. Loualidi, M.~Miskaoui, and S.~Nasri, ``{Radiative Neutrino Mass in a
  Nonholomorphic $T'$ Modular Invariant Model},''
  \href{http://arxiv.org/abs/2606.11346}{{\ttfamily arXiv:2606.11346
  [hep-ph]}}.

\bibitem{Li:2025kcr}
C.-C. Li and G.-J. Ding, ``{Lepton models from non-holomorphic
  ${A}_{5}{\prime}$ modular flavor symmetry},''
  \href{http://dx.doi.org/10.1007/JHEP01(2026)032}{{\em JHEP} {\bfseries 01}
  (2026) 032}, \href{http://arxiv.org/abs/2509.15183}{{\ttfamily
  arXiv:2509.15183 [hep-ph]}}.

\bibitem{Zhang:2026kyy}
X.~Zhang and Y.~Reyimuaji, ``{Neutrino Mass and Leptogenesis in the Non-SUSY
  Modular $A^\prime_5$ Inverse Seesaw},''
  \href{http://arxiv.org/abs/2603.19104}{{\ttfamily arXiv:2603.19104
  [hep-ph]}}.

\bibitem{Novichkov:2020eep}
P.~P. Novichkov, J.~T. Penedo, and S.~T. Petcov, ``{Double cover of modular
  $S_4$ for flavour model building},''
  \href{http://dx.doi.org/10.1016/j.nuclphysb.2020.115301}{{\em Nucl. Phys. B}
  {\bfseries 963} (2021) 115301},
  \href{http://arxiv.org/abs/2006.03058}{{\ttfamily arXiv:2006.03058
  [hep-ph]}}.

\bibitem{Liu:2020akv}
X.-G. Liu, C.-Y. Yao, and G.-J. Ding, ``{Modular invariant quark and lepton
  models in double covering of $S_4$ modular group},''
  \href{http://dx.doi.org/10.1103/PhysRevD.103.056013}{{\em Phys. Rev. D}
  {\bfseries 103} no.~5, (2021) 056013},
  \href{http://arxiv.org/abs/2006.10722}{{\ttfamily arXiv:2006.10722
  [hep-ph]}}.

\bibitem{Ding:2022nzn}
G.-J. Ding, X.-G. Liu, and C.-Y. Yao, ``{A minimal modular invariant neutrino
  model},'' \href{http://dx.doi.org/10.1007/JHEP01(2023)125}{{\em JHEP}
  {\bfseries 01} (2023) 125}, \href{http://arxiv.org/abs/2211.04546}{{\ttfamily
  arXiv:2211.04546 [hep-ph]}}.

\bibitem{Abe:2023ilq}
Y.~Abe, T.~Higaki, J.~Kawamura, and T.~Kobayashi, ``{Quark masses and CKM
  hierarchies from $S_4'$ modular flavor symmetry},''
  \href{http://dx.doi.org/10.1140/epjc/s10052-023-12303-2}{{\em Eur. Phys. J.
  C} {\bfseries 83} no.~12, (2023) 1140},
  \href{http://arxiv.org/abs/2301.07439}{{\ttfamily arXiv:2301.07439
  [hep-ph]}}.

\bibitem{Abe:2023qmr}
Y.~Abe, T.~Higaki, J.~Kawamura, and T.~Kobayashi, ``{Quark and lepton
  hierarchies from S4' modular flavor symmetry},''
  \href{http://dx.doi.org/10.1016/j.physletb.2023.137977}{{\em Phys. Lett. B}
  {\bfseries 842} (2023) 137977},
  \href{http://arxiv.org/abs/2302.11183}{{\ttfamily arXiv:2302.11183
  [hep-ph]}}.

\bibitem{Petcov:2026mdx}
S.~T. Petcov and M.~Tanimoto, ``{$S^\prime_4$ Quark Flavour Model in the
  Vicinity of the Fixed Point $\tau= i\infty$},''
  \href{http://arxiv.org/abs/2601.04529}{{\ttfamily arXiv:2601.04529
  [hep-ph]}}.

\bibitem{Fukugita:1986hr}
M.~Fukugita and T.~Yanagida, ``{Baryogenesis Without Grand Unification},''
  \href{http://dx.doi.org/10.1016/0370-2693(86)91126-3}{{\em Phys. Lett. B}
  {\bfseries 174} (1986) 45--47}.

\bibitem{Planck:2018vyg}
{\bfseries Planck} Collaboration, N.~Aghanim {\em et~al.}, ``{Planck 2018
  results. VI. Cosmological parameters},''
  \href{http://dx.doi.org/10.1051/0004-6361/201833910}{{\em Astron. Astrophys.}
  {\bfseries 641} (2020) A6}, \href{http://arxiv.org/abs/1807.06209}{{\ttfamily
  arXiv:1807.06209 [astro-ph.CO]}}. [Erratum: Astron.Astrophys. 652, C4
  (2021)].

\bibitem{Diamond:2005afc}
F.~Diamond and J.~Shurman, {\em A First Course in Modular Forms}.
\newblock Springer, 2005.

\bibitem{Esteban:2024eli}
I.~Esteban, M.~C. Gonzalez-Garcia, M.~Maltoni, I.~Martinez-Soler, J.~a.~P.
  Pinheiro, and T.~Schwetz, ``{NuFit-6.0: Updated global analysis of
  three-flavor neutrino oscillations},''
  \href{http://arxiv.org/abs/2410.05380}{{\ttfamily arXiv:2410.05380
  [hep-ph]}}.

\bibitem{Xing:2007fb}
Z.-z. Xing, H.~Zhang, and S.~Zhou, ``{Updated Values of Running Quark and
  Lepton Masses},'' \href{http://dx.doi.org/10.1103/PhysRevD.77.113016}{{\em
  Phys. Rev. D} {\bfseries 77} (2008) 113016},
  \href{http://arxiv.org/abs/0712.1419}{{\ttfamily arXiv:0712.1419 [hep-ph]}}.

\bibitem{Bilenky:1980cx}
S.~M. Bilenky, J.~Hosek, and S.~T. Petcov, ``{On Oscillations of Neutrinos with
  Dirac and Majorana Masses},''
  \href{http://dx.doi.org/10.1016/0370-2693(80)90927-2}{{\em Phys. Lett. B}
  {\bfseries 94} (1980) 495--498}.

\bibitem{King:2020qaj}
S.~J.~D. King and S.~F. King, ``{Fermion mass hierarchies from modular
  symmetry},'' \href{http://dx.doi.org/10.1007/JHEP09(2020)043}{{\em JHEP}
  {\bfseries 09} (2020) 043}, \href{http://arxiv.org/abs/2002.00969}{{\ttfamily
  arXiv:2002.00969 [hep-ph]}}.

\bibitem{Ding:2025mar}
G.-J. Ding, S.~F. King, J.-N. Lu, and M.-H. Weng, ``{Modular symmetry with
  weighton},'' \href{http://dx.doi.org/10.1007/JHEP10(2025)028}{{\em JHEP}
  {\bfseries 10} (2025) 028}, \href{http://arxiv.org/abs/2505.12916}{{\ttfamily
  arXiv:2505.12916 [hep-ph]}}.

\bibitem{Feroz:2007kg}
F.~Feroz and M.~P. Hobson, ``{Multimodal nested sampling: an efficient and
  robust alternative to MCMC methods for astronomical data analysis},''
  \href{http://dx.doi.org/10.1111/j.1365-2966.2007.12353.x}{{\em Mon. Not. Roy.
  Astron. Soc.} {\bfseries 384} (2008) 449},
\href{http://arxiv.org/abs/0704.3704}{{\ttfamily arXiv:0704.3704 [astro-ph]}}.

\bibitem{Feroz:2008xx}
F.~Feroz, M.~P. Hobson, and M.~Bridges, ``{MultiNest: an efficient and robust
  Bayesian inference tool for cosmology and particle physics},''
  \href{http://dx.doi.org/10.1111/j.1365-2966.2009.14548.x}{{\em Mon. Not. Roy.
  Astron. Soc.} {\bfseries 398} (2009) 1601--1614},
\href{http://arxiv.org/abs/0809.3437}{{\ttfamily arXiv:0809.3437 [astro-ph]}}.

\bibitem{DUNE:2020ypp}
{\bfseries DUNE} Collaboration, B.~Abi {\em et~al.}, ``{Deep Underground
  Neutrino Experiment (DUNE), Far Detector Technical Design Report, Volume II:
  DUNE Physics},'' \href{http://arxiv.org/abs/2002.03005}{{\ttfamily
  arXiv:2002.03005 [hep-ex]}}.

\bibitem{Hyper-KamiokandeProto-:2015xww}
{\bfseries Hyper-Kamiokande Proto-} Collaboration, K.~Abe {\em et~al.},
  ``{Physics potential of a long-baseline neutrino oscillation experiment using
  a J-PARC neutrino beam and Hyper-Kamiokande},''
  \href{http://dx.doi.org/10.1093/ptep/ptv061}{{\em PTEP} {\bfseries 2015}
  (2015) 053C02}, \href{http://arxiv.org/abs/1502.05199}{{\ttfamily
  arXiv:1502.05199 [hep-ex]}}.

\bibitem{Hyper-Kamiokande:2018ofw}
{\bfseries Hyper-Kamiokande} Collaboration, K.~Abe {\em et~al.},
  ``{Hyper-Kamiokande Design Report},''
  \href{http://arxiv.org/abs/1805.04163}{{\ttfamily arXiv:1805.04163
  [physics.ins-det]}}.

\bibitem{Ballett:2016daj}
P.~Ballett, S.~F. King, S.~Pascoli, N.~W. Prouse, and T.~Wang, ``{Sensitivities
  and synergies of DUNE and T2HK},''
  \href{http://dx.doi.org/10.1103/PhysRevD.96.033003}{{\em Phys. Rev. D}
  {\bfseries 96} no.~3, (2017) 033003},
  \href{http://arxiv.org/abs/1612.07275}{{\ttfamily arXiv:1612.07275
  [hep-ph]}}.

\bibitem{JUNO:2025gmd}
{\bfseries JUNO} Collaboration, A.~Abusleme {\em et~al.}, ``{First measurement
  of reactor neutrino oscillations at JUNO},''
  \href{http://arxiv.org/abs/2511.14593}{{\ttfamily arXiv:2511.14593
  [hep-ex]}}.

\bibitem{JUNO:2022mxj}
{\bfseries JUNO} Collaboration, A.~Abusleme {\em et~al.}, ``{Sub-percent
  precision measurement of neutrino oscillation parameters with JUNO},''
  \href{http://dx.doi.org/10.1088/1674-1137/ac8bc9}{{\em Chin. Phys. C}
  {\bfseries 46} no.~12, (2022) 123001},
  \href{http://arxiv.org/abs/2204.13249}{{\ttfamily arXiv:2204.13249
  [hep-ex]}}.

\bibitem{Euclid:2024imf}
{\bfseries Euclid} Collaboration, M.~Archidiacono {\em et~al.}, ``{Euclid
  preparation. Sensitivity to neutrino parameters},''
  \href{http://arxiv.org/abs/2405.06047}{{\ttfamily arXiv:2405.06047
  [astro-ph.CO]}}.

\bibitem{KamLAND-Zen:2024eml}
{\bfseries KamLAND-Zen} Collaboration, S.~Abe {\em et~al.}, ``{Search for
  Majorana Neutrinos with the Complete KamLAND-Zen Dataset},''
  \href{http://arxiv.org/abs/2406.11438}{{\ttfamily arXiv:2406.11438
  [hep-ex]}}.

\bibitem{LEGEND:2021bnm}
{\bfseries LEGEND} Collaboration, N.~Abgrall {\em et~al.}, ``{The Large
  Enriched Germanium Experiment for Neutrinoless $\beta\beta$ Decay}:
  {LEGEND-1000 Preconceptual Design Report},''
  \href{http://arxiv.org/abs/2107.11462}{{\ttfamily arXiv:2107.11462
  [physics.ins-det]}}.

\bibitem{nEXO:2021ujk}
{\bfseries nEXO} Collaboration, G.~Adhikari {\em et~al.}, ``{nEXO: neutrinoless
  double beta decay search beyond 10$^{28}$ year half-life sensitivity},''
  \href{http://dx.doi.org/10.1088/1361-6471/ac3631}{{\em J. Phys. G} {\bfseries
  49} no.~1, (2022) 015104}, \href{http://arxiv.org/abs/2106.16243}{{\ttfamily
  arXiv:2106.16243 [nucl-ex]}}.

\bibitem{KATRIN:2024cdt}
{\bfseries KATRIN} Collaboration, M.~Aker {\em et~al.}, ``{Direct neutrino-mass
  measurement based on 259 days of KATRIN data},''
  \href{http://dx.doi.org/10.1126/science.adq9592}{{\em Science} {\bfseries
  388} no.~6743, (2025) adq9592},
  \href{http://arxiv.org/abs/2406.13516}{{\ttfamily arXiv:2406.13516
  [nucl-ex]}}.

\bibitem{KATRIN:2021dfa}
{\bfseries KATRIN} Collaboration, M.~Aker {\em et~al.}, ``{The design,
  construction, and commissioning of the KATRIN experiment},''
  \href{http://dx.doi.org/10.1088/1748-0221/16/08/T08015}{{\em JINST}
  {\bfseries 16} no.~08, (2021) T08015},
  \href{http://arxiv.org/abs/2103.04755}{{\ttfamily arXiv:2103.04755
  [physics.ins-det]}}.

\bibitem{Project8:2022wqh}
{\bfseries Project 8} Collaboration, A.~A. Esfahani {\em et~al.}, ``{The
  Project 8 Neutrino Mass Experiment},'' in {\em {Snowmass 2021}}.
\newblock 3, 2022.
\newblock \href{http://arxiv.org/abs/2203.07349}{{\ttfamily arXiv:2203.07349
  [nucl-ex]}}.

\bibitem{Klinkhamer:1984di}
F.~R. Klinkhamer and N.~S. Manton, ``{A Saddle Point Solution in the
  Weinberg-Salam Theory},''
  \href{http://dx.doi.org/10.1103/PhysRevD.30.2212}{{\em Phys. Rev. D}
  {\bfseries 30} (1984) 2212}.

\bibitem{Arnold:1987mh}
P.~B. Arnold and L.~D. McLerran, ``{Sphalerons, Small Fluctuations and Baryon
  Number Violation in Electroweak Theory},''
  \href{http://dx.doi.org/10.1103/PhysRevD.36.581}{{\em Phys. Rev. D}
  {\bfseries 36} (1987) 581}.

\bibitem{Kuzmin:1985mm}
V.~A. Kuzmin, V.~A. Rubakov, and M.~E. Shaposhnikov, ``{On the Anomalous
  Electroweak Baryon Number Nonconservation in the Early Universe},''
  \href{http://dx.doi.org/10.1016/0370-2693(85)91028-7}{{\em Phys. Lett. B}
  {\bfseries 155} (1985) 36}.

\bibitem{Rubakov:1996vz}
V.~A. Rubakov and M.~E. Shaposhnikov, ``{Electroweak baryon number
  nonconservation in the early universe and in high-energy collisions},''
  \href{http://dx.doi.org/10.1070/PU1996v039n05ABEH000145}{{\em Usp. Fiz. Nauk}
  {\bfseries 166} (1996) 493--537},
  \href{http://arxiv.org/abs/hep-ph/9603208}{{\ttfamily arXiv:hep-ph/9603208}}.

\bibitem{Abada:2006ea}
A.~Abada, S.~Davidson, A.~Ibarra, F.~X. Josse-Michaux, M.~Losada, and
  A.~Riotto, ``{Flavour Matters in Leptogenesis},''
  \href{http://dx.doi.org/10.1088/1126-6708/2006/09/010}{{\em JHEP} {\bfseries
  09} (2006) 010}, \href{http://arxiv.org/abs/hep-ph/0605281}{{\ttfamily
  arXiv:hep-ph/0605281}}.

\bibitem{Abada:2006fw}
A.~Abada, S.~Davidson, F.-X. Josse-Michaux, M.~Losada, and A.~Riotto, ``{Flavor
  issues in leptogenesis},''
  \href{http://dx.doi.org/10.1088/1475-7516/2006/04/004}{{\em JCAP} {\bfseries
  04} (2006) 004}, \href{http://arxiv.org/abs/hep-ph/0601083}{{\ttfamily
  arXiv:hep-ph/0601083}}.

\bibitem{Nardi:2006fx}
E.~Nardi, Y.~Nir, E.~Roulet, and J.~Racker, ``{The Importance of flavor in
  leptogenesis},'' \href{http://dx.doi.org/10.1088/1126-6708/2006/01/164}{{\em
  JHEP} {\bfseries 01} (2006) 164},
  \href{http://arxiv.org/abs/hep-ph/0601084}{{\ttfamily arXiv:hep-ph/0601084}}.

\bibitem{Antusch:2006cw}
S.~Antusch, S.~F. King, and A.~Riotto, ``{Flavour-Dependent Leptogenesis with
  Sequential Dominance},''
  \href{http://dx.doi.org/10.1088/1475-7516/2006/11/011}{{\em JCAP} {\bfseries
  11} (2006) 011}, \href{http://arxiv.org/abs/hep-ph/0609038}{{\ttfamily
  arXiv:hep-ph/0609038}}.

\bibitem{Davidson:2008bu}
S.~Davidson, E.~Nardi, and Y.~Nir, ``{Leptogenesis},''
  \href{http://dx.doi.org/10.1016/j.physrep.2008.06.002}{{\em Phys. Rept.}
  {\bfseries 466} (2008) 105--177},
  \href{http://arxiv.org/abs/0802.2962}{{\ttfamily arXiv:0802.2962 [hep-ph]}}.

\bibitem{Covi:1996wh}
L.~Covi, E.~Roulet, and F.~Vissani, ``{CP violating decays in leptogenesis
  scenarios},'' \href{http://dx.doi.org/10.1016/0370-2693(96)00817-9}{{\em
  Phys. Lett. B} {\bfseries 384} (1996) 169--174},
  \href{http://arxiv.org/abs/hep-ph/9605319}{{\ttfamily arXiv:hep-ph/9605319}}.

\bibitem{Buchmuller:2004nz}
W.~Buchmuller, P.~Di~Bari, and M.~Plumacher, ``{Leptogenesis for
  pedestrians},'' \href{http://dx.doi.org/10.1016/j.aop.2004.02.003}{{\em
  Annals Phys.} {\bfseries 315} (2005) 305--351},
  \href{http://arxiv.org/abs/hep-ph/0401240}{{\ttfamily arXiv:hep-ph/0401240}}.

\bibitem{Buchmuller:2005eh}
W.~Buchmuller, R.~D. Peccei, and T.~Yanagida, ``{Leptogenesis as the origin of
  matter},''
  \href{http://dx.doi.org/10.1146/annurev.nucl.55.090704.151558}{{\em Ann. Rev.
  Nucl. Part. Sci.} {\bfseries 55} (2005) 311--355},
  \href{http://arxiv.org/abs/hep-ph/0502169}{{\ttfamily arXiv:hep-ph/0502169}}.

\bibitem{Giudice:2003jh}
G.~F. Giudice, A.~Notari, M.~Raidal, A.~Riotto, and A.~Strumia, ``{Towards a
  complete theory of thermal leptogenesis in the SM and MSSM},''
  \href{http://dx.doi.org/10.1016/j.nuclphysb.2004.02.019}{{\em Nucl. Phys. B}
  {\bfseries 685} (2004) 89--149},
  \href{http://arxiv.org/abs/hep-ph/0310123}{{\ttfamily arXiv:hep-ph/0310123}}.

\end{thebibliography}

\providecommand{\href}[2]{#2}\begingroup\raggedright\endgroup

\end{document}